\documentclass[11pt,draftcls,onecolumn]{IEEEtran}

\IEEEoverridecommandlockouts

\usepackage{amsmath,amsfonts,amssymb,amsthm}
\usepackage{mathrsfs}
\usepackage{comment,blkarray}
\usepackage{multirow,bigdelim}
\usepackage{mathrsfs}
\usepackage{cite}
\usepackage[ruled,commentsnumbered, vlined]{algorithm2e}
\usepackage{tikz}
\usepackage{verbatim}
\usepackage{graphicx}
\usepackage{subfigure}		
\usepackage{booktabs}	
\usepackage{xcolor}
\usepackage{colortbl}													
\usepackage{multirow}
\usepackage{CJK}
\usepackage{indentfirst}
\usepackage{cases}	

\usepackage[justification=centering]{caption}

\def\BibTeX{{\rm B\kern-.05em{\sc i\kern-.025em b}\kern-.08em
    T\kern-.1667em\lower.7ex\hbox{E}\kern-.125emX}}								

\includecomment{itwfull}
\excludecomment{removeEX4}
\excludecomment{itw2016}
\excludecomment{journalonly}

\usepackage{url}
\usepackage{subfigure}
\usepackage{multicol}

\usepackage{ifpdf}															%
\ifpdf																		%
	\usepackage{hyperref}
\else																		%
\fi	
\usetikzlibrary{arrows,decorations.pathmorphing,decorations.footprints,fadings,calc,
trees,mindmap,shadows,decorations.text,patterns,positioning,shapes,matrix,fit}

\makeatletter

\newcommand{\Rmnum}[1]{\expandafter\@slowromancap\romannumeral #1@}

\newif\if@borderstar
\def\bordermatrix{\@ifnextchar*{%
  \@borderstartrue\@bordermatrix@i}{\@borderstarfalse\@bordermatrix@i*}%
}
\def\@bordermatrix@i*{\@ifnextchar[{\@bordermatrix@ii}{\@bordermatrix@ii[()]}}
\def\@bordermatrix@ii[#1]#2{%
\begingroup
  \m@th\@tempdima8.75\p@\setbox\z@\vbox{%
    \def\cr{\crcr\noalign{\kern 2\p@\global\let\cr\endline }}%
    \ialign {$##$\hfil\kern 2\p@\kern\@tempdima & \thinspace %
    \hfil $##$\hfil && \quad\hfil $##$\hfil\crcr\omit\strut %
    \hfil\crcr\noalign{\kern -\baselineskip}#2\crcr\omit %
    \strut\cr}}%
  \setbox\tw@\vbox{\unvcopy\z@\global\setbox\@ne\lastbox}%
  \setbox\tw@\hbox{\unhbox\@ne\unskip\global\setbox\@ne\lastbox}%
  \setbox\tw@\hbox{%
    $\kern\wd\@ne\kern -\@tempdima\left\@firstoftwo#1%
    \if@borderstar\kern2pt\else\kern -\wd\@ne\fi%
    \global\setbox\@ne\vbox{\box\@ne\if@borderstar\else\kern 2\p@\fi}%
    \vcenter{\if@borderstar\else\kern -\ht\@ne\fi%
    \unvbox\z@\kern -\if@borderstar2\fi\baselineskip}%
    \if@borderstar\kern -2\@tempdima\kern2\p@\else\,\fi\right\@secondoftwo#1 $%
  }\null \;\vbox{\kern\ht\@ne\box\tw@}%
\endgroup
}
\makeatother

\newtheorem{theorem}{Theorem}
\newtheorem{lemma}{Lemma}
\newtheorem{cor}{Corollary}

\newtheorem{prop}{Proposition}
\newtheorem{example}{Example}
\newtheorem{definition}{Definition}

\allowdisplaybreaks

\newcommand{\mA}{\mathcal{A}}

\newcommand{\mN}{\mathcal{N}}

\newcommand{\mE}{\mathcal{E}}

\newcommand{\mP}{\mathcal{P}}
\newcommand{\mX}{\mathcal{X}}

\newcommand{\mbC}{\mathbf{C}}
\newcommand{\cl}{{\mathrm{cl}}}
\newcommand{\Cl}{{\mathrm{Cl}}}

\def\Im{\mathrm{Im}}

\usetikzlibrary{decorations.markings}

\newcommand{\adjustedbar}[1]{%
  \overline{\mkern-0.9mu #1 \mkern-0.9mu}
}

\tikzstyle{sum} = [draw, fill=blue!20, circle, node distance=1cm]
\tikzstyle{input} = [coordinate]
\tikzstyle{output} = [coordinate]
\tikzstyle{pinstyle} = [pin edge={to-,thin,black}]

\tikzstyle{vertex}=[draw,circle,fill=gray!30,minimum size=6pt, inner sep=0pt]

\begin{document}
%
\title{Uniquely-Decodable Coding for  \\ Zero-Error Network Function Computation}
\author{Xuan~Guang,~Jihang~Yang,~and~Ruze~Zhang
}

\maketitle

\begin{abstract}
We consider
uniquely-decodable coding for zero-error network function computation, where in  a directed  acyclic graph, the single sink node
is required to compute with zero error  a target function  multiple times, whose arguments are the information sources   generated at a set of source nodes.
We are  interested in the  \emph{computing capacity} from the information theoretic point of view, which is
 defined as the  infimum of the maximum expected number of bits transmitted on all the edges
 for computing the target function once on average.
We first prove some new results on clique entropy, in particular, the substitution lemma of clique entropy for probabilistic graphs with a certain condition.
 With them, we prove a  lower bound on the computing capacity associated with  clique entropies of the induced characteristic graphs, where the obtained lower bound is applicable to  arbitrary network topologies, arbitrary  information sources, and arbitrary target functions.
By refining the probability distribution of information sources, we further strictly improve the obtained lower bound. In addition, we  compare uniquely-decodable network function-computing coding and fixed-length network function-computing coding, and show that the former indeed outperforms the latter in terms of the computing capacity. Therein, we provide a novel graph-theoretic explanation of the key parameter in the best known bound on the computing capacity for fixed-length network function-computing codes, which would be helpful to improve the existing results.
\end{abstract}

\section{Introduction}\label{Sec:Introduction}

In this paper, we investigate uniquely-decodable coding for {network function computation}, in which the following model is considered. In  a directed  acyclic graph $\mathcal{G}$, the single sink node $\rho$
is required to compute with zero error  a target function $f$ multiple times, whose arguments are the information sources~$X_S$ generated at a set of source nodes $S$.
The graph~$\mathcal{G}$, together with $S$ and $\rho$, forms a
\emph{network~$\mathcal{N}$},  and we use the triple $(\mathcal{N},X_S,f)$ to denote this model of network function computation. We consider uniquely-decodable network function-computing codes for the model $(\mathcal{N},X_S,f)$, where for such a code, the image set of the encoding function for each edge is a uniquely-decodable code. From the information theoretic point of view, we are
 interested in characterizing the  \emph{computing capacity} for $(\mathcal{N},X_S,f)$, which is
 defined as the  infimum of the maximum expected number of bits transmitted on all the edges
 for computing the target function once on average. This computing capacity  measures
 the efficiency of computing the target function over the network in the average case.
 Rather than the fixed-length network function-computing codes for the model $(\mathcal{N},X_S,f)$ widely investigated
 in the literature, e.g.,~\cite{Huang_Comment_cut_set_bound,
Appuswamy13,Ramamoorthy-Langberg-JSAC13-sum-networks,
Appuswamy14,Tripathy-Ramamoorthy-IT18-sum-networks,Appuswamy11,
Guang_Improved_upper_bound,Kowshik12,Li_Xu_vector_linear_diamond,
Yao-Jafar-3user,Guang_Zhang_Arithmetic_sum_Sel_Areas}, the corresponding computing capacity, namely that the infimum of the number  of bits transmitted on all the edges
 for computing the target function once on average,  measures
 the efficiency of computing the target function over the network in the worst case.\footnote{Clearly, the  fixed-length network function-computing codes  can be regarded as  a special type of
  uniquely-decodable network function-computing codes.}
  In general, uniquely-decodable coding outperforms fixed-length coding in terms of capacity.

 %
%
%
%
%

\subsection{Related Works}

Source coding problem with uniquely-decodable  codes was launched by   Shannon in his celebrated work~\cite{Shannon48}, in which  a uniquely-decodable source code, called {\em the Shannon Code}, was given to compress an information source. Multi-terminal source coding problems with uniquely-decodable  codes were also  studied  in the literature.
Alon and Orlitsky~\cite{Alon-Orlitskx_source_cod_graph_entropies} studied  the  source coding problem with side information at the decoder, where the encoder compresses a source~$X$ by a uniquely-decodable code; and the decoder, which not only receives the output of the encoder but also observes another source  $Y$ correlated with~$X$, is required to reconstruct $X$ with zero error.  Alon and Orlitsky in the paper developed a graph coloring approach to lower and upper bound the  infimum coding rate for uniquely-decodable codes and proved some interesting results  on chromatic entropy, graph entropy and clique entropy.
 The zero-error distributed source coding problem with uniquely-decodable codes was considered by Koulgi \textit{et al}.~\cite{Koulgi_zero-error-cod_cor_inf_sour}, in which two correlated sources are compressed separately by two encoders using uniquely-decodable codes; and the decoder  receives the outputs of both  encoders to reconstruct with zero error the two sources.

As a special type of uniquely-decodable coding, fixed-length coding was studied for a plenty of  information-theoretic problems in the literature.
Witsenhausen~\cite{Witsenhausen-IT-76} considered (nonsingular) fixed-length codes for the zero-error source coding problem with side information at the decoder, rather than the uniquely-decodable codes for the same model studied in~\cite{Alon-Orlitskx_source_cod_graph_entropies} as mentioned above. Orlitsy and Roche~\cite{Orlitsky-Roche_general_side_inf_model_rat_reg}  studied fixed-length codes for the function compression problem with side information at the decoder, where  the encoder compresses a source $X$, and the decoder, which receives the output of the encoder and also observes another source $Y$ correlated with $X$, is required to compute an arbitrary function of~$X$ and $Y$ with asymptotically zero error. For the problem, Malak~\cite{Malak-fun-compre-side-infor} further  presented a fixed-length  coding scheme by using a fractional coloring approach.
In~\cite{Slepian-Wolf-IT73}, the well-known multi-terminal source coding problem, the Slepian-Wolf problem, was studied where two correlated sources $X$ and $Y$ are compressed separately by two encoders using fixed-length codes; and the decoder receives the outputs of both  encoders for reconstructing the two sources $X$ and $Y$  with asymptotically zero error. The  problem in~\cite{Koulgi_zero-error-cod_cor_inf_sour} as mentioned above can be regarded as a zero-error version of the Slepian-Wolf problem for uniquely-decodable codes. K\"{o}rner and Marton~\cite{Korner-Marton-IT73} investigated the computation of the modulo $2$ sum of two correlated binary sources by using fixed-length codes.
Doshi \textit{et al}.~\cite{Doshi_fun_comp_graph_color_sch}
generalized the K{\"o}rner-Marton problem to consider compressing with asymptotically zero error an arbitrary function of two correlated sources by using fixed-length codes.
 Feizi and M{\'e}dard~\cite{Feizi-Medard} further considered the function compression problem over a tree network by using fixed-length codes. Recently, Guang and Zhang~\cite{Guang_Zero-Fun-Compre} investigated the zero-error distributed compression problem of binary arithmetic sum by using fixed-length codes, where the authors developed a novel  function-estimation approach of  the minimum chromatic number.

\subsection{Contributions and Organization of the Paper}
    The main contributions and organization of the paper are given as follows.
    \begin{itemize}
    \item In Section \ref{Sec:Net-Fun-Compre-model}, we formally present the model of zero-error network function computation, and define the uniquely-decodable   network function-computing codes  and the computing capacity from the viewpoint of  compression for the model.
    \item In Section \ref{Preliminaries},
    we present some graph-theoretic notations, definitions and some necessary results about chromatic entropy, graph entropy and clique entropy. Then  we prove several new results on clique entropy, in particular, the substitution lemma of clique entropy for probabilistic graphs with a certain condition.
    \item In Section~\ref{sec-lower-bound},  we  prove a  lower bound on the computing capacity associated with clique entropies of the induced characteristic graphs, which  is applicable to  arbitrary network topologies, arbitrary  information sources and arbitrary target functions.
      We  also illustrate the obtained lower bound by a typical model of computing the arithmetic sum of three independent and identically distributed~(i.i.d) binary information sources  over the diamond network.
 \item In Section~\ref{sec-improve-bound}, by refining the probability distribution of information sources, we  prove an improved  lower bound, which strictly enhances the  lower bound we have obtained. Furthermore, we  compare uniquely-decodable network function-computing coding and fixed-length network function-computing coding, and show that the former indeed outperforms the latter in terms of the computing capacity.
In particular, we provide a novel graph-theoretic explanation of the key parameter in  the best known bound on the computing capacity  for   fixed-length network function-computing codes, which would be helpful to improve the existing results.
\item
Concluding remarks including the future research are in Section~\ref{sec-discussion}.
    \end{itemize}

\section{Preliminaries}\label{Sec:Net-Fun-Compre-model}

%
%
%
%
%
%
%
%
%
%
%
%
%
%
%
%
%
%

\subsection{Model of Network Function Computation}

Let $\mathcal{G}=(\mathcal{V},\mathcal{E})$ be a directed acyclic graph, where $\mathcal{V}$  is a finite node set and $\mathcal{E}$ is a finite  edge set. For an edge $e\in\mathcal{E}$ connecting a node $u\in\mathcal{V}$ to another node $v\in\mathcal{V}$, we use $\text{tail}(e)$ and $\text{head}(e)$ to denote  the \emph{tail} node and the \emph{head} node of $e$, respectively, i.e., $u=\text{tail}(e)$ and $v=\text{head}(e)$. Accordingly, for a node
$v\in \mathcal{V}$, let
\begin{equation*}
\mathrm{In}(v)=\big\{e\in \mathcal{E}:~ \text{head}(e) = v\big\}\quad\text{and}
\quad \mathrm{Out}(v)=\big\{e\in \mathcal{E}:~\text{tail}(e) = v\big\},
\end{equation*}  the set of input edges of $v$ and the set of
 output edges of $v$, respectively.
Let $S\subset \mathcal{V}$ be the set of $s$ \textit{source nodes} $\sigma_{1},\sigma_{2},\dots,\sigma_{s} $, and $\rho\in \mathcal{V}\setminus S$ be the single \textit{sink node}, where  each source node has no input edges and the sink node~$\rho$ has no output edges, i.e.,
$\mathrm{In}(\sigma_i) = \mathrm{Out}(\rho) = \emptyset$  for $i=1,2,\cdots, s$. Without loss of generality, we assume that for each $v\in \mathcal{V}\setminus\{\rho\}$, there always exists a directed path from $v$ to $\rho$ in $\mathcal{G}$. The graph~$\mathcal{G}$, together with $S$ and $\rho$, forms a
network $\mathcal{N}$, i.e., $\mathcal{N}= (\mathcal{G}, S, \rho)$.

Let
$
f\colon \mathcal{A}^s\to\Im f$
be a nonconstant function, called the  \emph{target function}, where $\mathcal{A}$ is a finite alphabet, and thus $\Im f$, denoting the image set of $f$, is also finite.
We assume that the $i$th
argument  of  $f$ is  generated at the $i$th  source node $\sigma_i$ for $i=1,2,\cdots, s$. The target function $f$  is
required to be computed with zero error at the sink node $\rho$ over the network $\mathcal{N}$.
More precisely, for each $i$, we let the \emph{information source} at the $i$th source node~$\sigma_i$ be a random  variable $X_i$  taking values in the finite set $\mathcal{A}$, and let $X_S\triangleq (X_1,X_2,\cdots,X_s)$. We further let $P_{X_S}$ be the joint probability distribution of all the information sources that satisfies
 \begin{equation*}
P_{X_S}(x_S)\triangleq{\rm{Pr}}\big\{X_S=x_S\big\}>0,\quad~\forall~ x_S\triangleq (x_1,x_2,\cdots,x_s)\in\mathcal{A}^s.
 \end{equation*}

 In our model, we consider computing the target function $f$ multiple times on the network $\mathcal{N}$. To be specific, for a positive integer $k$,
 each source node $\sigma_i$ generates $k$ i.i.d. random variables $X_{i,1}, X_{i,2}, \cdots , X_{i,k}$
with generic random
variable $X_i$, and then we write
 \begin{equation*}\boldsymbol{X}_i = \big(X_{i,1}, X_{i,2}, \cdots , X_{i,k}\big)^{\top}.
 \end{equation*}
We let $\boldsymbol{x}_i=(x_{i,1}, x_{i,2}, \cdots ,
x_{i,k})^{\top}\in\mathcal{A}^k$
 be a  value taken by $\boldsymbol{X}_i$, called the \emph{source message} generated by $\sigma_i$. For a subset $I\subseteq S$, we further let
 \begin{equation*}
 \boldsymbol{X}_I\triangleq (\boldsymbol{X}_i:\,\sigma_i\in I)\quad\text{and}\quad\boldsymbol{x}_I\triangleq(\boldsymbol{x}_i:\,\sigma_i\in I),
  \end{equation*}
  and use $\mathcal{A}^{k\times I}$ (instead of $\mathcal{A}^{k\times |I|}$ for notational
simplicity) to denote the set of all possible $k\times|I|$ matrices
taken by~$\boldsymbol{x}_I$.\footnote{When $k=1$, we write  $\boldsymbol{x}_I\in\mathcal{A}^{1\times I}$ as $x_I\in \mathcal{A}^I$ for notational
simplicity.}
In particular, when $I=S$, we have
 $\boldsymbol{X}_S = (\boldsymbol{X}_1, \boldsymbol{X}_2, \cdots , \boldsymbol{X}_s)$ and $\boldsymbol{x}_S = (\boldsymbol{x}_1, \boldsymbol{x}_2, \cdots , \boldsymbol{x}_s).$
%
 For any  $\boldsymbol{x}_S\in \mathcal{A}^{k\times S}$
taken by $\boldsymbol{X}_S$, the $k$ values of the target function
\begin{equation*}
f(\boldsymbol{x}_{ S})\triangleq
\big(f(x_{1,j},x_{2,j},\cdots,x_{s,j}):~j=1,2,\cdots,k\big)^{\top}
\end{equation*}
are required to be computed with zero error at the sink node $\rho$.
Then we have specified our network function computation model, denoted by
$(\mathcal{N},X_S,f)$.

\subsection{Uniquely-Decodable   Network Function-Computing Coding}\label{Ud-code}

In this subsection, we will define a {\em $k$-shot uniquely-decodable   (network function-computing) code} for the model $(\mathcal{N},X_S,f)$, of which the purpose is to compute $f$, the target function, $k$ times with zero error at~$\rho$ over the network $\mathcal{N}$.
We continue to consider the source message  $\boldsymbol{x}_i\in\mathcal{A}^k$ generated by $\sigma_i$ for~$i=1,2,\cdots,s$.
A $k$-shot uniquely-decodable (network function-computing) code for $(\mathcal{N},X_S,f)$, denoted by  $\mathbf{C}=\{\theta_{e}:e\in \mathcal{E};~\psi\}$, consists of
\begin{itemize}
\item
a \emph{local encoding function} $\theta_{e}$ for each edge $e\in\mathcal{E}$ given by
\begin{subnumcases} { \theta_{e}:}
\quad\;\;\mathcal{A}^{k}\rightarrow \Im\,\theta_{e} & if $e\in \mathrm{Out}(\sigma_{i})$ for some source node $\sigma_i$,\label{wzero}\\
\displaystyle\prod_{d\in \mathrm{In}\left(\text{tail}(e)\right)} \Im\,\theta_{d}\rightarrow \Im\,\theta_{e} & otherwise,\label{weqn}
\end{subnumcases}
where  for each $e\in\mathcal{E}$, $\Im\,\theta_{e}$, the image set of $\theta_{e}$, is a binary uniquely-decodable code,\footnote{We refer to an image set
 $\Im\,\theta_e\subseteq\{0,1\}^*$  as a binary uniquely-decodable code if $\Im\,\theta_e$, regarded as a binary source code, is  uniquely-decodable,
 where $\{0,1\}^*$ is the set of all binary sequences of finite lengths, i.e., $\{0,1\}^{*}=\bigcup_{i\geq 1}\{0,1\}^{i}$ with $\{0,1\}^i$ being the set of  binary sequences of length $i$.}
\item a \emph{decoding function} at the sink node $\rho$ given by
 \begin{align*}
\psi:\;\prod_{e\in\mathrm{In}(\rho)} \Im\,\theta_{e}\rightarrow \big({\Im f}\big)^{k},
\end{align*}
which is used to compute the $k$ function values $f(\boldsymbol{x}_S)$ with zero error by the messages received by~$\rho$.
\end{itemize}

For the  code $\mbC$, the local encoding functions $\theta_{e},\,e\in\mathcal{E}$ are used to compute the messages transmitted on the edges and are executed following a given topological order on the edges in $\mathcal{E}$. By~\eqref{wzero} and \eqref{weqn}, we see that when~$e$ is an output edge of a source node $\sigma_{i}\in S$, $\theta_{e}$ takes the source message~$
\boldsymbol{x}_i$ as its inputs; and
when $e$ is an output edge of a non-source node~$u\triangleq \text{tail}(e)$, $\theta_{e}$ takes the messages received by~$u$ through the input edges as its inputs. Here, the uniquely-decodable property for $\Im\,\theta_{e},e\in\mathcal{E}$ in~$\mbC$ is necessary to be satisfied such that the code $\mbC$ can be used multiple times to guarantee  the target function~$f$ to be always computed with zero error.

Further, we let ${\boldsymbol{y}_e}\in \{0,1\}^*$ be the message transmitted on each edge $e\in \mathcal{E}$ when the code $\mathbf{C}$ is used and  the source messages are $\boldsymbol{x}_{ S}$.
With the encoding mechanism as described in~\eqref{wzero} and \eqref{weqn}, we can readily see that ${\boldsymbol{y}_e}$ is a function of~$\boldsymbol{x}_S$, denoted by $g_{e}(\boldsymbol{x}_{ S})$ (i.e., ${\boldsymbol{y}_e}=g_{e}(\boldsymbol{x}_{ S})$), where $g_{e}$ can be obtained by recursively applying   the local encoding
functions $\theta_{e}, e\in\mathcal{E}$.
More precisely, for each edge $e\in\mathcal{E}$, we have
\begin{equation*}
\begin{split}
g_{e}(\boldsymbol{x}_S)=\left\{
\begin{array}{ll}
\theta_e(\boldsymbol{x}_i)                   &\;\text{if $e\in \mathrm{Out}(\sigma_{i})$ for some source node $\sigma_{i}$,}\\[2mm]
\theta_e\big(g_{\mathrm{In}(u)}(\boldsymbol{x}_S)\big) &\; \text{otherwise,}
\end{array}
\right.
\end{split}
\end{equation*}
where $u\triangleq\mathrm{tail}(e)$ and $g_{C}(\boldsymbol{x}_{ S})\triangleq\big(g_{e}(\boldsymbol{x}_{ S}):e\in C\big)$ for an
edge set $C\subseteq \mathcal{E}$.
We call $g_{e}$ the \emph{global encoding function} for  an edge $e$ of the  code $\mbC$.

We say that such a $k$-shot  code $\mathbf{C}$ for $(\mathcal{N},X_S,f)$ is \emph{admissible} if
\begin{equation}\nonumber
{\rm{Pr}}\big\{\psi\big(g_{\mathrm{In}(\rho)}(\boldsymbol{X}_{ S})\big)=f\big(\boldsymbol{X}_{ S}\big)\big\}=1,
\end{equation}
i.e.,  the target function $f$ can be computed with zero error $k$ times   at $\rho$ by using  $\mathbf{C}$.  Further,
for this  admissible $k$-shot  code $\mbC$, we define
%
\begin{align*}
L_e(\mbC)=E\big[\ell\big(g_e(\boldsymbol{X}_S)\big)\big],\quad\forall~e\in\mathcal{E},
\end{align*}
where $\ell(\boldsymbol{y})$ stands for the length of a binary sequence $\boldsymbol{y}$. Then $L_e(\mbC)$
is the expected number of bits transmitted on an edge $e$ by using the code $\mbC$. We further let
\begin{align}
R_{e}(\mbC)\triangleq\frac{L_e(\mbC)}{k},\quad~\forall~ e\in \mathcal{E},\label{rate-def-on-edge}
\end{align}
which is the average expected number of bits transmitted on the edge $e$ for computing the target function~$f$ once  by using the code $\mbC$, called the \emph{expected bit rate} on the edge $e$. With this, we define the \emph{coding rate} of the code~$\mbC$ by
 \begin{align}\label{def-RC}
R(\mbC)\triangleq\max_{e\in\mathcal{E}}\;R_e(\mbC),
\end{align}
 which is  the maximum expected bit rate on all the edges in $\mathcal{E}$, or
  the maximum average expected number of bits transmitted on  the edges in $\mathcal{E}$,  for computing the target function $f$ once by using the code $\mbC$.

 In addition, a  nonnegative real
 $R$ is called \emph{(asymptotically) achievable}  if for any $\epsilon> 0$, there
exists for  a positive integer $k$, an admissible $k$-shot uniquely-decodable  code $\mbC$    for  the model $(\mathcal{N},X_S,f)$ such that
\begin{equation*}
R(\mbC)<R+\epsilon.
\end{equation*}
Consequently, the \emph{computing capacity}  for the model $(\mathcal{N},X_{S},f)$ is defined as
\begin{align*}
\mathcal{C}(\mathcal{N},X_{S},f)=\inf\Big\{R:~\text{$R$ is achievable for $(\mathcal{N},X_{S},f)$}\Big\}.
\end{align*}

\section{Preparatory Results}\label{Preliminaries}
 Consider an undirected graph $G=(V,E)$ without loop,\footnote{In an undirected graph $G$, an edge $e$ between two vertices $u$ and $v$  is a loop if $u=v$. All the undirected graphs considered in the paper are assumed to have no loop.} where $V$ and $E$ are the vertex set and the edge set, respectively. In  $G$, we say that two distinct vertices are \emph{connected} if  there exists an edge between them.
A \emph{clique} in  $G$  is a subset of vertices in which every
two distinct vertices are connected, namely that a clique in $G$
is the vertex set of a complete subgraph of $G$.
We denote  by $\Omega(G)$ the family of all the cliques in~$G$. Rather, an \emph{independent set} in $G$ is a subset of vertices in which any
two distinct vertices are not connected, and we denote by
 $\Gamma(G)$  the family of all the independent  sets in~$G$.
A vertex~$v\in V$ is called an \emph{isolated vertex}  if $v$ and each vertex in $V\setminus\{v\}$ are not connected.
The complement of the graph $G$, denoted by~$G^{\textup{c}}$, is also an undirected graph without loop, in which $G^{\textup{c}}$ has the
same vertex set~$V$ as in $G$, and any two distinct vertices are connected in $G^{\textup{c}}$ if and
only if they are not connected in $G$.
Further,  we call a mapping $c$ defined on the vertex set $V$ to be a \emph{coloring} of  $G$ if $c(u)\neq c(v)$ for any two distinct vertices $u$ and $v$ in $G$ that are connected.

Further, we let $k$ be a positive integer and consider $k$ undirected graphs $G_{i}=(V_i,E_i)$,  $1\leq i\leq k$.
We define the \emph{AND product} of $G_1,G_2,\cdots,G_k$, denoted by $\land^{k}_{i=1}G_{i}$, to be an undirected graph where the
 vertex set is $$\prod_{i=1}^{k}V_{i}= V_{1}\times V_{2}\times\cdots\times V_{k},\footnote{For the convenience of discussion in the rest of the paper, we write vertices as column vectors in the AND product and OR product (to be clear later) of multiple undirected graphs.}$$
and two distinct vertices $(u_{1},u_{2},\cdots,u_{k})^\top$ and $(v_{1},v_{2},\cdots,v_{k})^\top$ in $\prod_{i=1}^kV_i$
are connected if and only if~$u_{i}$ and  $v_{i}$  are connected in~$G_{i}$ for each $i\in \{1,2\cdots, k\}$ with $u_{i}\neq v_{i}$.
Further, we define the \emph{OR product} of $G_{1},G_{2},\cdots,G_{k}$, denoted by $\vee^{k}_{i=1}G_{i}$,  to be an undirected graph where the vertex set is also $\prod_{i=1}^{k}V_{i}$ and two distinct
 vertices $(u_{1},u_{2},\cdots,u_{k})^\top$ and $(v_{1},v_{2},\cdots,v_{k})^\top$  are connected
if and only if~$u_{i}$ and $v_{i}$ are connected in $G_{i}$ for some $i\in \{1,2\cdots,k\}$ with $u_{i}\neq v_{i}$.
In particular, if the $k$ undirected graphs $G_i$, $1\leq i\leq k$ are identical to an undirected  graph~$G$, the AND product and the OR product of the $k$ identical graphs $G$ are, respectively, denoted by
 $G^{\land k}$ and $G^{\vee k}$, and called the \emph{$k$-fold AND product} of~$G$ and the \emph{$k$-fold OR product} of~$G$.

{\subsection{Chromatic Entropy, Graph Entropy and Clique Entropy}
}\label{subsec:pre-results}


Let $G=(V,E)$ be an undirected graph and $Z$ be a random variable taking values in the
 vertex set~$V$ of $G$. The pair $(G,Z)$ is called
a \emph{probabilistic graph}.
The \emph{chromatic entropy} of the probabilistic graph~$(G, Z)$ is
defined as
\begin{equation}\label{def:chrom-entro}
H_{\chi}(G, Z)\triangleq\min\big\{H\big(c(Z)\big):\;\text{$c$ is a coloring of $G$}\big\}.
\end{equation}

K\"{o}rner in \cite{Korner_graph-entropx_def} defined the \emph{graph entropy} of a probabilistic graph $(G,Z)$ by
\begin{align*}
H_{\kappa}(G,Z)\triangleq\min_{Z\in W\in\Gamma(G)} I(W;Z),
\end{align*}
where  the notation $Z\in W\in\Gamma(G)$ is explained as follows.
 Here, $W$ is a random variable taking  all the independent sets in $\Gamma(G)$ as values and
 the joint probability distribution $P_{ZW}$  of $Z$ and $W$  satisfies
\begin{align}
P_{ZW}(z,w)=0,\quad~\forall~(z,w)\in V\times\Gamma(G)\;\text{with}\; z\notin w.\label{graph-entrop-condi}
\end{align}
 As such,  $Z\in W\in\Gamma(G)$ can be viewed as a constraint on the random variable $W$
 taking values in $\Gamma(G)$ and satisfying the condition~\eqref{graph-entrop-condi}.
We further remark that  for the probability distribution  $P_{Z}$  of~$Z$, we have
\begin{equation*}
P_{Z}(z)=\sum\limits_{w\in\Gamma(G)}P_{ZW}(z,w)=\sum\limits_{w\in\Gamma(G)~\text{s.t.}~z\in w}P_{ZW}(z,w),
\qquad\forall~z\in V.
\end{equation*}

 Alon and Orlitsky in~\cite{Alon-Orlitskx_source_cod_graph_entropies} further defined  the \emph{clique entropy} of a probabilistic graph $(G, Z)$
as
\begin{align}\label{def:clique-entro}
H_{\omega}(G,Z)\triangleq\max_{Z\in W\in\Omega(G)} H(Z|W),
\end{align}
where, similarly,  $Z\in W\in\Omega(G)$ represents that the random variable $W$ takes all the cliques in $\Omega(G)$ as values  and  the joint probability distribution $P_{ZW}$  of $Z$ and $W$ satisfies that
\begin{equation}
P_{ZW}(z,w)=0,\quad~\forall~(z,w)\in V\times\Omega(G)\;\text{with}\; z\notin w.\label{cond-X-in-W-Omega(G)}
\end{equation}
This similarly shows that for each $z\in V$,
\begin{align*}
P_Z(z)=\sum_{\omega\in\Omega(G)~\textup{s.t.}~z\in\omega} P_{ZW}(z,\omega).
\end{align*}

In the following proposition, we exactly calculate the clique entropy $H_{\omega}(G,Z)$ for
 two special cases that $G$  is a complete graph and $G$ is an empty graph (defined as the complement of a complete graph).
\begin{prop}\label{prop:clique-entro-property}
For a probabilistic graph $(G,Z)$,
\begin{align*}
H_{\omega}(G,Z)=
\begin{cases}
0 & \textup{if $G$ is an empty graph, }  \\
H(Z) & \textup{if $G$ is a complete graph.}
\end{cases}
\end{align*}
\end{prop}
\begin{IEEEproof}
See Appendix~\ref{appendix-pf-prop:clique-entro-property}.
\end{IEEEproof}

We now present three lemmas on graph entropy, clique entropy and chromatic entropy, which were proved by  Alon and Orlitsky in~\cite{Alon-Orlitskx_source_cod_graph_entropies}.
\begin{lemma}[{\!\!\cite[Lemma~9]{Alon-Orlitskx_source_cod_graph_entropies}}]\label{lemma:OR-pro-grp-entr=sum-g-en}
Consider $k$ probabilistic
graphs
 $(G_{1},Z_{1})$, $(G_{2},Z_{2}), \cdots, (G_{k},Z_{k})$, in which $Z_{i},1\leq i\leq k$ are mutually independent. Let $\boldsymbol{G}_{\vee}={\vee}^{k}_{i=1}G_{i}$ and $\boldsymbol{Z}=(Z_{1},Z_{2},\cdots,Z_{k})^{\top}$. Then
\begin{equation*}
H_{\kappa}(\boldsymbol{G}_{\vee},\boldsymbol{Z})=\sum_{i=1}^{k}H_{\kappa}(G_{i},Z_{i}).
\end{equation*}
\end{lemma}

\begin{lemma}[\!{\!\cite[Lemma~12]{Alon-Orlitskx_source_cod_graph_entropies}}]\label{lemma:graph_clique_entropy_relation}
Consider a probabilistic graph $(G, Z)$. Then
\begin{equation*}
H_{\kappa}(G^{\textup{c}},Z)=H(Z)-H_{\omega}(G,Z),
\end{equation*}
or equivalently,
\begin{equation*}
H_{\kappa}(G^{\textup{c}},Z)+H_{\omega}(G,Z)=H(Z).
\end{equation*}
\end{lemma}

\begin{lemma}[\!{\!\cite[Lemmas 13 and 14]{Alon-Orlitskx_source_cod_graph_entropies}}]\label{lemma:graph-chrom-entro-rel}
Consider a probabilistic graph $(G, Z)$. Then
\begin{equation*}
H_{\omega}(G,Z) \leq H_{\kappa}(G,Z)\leq H_{\chi}(G,Z).
\end{equation*}
\end{lemma}
%

We continue to consider an undirected graph $G=(V,E)$. A vertex subset $U\subseteq V$  is said to be {\emph{autonomous}} in $G$ if each vertex $v$ in $V\setminus U$  is either connected to all of the vertices in~$  U$ or none of the vertices in $U$. In other words, all the vertices in the autonomous vertex subset $  U$ have the same neighborhood in $V \setminus U$.
Consider  such an autonomous  vertex subset $U\subseteq V$ in $G$ and  a new vertex $u$ not in~$V$. We define the operation of \emph{replacing~$U$ by $u$ in $G$} as follows:
\romannumeral1) removing  all the vertices in~$ U $ from $V$ and adding the new vertex $u$ to $V$,
 and \romannumeral2)   connecting~$u$ to the vertices in $V\setminus U$ that are connected to (adjacent to)  all the vertices of $ U $ in $G$.
We denote the resulting graph by $G|_{ U \rightarrow u}$.
We further define the \emph{projection} of $G$  onto $ U $, denoted by $G|_{ U }$,
 which is a subgraph of $G$
with the vertex set $ U $ where  two vertices are connected to each other in $G|_{ U }$ if  they are connected to each other in $G$.
 The above concepts can be extended to probabilistic graphs by the same way. To be specific, consider a probabilistic graph~$(G,Z)$.
 Let $U\subseteq V$ be an autonomous  vertex subset and $u$ be a new-added vertex  not in $V$. Let $Z|_{ U \rightarrow u}$  be the random variable taking  values on the vertex set  of $G|_{ U \rightarrow u}$, i.e., $(V\setminus U) \cup\{u\}$, according to the following probability distribution on the vertex set $(V\setminus U) \cup\{u\} $:
\begin{equation}\label{def:random-var-X_U-u}
\begin{split}
P_{Z|_{ U \rightarrow u}}(z)=\left\{
\begin{array}{ll}
P_{Z}(z)                   &\;\text{if $z\in V\setminus U$,}\\
P_{Z}( U )\triangleq\sum\limits_{v\in  U }P_{Z}(v) &\; \text{if $z=u$,}
\end{array}
\right.
\end{split}
\end{equation}
where $P_{Z}$ is the probability distribution of the random variable $Z$ which takes values in $V$.
Similarly, we use $Z|_{ U }$ to denote the projection of
$Z$ onto $ U $, which is a random variable taking values in $ U $ according to the probability distribution
\begin{equation}\label{def:random-var-X_U}
P_{Z|_ U }(z)\triangleq\frac{P_Z(z)}{P_{Z}( U )},\quad\forall~z\in U .
\end{equation}
Further, we can readily see that  $(G|_{ U \rightarrow u},Z|_{ U \rightarrow u})$ and $(G|_{ U},Z|_{ U })$ are also probabilistic graphs.
To end this subsection, we present the following \emph{substitution lemma} of  graph entropy proved by K\"{o}rner \textit{et al}.~\cite{Korner-group-property-of_graph-entropy}.
\begin{lemma}[Substitution Lemma]\label{lemma:graph_entropy_grouping_property}
Let $(G,Z)$ be a probabilistic graph, where $G=(V,E)$ is an undirected graph and $Z$ is a random variable taking  values on the vertex set $V$.   For  an autonomous vertex subset $U\subseteq V$ and  a new vertex $u$ not in $V$, the following equality holds:
\begin{align*}
H_{\kappa}(G,Z)=H_{\kappa}\big(G|_{ U \rightarrow u},Z|_{ U \rightarrow u}\big)+P_{Z}( U )\cdot H_{\kappa}\big(G|_{ U },Z|_{ U }\big).
\end{align*}
\end{lemma}

\subsection{Some New Results on Clique Entropy}

In this subsection, we will prove several results on clique entropy, which are useful in the rest of the paper.
We first present  three lemmas on clique entropy of which the proofs are deferred to Appendices~\ref{pf-lemma-cli-upper-bound},~\ref{appendix:pf-clique_entropx_grouping_property}, and~\ref{appendix:pf-AND-pro=sum-g-en}, respectively.

\begin{lemma}\label{lemma-cli-upper-bound}
Let $(G,Z)$ be a probabilistic graph, where $G=(V,E)$ is an undirected graph and $Z$ is a random variable taking  values on the vertex set $V$. Then,
\begin{align*}
H_{\omega}(G,Z)\leq \log\omega(G),\footnotemark
\end{align*}
where $\omega(G)$ is the clique number of $G$, defined as the number of vertices in a maximum clique in~$G$, i.e.,
\begin{equation*}
\omega(G)\triangleq\max\big\{|w|:~\text{$w\in \Omega(G)$}\big\}.
\end{equation*}
\end{lemma}
\footnotetext{In this paper, we always use ``$\log$'' to denote ``$\log_2$'', the logarithm with base $2$, for notational simplicity.}

\begin{lemma}\label{lemma:clique_entropy_grouping_property}
Let $(G,Z)$ be a probabilistic graph, where $G=(V,E)$ is an undirected graph and $Z$ is a random variable taking  values on the vertex set $V$.   For an autonomous vertex subset $ U \subseteq V$  and a new vertex $u$ not in $V$, the following equality holds:
\begin{align*}
H_{\omega}(G,Z)=H_{\omega}\big(G|_{ U \rightarrow u},Z|_{ U \rightarrow u}\big)+P_{Z}( U )\cdot H_{\omega}\big(G|_{ U },Z|_{ U }\big).
\end{align*}
\end{lemma}


\begin{lemma}\label{lemma:AND-pro=sum-g-en}
Consider $k$ probabilistic
graphs
 $(G_{1},Z_{1})$, $(G_{2},Z_{2}), \cdots, (G_{k},Z_{k})$, where  $Z_{i},1\leq i\leq k$ are mutually independent. Let $\boldsymbol{G}_{\land}={\land}^{k}_{i=1}G_{i}$ and $\boldsymbol{Z}=(Z_{1},Z_{2},\cdots,Z_{k})^{\top}$. Then
\begin{align*}
H_{\omega}(\boldsymbol{G}_{\land},\boldsymbol{Z})=\sum_{i=1}^k H_{\omega}(G_i,Z_i).
\end{align*}

\end{lemma}
\medskip


 We continue to consider a probabilistic graph $(G,Z)$, where $G=(V,E)$ is an undirected graph and~$Z$ is a random variable taking values in the
 vertex set $V$ of $G$.
  Let $U_1, U_2, \cdots, U_{t} \subseteq V$ be $t$  disjoint autonomous vertex subsets in $G$ and  $u_1, u_2, \cdots, u_{t}$ be $t$ new-added vertices  not in $V$.
  With them, we can take a sequence of operations of replacing $U_j$ to $u_j$ from $j=1$ to $j=t$. To be specific, we first replace~$U_1$ by $u_1$ in $G$ and then obtain the resulting graph $G|_{ U_1 \rightarrow u_1}$.
  Next, we note that the vertex subset  $U_2$ is also autonomous in $G|_{ U_1 \rightarrow u_1}$ because $U_1\cap U_2=\emptyset$. Thus, we are able to replace $U_2$ by $u_2$ in $G|_{ U_1 \rightarrow u_1}$ and obtain the resulting  graph $G|_{ U_1 \rightarrow u_1}|_{ U_2 \rightarrow u_2}$, so on and so forth.
  The final resulting graph is $G|_{ U_1 \rightarrow u_1}|_{ U_2 \rightarrow u_2}\cdots|_{ U_t \rightarrow u_t}$, which we denote by $G|_{\uplus_{j=1}^{t} U_j\to u_j}$ for notational simplicity, i.e.,
  \begin{align*}
 G|_{\uplus_{j=1}^{t} U_j\to u_j}\triangleq  G|_{ U_1 \rightarrow u_1}|_{ U_2 \rightarrow u_2}\cdots|_{ U_t \rightarrow u_t}.
 \end{align*}
 With the above elaboration, we see  that $G|_{\uplus_{j=1}^{t} U_j\to u_j}$ is well-defined and is independent of the order of the operations, namely that    for each permutation $\tau$ of $\{1, 2, \cdots, t\}$,
$$
G|_{\uplus_{j=1}^{t} U_j\to u_j} =  G|_{\uplus_{j=1}^{t} U_{\tau(j)}\to u_{\tau(j)}}.
$$
Similarly, we can also obtain a new random variable
    \begin{align*}
 Z|_{\uplus_{j=1}^{t} U_j\to u_j}\triangleq  Z|_{ U_1 \rightarrow u_1}|_{ U_2 \rightarrow u_2}\cdots|_{ U_t \rightarrow u_t}
 \end{align*}
according to  the probability distribution
\begin{equation}\label{def:X-multi-replace2-prob}
P_{ Z|_{\uplus_{j=1}^{t} U_j\to u_j}}(z)=\left\{
\begin{array}{ll}
P_{Z}(z)                   &\;\text{if $z\in V\setminus(\cup_{j=1}^t U_j) $,}\\
P_{Z}( U_j )\triangleq\sum\limits_{v\in  U_j }P_{Z}(v) &\; \text{if $z=u_j$,  $1\leq j\leq t$.}
\end{array}
\right.
\end{equation}
We also have
\begin{align*}
 Z|_{\uplus_{j=1}^{t} U_j\to u_j} =  Z|_{\uplus_{j=1}^{t} U_{\tau(j)}\to u_{\tau(j)}},
\end{align*}
for each permutation $\tau$ of $\{1, 2, \cdots, t\}$. Clearly, $\big(G|_{\uplus_{j=1}^{t} U_j\to u_j},Z|_{\uplus_{j=1}^{t} U_j\to u_j}\big)$ is a probabilistic graph, too.
Now, we generalize Lemma~\ref{lemma:clique_entropy_grouping_property} to obtain the following consequence.

\begin{lemma}\label{lemma:clique_entropy_grouping_property-autonomous}
Let $(G,Z)$ be a probabilistic graph, where $G=(V,E)$ is an undirected graph and $Z$ is a random variable taking  values on the vertex set $V$. Let $U_1, U_2, \cdots, U_{t} \subseteq V$ be $t$ disjoint autonomous vertex subsets in $G$, and  let $u_1, u_2, \cdots, u_{t}$ be $t$ new-added vertices not in $V$.
Then, the following equality holds:
\begin{align}
H_{\omega}(G,Z)=H_{\omega}\big(G|_{\uplus_{j=1}^{t} U_j\to u_j},Z|_{\uplus_{j=1}^{t} U_j\to u_j}\big)+\sum_{j=1}^{t} P_{Z}( U_j)\cdot H_{\omega}\big(G|_{ U _j},Z|_{ U _j}\big).
\label{eq:clique_entropy_grouping_property-autonomous}
\end{align}
\end{lemma}

\begin{IEEEproof}
We will prove the lemma by induction on $t$. When $t=1$, the equality~\eqref{eq:clique_entropy_grouping_property-autonomous} is true by Lemma~\ref{lemma:clique_entropy_grouping_property}.
 We now assume that the equality~\eqref{eq:clique_entropy_grouping_property-autonomous} is true for some $t\geq 1$, and  consider the case of $t+1$. Here, we have $t+1$  disjoint autonomous vertex subsets $U_1, U_2, \cdots, U_{t+1} \subseteq V$ in $G$, and $t+1$ new-added vertices $u_1, u_2, \cdots, u_{t+1}$ not in $V$.
First, by the induction hypothesis for the case of $t$, we have
\begin{align}
H_{\omega}(G,Z)=H_{\omega}\big(G|_{\uplus_{j=1}^{t} U_j\to u_j},Z|_{\uplus_{j=1}^{t} U_j\to u_j}\big)+\sum_{j=1}^{t} P_{Z}( U_j)\cdot H_{\omega}\big(G|_{ U _j},Z|_{ U _j}\big).\label{pf-clique-property-eq2}
\end{align}

Now, we consider the probabilistic graph $\big(G|_{\uplus_{j=1}^{t} U_j\to u_j},Z|_{\uplus_{j=1}^{t} U_j\to u_j}\big)$.
Note that the vertex subset
 $U_{t+1}$ is   autonomous  in  $G|_{\uplus_{j=1}^{t} U_j\to u_j}$ and $u_{t+1}$ is  a new-added vertex  not in the vertex set of  $G|_{\uplus_{j=1}^{t} U_j\to u_j}$.
%
 By Lemma~\ref{lemma:clique_entropy_grouping_property}, we immediately obtain the following equality by regarding $(G,Z)$ and $U$ in Lemma~\ref{lemma:clique_entropy_grouping_property} by    $\big(G|_{\uplus_{j=1}^{t} U_j\to u_j},Z|_{\uplus_{j=1}^{t} U_j\to u_j}\big)$ and $U_{t+1}$:
 \begin{align}
 &H_{\omega}\big(G|_{\uplus_{j=1}^{t} U_j\to u_j},Z|_{\uplus_{j=1}^{t} U_j\to u_j}\big)\nonumber
 \\&=
 H_{\omega}\big(G|_{\uplus_{j=1}^{t} U_j\to u_j}|_{U_{t+1}\to u_{t+1}},Z|_{\uplus_{j=1}^{t} U_j\to u_j}|_{U_{t+1}\to u_{t+1}}\big)
 \nonumber\\&~~~~+P_{ Z|_{\uplus_{j=1}^{t} U_j\to u_j}}(U_{t+1})\cdot
 H_{\omega}\big(G|_{\uplus_{j=1}^{t} U_j\to u_j}|_{U_{t+1}},Z|_{\uplus_{j=1}^{t} U_j\to u_j}|_{U_{t+1}}\big)\nonumber\\
 &=
 H_{\omega}\big(G|_{\uplus_{j=1}^{t+1} U_j\to u_j},Z|_{\uplus_{j=1}^{t+1} U_j\to u_j}\big)
 +P_{ Z|_{\uplus_{j=1}^{t} U_j\to u_j}}(U_{t+1})\cdot
 H_{\omega}\big(G|_{U_{t+1}},Z|_{U_{t+1}}\big),\label{pf-clique-property-eq3-2}
 \end{align}
 where the equality~\eqref{pf-clique-property-eq3-2} follows from the fact that
 \begin{align*}
 G|_{\uplus_{j=1}^{t} U_j\to u_j}|_{U_{t+1}}=G|_{U_{t+1}}\quad\text{and}\quad
 Z|_{\uplus_{j=1}^{t} U_j\to u_j}|_{U_{t+1}}=Z|_{U_{t+1}}
 \end{align*}
 because $U_1, U_2, \cdots, U_{t+1}$ are disjoint.
Further, by definition, we can readily see that
 \begin{align}
 P_{ Z|_{\uplus_{j=1}^{t} U_j\to u_j}}(U_{t+1})&=\sum_{z\in U_{t+1}} P_{ Z|_{\uplus_{j=1}^{t} U_j\to u_j}}(z)\nonumber\\
 &=\sum_{z\in U_{t+1}}P_Z(z)\label{pf-clique-property-eq4-1}\\
 &=P_{Z}(U_{t+1}),\label{pf-clique-property-eq4-2}
 \end{align}
 where the equality \eqref{pf-clique-property-eq4-1} follows from  $z\in U_{t+1}\subseteq V\setminus(\cup_{j=1}^t U_j)$ and \eqref{def:X-multi-replace2-prob}. With \eqref{pf-clique-property-eq4-2}, we continue  considering \eqref{pf-clique-property-eq3-2} to obtain that
\begin{align*}
 &H_{\omega}\big(G|_{\uplus_{j=1}^{t} U_j\to u_j},Z|_{\uplus_{j=1}^{t} U_j\to u_j}\big)
 \\&=
 H_{\omega}\big(G|_{\uplus_{j=1}^{t+1} U_j\to u_j},Z|_{\uplus_{j=1}^{t+1} U_j\to u_j}\big)
+P_{Z}(U_{t+1})\cdot
 H_{\omega}\big(G|_{U_{t+1}},Z|_{U_{t+1}}\big).\nonumber
 \end{align*}
Together with \eqref{pf-clique-property-eq2}, we have
\begin{align*}
H_{\omega}(G,Z)&=H_{\omega}\big(G|_{\uplus_{j=1}^{t} U_j\to u_j},Z|_{\uplus_{j=1}^{t} U_j\to u_j}\big)+\sum_{j=1}^{t} P_{Z}( U_j)\cdot H_{\omega}\big(G|_{ U _j},Z|_{ U _j}\big)\\
&=H_{\omega}\big(G|_{\uplus_{j=1}^{t+1} U_j\to u_j},Z|_{\uplus_{j=1}^{t+1} U_j\to u_j}\big)+\sum_{j=1}^{t+1} P_{Z}( U_j)\cdot H_{\omega}\big(G|_{ U _j},Z|_{ U _j}\big).
\end{align*}
The lemma is thus proved.
\end{IEEEproof}

Next, we say that an autonomous vertex subset $U\subseteq V$ in $G$ is \emph{isolated} if in the graph $G|_{ U \rightarrow u}$ of replacing $U$ by a new vertex $u$,
the new-added vertex~$u$  is an isolated vertex, or equivalently,
each vertex in
$ U $ is connected to none of the vertices in $V\setminus U $.
We say that an autonomous vertex subset $U\subseteq V$ in $G$ is \emph{completely-connected}  if in $G|_{ U \rightarrow u}$, the new-added vertex $u$ replacing $U$ is connected to all of the vertices in $V\setminus U $, namely that, in the original graph $G$,  each vertex in
$ U $ is  connected to all of the vertices in $V\setminus U $. Now, we can   obtain the corollary below following Proposition~\ref{prop:clique-entro-property} and
Lemma~\ref{lemma:clique_entropy_grouping_property-autonomous}.

\begin{cor}\label{cor-clique-entropy-pro1}
 Consider  a probabilistic graph $(G,Z)$ with $G=(V,E)$ being an undirected graph and~$Z$ being a random variable taking  values on the vertex set $V$, where $U_1, U_2, \cdots, U_{t} \subseteq V$ are $t$ disjoint autonomous vertex subsets in $G$ that constitute  a partition of $V$. Let  $u_1, u_2, \cdots, u_{t}$ be $t$ vertices not in~$V$.
\begin{itemize}
\item If all $U_j,1\leq j\leq t $ are isolated, then $G|_{\uplus_{j=1}^{t} U_j\to u_j}$ is an empty graph and then
    \begin{align*}
H_{\omega}(G,Z)=\sum_{j=1}^{t} P_{Z}( U_j)\cdot H_{\omega}\big(G|_{  U_j},Z|_{ U_j}\big).
\end{align*}
\item If all $U_j,1\leq j\leq t $ are completely-connected, then $G|_{\uplus_{j=1}^{t} U_j\to u_j}$ is a complete graph and then
   \begin{align*}
H_{\omega}(G,Z)=\sum_{j=1}^{t} P_{Z}( U_j)\cdot\Big[H_{\omega}\big(G|_{ U_j},Z|_{ U_j}\big)-\log P_{Z}( U_j)\Big].
\end{align*}
\end{itemize}
\end{cor}
\begin{IEEEproof}
We first consider the case that all the $t$ disjoint autonomous vertex subsets $U_1,U_2,\cdots,U_t$ in~$G$ are isolated.
Immediately, we can readily see that the graph
$G|_{\uplus_{j=1}^{t} U_j\to u_j}$ is an empty graph. It follows from Proposition~\ref{prop:clique-entro-property} and Lemma~\ref{lemma:clique_entropy_grouping_property-autonomous} that
\begin{align}
H_{\omega}(G,Z)&=H_{\omega}\big(G|_{\uplus_{j=1}^{t} U_j\to u_j},Z|_{\uplus_{j=1}^{t} U_j\to u_j}\big)+\sum_{j=1}^{t} P_{Z}( U_j)\cdot H_{\omega}\big(G|_{ U _j},Z|_{ U _j}\big)\nonumber\\
&=\sum_{j=1}^{t} P_{Z}( U_j)\cdot H_{\omega}\big(G|_{ U _j},Z|_{ U _j}\big),\label{pfcor-clique-entropy-pro-eq1}
\end{align}
where the equality \eqref{pfcor-clique-entropy-pro-eq1} follows from $$H_{\omega}\big(G|_{\uplus_{j=1}^{t} U_j\to u_j},Z|_{\uplus_{j=1}^{t} U_j\to u_j}\big)=0$$ because $G|_{\uplus_{j=1}^{t} U_j\to u_j}$ is an empty graph.

For the case that all the autonomous vertex subsets $U_1,U_2,\cdots,U_t$
 are completely-connected, we see that
$G|_{\uplus_{j=1}^{t} U_j\to u_j}$ is a complete graph. By Proposition~\ref{prop:clique-entro-property} and Lemma~\ref{lemma:clique_entropy_grouping_property-autonomous}, we have
\begin{align}
H_{\omega}(G,Z)&=H_{\omega}\big(G|_{\uplus_{j=1}^{t} U_j\to u_j},Z|_{\uplus_{j=1}^{t} U_j\to u_j}\big)+\sum_{j=1}^{t} P_{Z}( U_j)\cdot H_{\omega}\big(G|_{ U _j},Z|_{ U _j}\big)\nonumber\\
&=H\big(Z|_{\uplus_{j=1}^{t} U_j\to u_j}\big)+\sum_{j=1}^{t} P_{Z}( U_j)\cdot H_{\omega}\big(G|_{ U _j},Z|_{ U _j}\big)\label{pfcor-clique-entropy-pro-eq2}\\
&=\sum_{j=1}^{t} P_{Z}( U_j)\cdot\Big[-\log P_{Z}( U_j) +H_{\omega}\big(G|_{ U_j},Z|_{ U_j}\big)\Big],\label{pfcor-clique-entropy-pro-eq3}
\end{align}
where the equality \eqref{pfcor-clique-entropy-pro-eq2} follows from $$H_{\omega}\big(G|_{\uplus_{j=1}^{t} U_j\to u_j},Z|_{\uplus_{j=1}^{t} U_j\to u_j}\big)=H\big(Z|_{\uplus_{j=1}^{t} U_j\to u_j}\big)$$ because $G|_{\uplus_{j=1}^{t} U_j\to u_j}$ is a complete graph, and the equality \eqref{pfcor-clique-entropy-pro-eq3} follows because the probability distribution of the random variable
 $Z|_{\uplus_{j=1}^{t} U_j\to u_j}$, which  takes  values on the vertex set $\{u_1,u_2,\cdots,u_t\}$  of $G|_{\uplus_{j=1}^{t} U_j\to u_j}$, is
\begin{equation*}
\begin{split}
P_{ Z|_{\uplus_{j=1}^{t} U_j\to u_j}}(u_i)=P_{Z}( U_i )=\sum\limits_{v\in  U_i }P_{Z}(v), &\qquad \forall ~ 1\leq i\leq t.
\end{split}
\end{equation*}
The corollary is thus proved.
\end{IEEEproof}

\section{Lower Bound on the Computing Capacity $\mathcal{C}(\mathcal{N},X_{S},f)$}\label{sec-lower-bound}

In this section, we will prove a general lower bound on the computing capacity $\mathcal{C}(\mathcal{N},X_{S},f)$, where ``general'' means that the lower bound is applicable to  arbitrary network topologies, arbitrary  information sources and arbitrary target functions.
Before discussing further,
we follow from \cite{Guang_Improved_upper_bound} to present some graph-theoretic notations and definitions.
Consider a
network $\mathcal{N}= (\mathcal{G}, S, \rho)$, where we recall that $\mathcal{G}=(\mathcal{V},\mathcal{E})$ is a directed acyclic graph.
For two nodes $u$ and $v$ in $\mathcal{V}$, we write  $u\to v$ if there exists a directed path from $u$ to $v$; and otherwise we say that $v$ is \emph{separated from} $u$, denoted by $u \nrightarrow v$.
  Given an edge subset $C \subseteq \mE$, we define three subsets of source nodes:
  \begin{equation}\label{def:KC-IC-JC}
\begin{aligned}
 K_C & =  \big\{ \sigma\in S:~\exists\ e\in C \text{ s.t. } \sigma\rightarrow \textup{tail}(e) \big\},\\
 I_C & = \big\{ \sigma\in S:~ \sigma \nrightarrow \rho \text{ upon deleting the edges in $C$ from $\mathcal{E}$} \big\},\\
 J_C & = K_C \setminus  I_C.
\end{aligned}
\end{equation}
 Here,~$J_C$ is the subset of the source nodes $\sigma$ satisfying that there exists not only a path from $\sigma$ to $\rho$ passing through an edge in~$C$ but also a path from $\sigma$ to $\rho$ not passing through any edge in~$C$. We can readily see that $I_{C}\subseteq K_{C}$, $K_C= I_C\cup J_C$ and $I_C\cap J_C = \emptyset$.
 Further, we say that an edge set~$C$ is  a {\em cut set} if and only if $I_C\neq \emptyset$. Let $\Lambda(\mathcal{N})$ be the family of all the cut sets in the network~$\mathcal{N}$, i.e.,
\begin{align*}
\Lambda(\mathcal{N})=\big\{ C\subseteq \mE:\ I_C \neq \emptyset \big\}.
\end{align*}
In particular, we say that a cut set $C$ is a {\em global cut set} if $I_C=S$.


\begin{definition}[\!\!{\cite[Definition~2]{Guang_Improved_upper_bound} and \cite[Definition~3]{Guang_Zero-Fun-Compre}}]\label{def:strong_partition}
	Let $C\in\Lambda(\mathcal{N})$ be a cut set and $\mathcal{P}_{C}=\{C_{1}, C_{2},\cdots,C_{t}\}$
	be a partition of the cut set $C$. The partition $\mathcal{P}_{C}$ is said to be a strong
	partition of $C$ if the following two conditions are satisfied:
	\begin{enumerate}
		\item $I_{C_{i}}\neq\emptyset$,\quad $\forall~ 1\leq i\leq t;$
		\item $I_{C_{i}}\cap K_{C_{j}}=\emptyset$,\quad$\forall~ 1\leq i,j\leq t $ and $i\neq j$.\footnote{There is a typo in the original definition of strong partition {\cite[Definition~2]{Guang_Improved_upper_bound}}, where in 2), ``$I_{C_i}\cap I_{C_j}=\emptyset$'' in~{\cite[Definition~2]{Guang_Improved_upper_bound}} should be ``$I_{C_{i}}\cap K_{C_{j}}=\emptyset$'' as stated in \cite[Definition~3]{Guang_Zero-Fun-Compre}.}
	\end{enumerate}
\end{definition}

For a cut set $C$ in $\Lambda(\mathcal{N})$,  the partition $\{C\}$ is called the
\emph{trivial strong partition} of $C$.
We now consider a cut set $C\subseteq\Lambda(\mathcal{N})$ and  a strong partition $\mathcal{P}_{C} = \{C_{1}, C_{2}, \cdots, C_{m}\}$  of $C$.  Let $I=I_C$, $J=J_C$ and $I_{\ell}=I_{C_{\ell}}$ for $1\leq \ell \leq m$ for notational simplicity. By~\eqref{def:KC-IC-JC}, it is clear that $I_{{\ell}}\subseteq I$ for  $1\leq \ell \leq m.$  Thus, we have
$\cup^m_{\ell=1}I_{\ell}\subseteq I,$ and accordingly let $L=I\setminus (\cup^m_{\ell=1}I_{\ell})$, which consists of the source nodes from each of which all the paths (not necessarily edge-disjoint) to the sink node $\rho$ pass through at least two  cut sets in the strong partition $\mathcal{P}_{C}$ of~$C$. Clearly, we further see that $\{I_{1},I_{2},\cdots,I_{m},L\}$ forms a partition of $I$.

Following from \cite{Guang_Improved_upper_bound}, we suppose that the argument of the target function~$f$ with subscript $i$ always stands for the symbol generated by the $i$th source node $\sigma_{i}$, and so we  ignore the order of the arguments of the target function $f$.
For example, let $S=\{\sigma_1,\sigma_2,\sigma_3,\sigma_4\}$, $I=\{\sigma_2,\sigma_4\}$, and
$J=\{\sigma_1,\sigma_3\}$. Then we
regard $f(x_I,x_J)$ and $f(x_J,x_I)$ as being the same as $f(x_S)$,
i.e.,
$$f (x_2, x_4, x_1, x_3) = f (x_1, x_3, x_2, x_4) = f (x_1, x_2, x_3, x_4).$$
This abuse of notation should cause no ambiguity and would
greatly simplify the notation.

We now present two equivalence relations, which will be used to define the characteristic graph in the following subsection.
 \begin{definition}[\!\!{\cite[Definition~1]{Guang_Improved_upper_bound}}]\label{def:I_a_j_equiv'}
	Consider two disjoint sets $I,J\subseteq S$ and a fixed ${a}_{J}\in \mathcal{A}^{J}$. For any ${b}_{I}$ and ${b}'_{I}\in \mathcal{A}^{I}$, we say ${b}_{I}$ and ${b}'_{I}$ are $(I,{a}_{J})$-equivalent if
	\begin{equation}\nonumber
		f({b}_{I},{a}_{J},{d})=	f({b}'_{I},{a}_{J},{d}),\quad \forall~ {d}\in\mathcal{A}^{ S\setminus(I\cup J)}.
	\end{equation}
\end{definition}
\begin{definition}[\!\!{\cite[Definition~3]{Guang_Improved_upper_bound}}]\label{def:I_a_L_a_j_equiv'}
	Let $I$ and $J$ be two disjoint subsets of $S$. Let
	$I_{\ell},~\ell= 1, 2,\cdots, m,$ be~$m$ disjoint subsets of $I$ and let $L = I\setminus(\cup_{\ell=1}^{m}I_{\ell})$. For given $a_{J}\in\mathcal{A}^{J}$ and $a_{L}\in\mathcal{A}^{L}$, we say that
	$b_{I_{\ell}}$	 and $b_{I_{\ell}}'$	 in $\mathcal{A}^{I_{\ell}}$	 are $(I_{\ell}, a_{L}, a_{J} )$-equivalent for $1\leq\ell \leq m$,
	if for each $c_{I_{j}}\in\mathcal{A}^{I_{j}} $ with $1\leq j\leq m$ and $j\neq \ell$,
	\begin{align*}
\big(b_{I_{\ell}}\,,a_{L},\,\{c_{I_{j}},\,1\leq j\leq m,\, j\neq \ell\}\big)\in\mathcal{A}^{I}\quad\text{and}\quad
	\big(b_{I_{\ell}}',\,a_{L},\,\{c_{I_{j}},\,1\leq j\leq m,\, j\neq \ell\}\big)\in\mathcal{A}^{I}\end{align*}
		are $(I, a_{J})$-equivalent.
	\end{definition}

For a fixed ${a}_{J}\in \mathcal{A}^{J}$,  the $(I,{a}_{J})$-equivalence
relation induces a partition of $\mathcal{A}^{I}$  and the blocks in the partition  are called \emph{$(I,{a}_{J})$-equivalence classes}. We use $\mathrm{Cl}[a_J ]$ to denote an $(I, a_J )$-equivalence class. In particular, when $a_{J} $ is clear from the context, we use $\mathrm{Cl}$ to denote $\mathrm{Cl}[a_J ]$ for notational simplicity.
In addition, we further fix $a_{L}\in\mathcal{A}^{L}$. Then, the $(I_{\ell}, a_{L}, a_{J} )$-equivalence relation  for each $1\leq\ell\leq m$ partitions~$\mathcal{A}^{I_{\ell}}$
into \emph{$(I_{\ell}
, a_{L}, a_{J})$-equivalence
classes}.
We use $\mathrm{cl}_{I_{\ell}}[a_L,a_J]$ to denote an $(I_{\ell}, a_{L}, a_{J} )$-equivalence class. Similarly, when $a_{L}$ and $a_{J} $ are clear from the context, we use $\mathrm{cl}_{I_{\ell}}$ to denote $\mathrm{cl}_{I_{\ell}}[a_L,a_J]$ for notational simplicity.

\subsection{Lower Bound}\label{sub-sec:Lower-bound}

We continue to consider a cut set
$C\in\Lambda(\mathcal{N})$  and a strong partition $\mathcal{P}_{C} = \{C_{1}, C_{2}, \cdots, C_{m}\}$ of $C$.
We now present the definition of the characteristic graph associated with the strong partition $\mP_C$ and the target function $f$.
\begin{definition}\label{def-char-graph-k=1}
Let $C\in\Lambda(\mathcal{N})$ be a cut set and $\mathcal{P}_{C} = \{C_{1}, C_{2}, \cdots, C_{m}\}$ be a strong partition of $C$. Let $I=I_{C}$, $J=J_{C}$ and $I_{\ell}=I_{C_{\ell}}$ for $1\leq\ell \leq m$,  and accordingly $L=I\setminus(\cup_{\ell=1}^{m} I_{\ell})$.
Then, the characteristic graph $G_{\mP_C,f}=(V_{\mP_C,f},E_{\mP_C,f})$ associated with the strong partition $\mathcal{P}_{C}$ and the target function~$f$ is defined to be an undirected graph, where the vertex set $V_{\mP_C,f}$
 consists of  all the vectors of source messages in $\mathcal{A}^{I\cup J}$ ($=\mathcal{A}^{K_C}$), and
  two vertices
\begin{align*}
(x_{I},x_{J})=\big(\{x_{I_\ell},\,1\leq \ell\leq m\},\,x_L,\,x_J\big)\in \mathcal{A}^{I\cup J}~~\text{and}~~
(x_{I}',x_{J}')=\big(\{x_{I_\ell}',\,1\leq \ell\leq m\},\,x_L',\,x_J'\big)\in\mathcal{A}^{I\cup J}
\end{align*}
 are connected  if  $
    x_J=x_J'
    $ and  one of the following two conditions are satisfied:
\begin{enumerate}
		\item ${x}_{I}$ and $x_I'$ are not $(I,{x}_{J})$-equivalent;

    \item ${x}_{I}$ and $x_I'$ are  $(I,{x}_{J})$-equivalent, but
      $x_L=x_L'$ and there exists an index $\ell$ for $1\leq\ell\leq m$ such that $x_{I_\ell}$ and $x_{I_\ell}'$ are not $(I_{\ell}, x_{L}, x_{J} )$-equivalent.
     \end{enumerate}
 \end{definition}

We note that the characteristic graph $G_{\mP_C,f}$ only depends on the strong partition $\mathcal{P}_{C}$ and the target function~$f$.
To  simplify the notation, when  $f$ is clear from the context,  we write $G_{\mP_C}=(V_{\mP_C},E_{\mP_C})$ instead of $G_{\mP_C,f}=(V_{\mP_C,f},E_{\mP_C,f})$, and further, when  $f$ and $\mP_C$ are both clear from the context, we write $G=(V,E)$ to replace $G_{\mP_C,f}=(V_{\mP_C,f},E_{\mP_C,f})$.
 We further remark that  $(G_{\mathcal{P}_C},X_{I_C\cup J_C})$ is a probabilistic graph.
The following example is given to illustrate such a characteristic graph $G_{\mP_C}$.

\begin{example}\label{ex:G_X-G_XY_a_nontrival_examp}
\begin{figure}
	\centering
\begin{minipage}[b]{0.5\textwidth}
	\centering
\tikzstyle{format}=[draw,circle,fill=gray!30,minimum size=6pt, inner sep=0pt]
	\begin{tikzpicture}
		\node[format](a1)at(-3,3.5){};
		\node[format](a2)at(0,3.5){};
		\node[format](a3)at(3,3.5){};
		\node[format](r1)at(-1.5,1.75){};
		\node[format](r2)at(1.5,1.75){};
		\node[format](p)at(0,0){};
		\draw[->,>=latex](a1)--(r1) node[midway, auto,swap, left=0mm] {\small{$e_1$}};
		\draw[->,>=latex](a2)--(r1) node[midway, auto,swap, left=0mm] {\small{$e_{2}$}};
		\draw[->,>=latex](a2)--(r2)node[midway, auto,swap, right=0mm] {\small{$e_{3}$}};
		\draw[->,>=latex](r1)--(p)  node[midway, auto,swap, left=0mm] {\small{$e_5$}};
		\draw[->,>=latex](r2)--(p)  node[midway, auto,swap, right=0mm] {\small{$e_6$}};
		\draw[->,>=latex](a3)--(r2) node[midway, auto,swap, right=0mm] {\small{$e_{4}$}};
		\node at (-3,3.9){ $\sigma_{1}$};
		\node at (0,3.9){ $\sigma_{2}$};
		\node at (3,3.9){ $\sigma_{3}$};
		\node at (-1.75,1.45){ $v_{1}$};
		\node at (1.75,1.45){ $v_{2}$};
		\node at (0,-0.4){ $\rho$};
	\end{tikzpicture}
	\caption{The network function computation model $(\mathcal{N},X_S,f)$.}
	\label{fig:N_1}
\end{minipage}%
\begin{minipage}[b]{0.5\textwidth}
\centering
\begin{tikzpicture}

\tikzstyle{n0} = [circle,minimum size=2mm,text centered]
\tikzstyle{v} = [draw,circle,fill=gray!30,minimum size=6pt, inner sep=0pt]

\node (a) [n0] {};
\node (b)[n0,right of=a,xshift=2cm]{};
\node (c)[n0,below of=a,yshift=-0.8cm]{};
\node (d)[n0,below of=b,yshift=-0.8cm]{};
\node(1) [v,above of=a,yshift=0.3cm]{};\node [above of=1,yshift=-0.6cm]{\small{$(0,0,0)$}};
\node(2) [v,above of=b,yshift=0.3cm]{};\node [above of=2,yshift=-0.6cm]{\small{$(0,0,1)$}};
\node(3) [v,left of=a,xshift=0.05cm]{};\node [left of=3,xshift=0.2cm]{\small{$(1,0,0)$}};
\node(4) [v,right of=b,xshift=0.05cm]{};\node [right of=4,xshift=-0.2cm]{\small{$(1,0,1)$}};
\node(5) [v,left of=c,xshift=0.05cm]{};\node [left of=5,xshift=0.2cm]{\small{$(0,1,0)$}};
\node(6) [v,right of=d,xshift=0.05cm]{};\node [right of=6,xshift=-0.cm]{\small{$(0,1,1)$}};
\node(7) [v,below of=c,yshift=-0.3cm]{};\node [below of=7,yshift=0.6cm]{\small{$(1,1,0)$}};
\node(8) [v,below of=d,yshift=-0.3cm]{};\node [below of=8,yshift=0.6cm]{\small{$(1,1,1)$}};
\draw[thick](1)--(2);
\draw[thick](1)--(3);\draw[thick](1)--(4);
\draw[thick](1)--(5);\draw[thick](1)--(6);\draw[thick](1)--(7);\draw[thick](1)--(8);
\draw[dashed, thick](2)--(3);\draw[thick](2)--(4);\draw[thick](2)--(6);\draw[thick](2)--(7);
\draw[thick](2)--(8);\draw[thick](3)--(4);\draw[thick](3)--(6);\draw[thick](3)--(7);
\draw[thick](3)--(8);
\draw[thick](4)--(8);\draw[thick](4)--(5);
\draw[thick](5)--(6);\draw[thick](5)--(7);\draw[thick](5)--(8);
\draw[dashed,,thick](6)--(7);\draw[thick](6)--(8);\draw[thick](7)--(8);
\end{tikzpicture}
\caption{The characteristic graph $G_{\mathcal{P}_C}$.}\label{fig:multi_char_G_XY}
\vspace{7mm}
\end{minipage}
\end{figure}

Consider a network function computation model $(\mathcal{N},X_S,f)$ as depicted in Fig.\,\ref{fig:N_1}, where
in the network $\mathcal{N}$,
there are three source nodes
$\sigma_{1},\sigma_{2},\sigma_{3}$ and a single sink node $\rho$;
the information sources~$X_1, X_2, X_3$ are independent and identically distributed (i.i.d.) random variables according to the uniform distribution over the alphabet $\mathcal{A}=\{0,1\}$; and the target function~$f$ is the  arithmetic sum
$
f(x_{1},x_{2},x_3)=x_{1}+x_{2}+x_{3}
$ with $\Im f=\{0,1,2,3\}$.\footnote{This specified model has been investigated deeply for the case of the fixed-length network function-computing coding in the literature, e.g., \cite{Guang_Improved_upper_bound,Appuswamy11,Huang_Comment_cut_set_bound}.}

We consider the global cut set  $C=\{e_5,e_6\}$, which has a unique nontrivial strong partition $\mathcal{P}_C=\big\{C_1=\{e_5\},C_2=\{e_6\}\big\}$.
We can see that $I\triangleq I_{C}=S=\{\sigma_1,\sigma_2,\sigma_3\}$, $J\triangleq J_{C}=\emptyset$, $I_{1}\triangleq I_{C_1}=\{\sigma_1\}$, $I_{2}\triangleq I_{C_2}=\{\sigma_3\}$,  and accordingly $L\triangleq I\setminus(I_{1}\cup I_2)=\{\sigma_2\}$. By Definition~\ref{def-char-graph-k=1}, for
the characteristic graph $G_{\mP_C}=(V_{\mP_C},E_{\mP_C})$ associated with the strong partition $\mP_C$ and the  arithmetic sum~$f$,
 the vertex set is
$$
V_{\mathcal{P}_C}=\mathcal{A}^{I\cup J}=\mathcal{A}^S=\{0,1\}^3.
$$
See Fig.\,\ref{fig:multi_char_G_XY}.
We now specify the edge set $E_{\mathcal{P}_C}$.
Consider two vertices $(x_{I},x_{J})$ and $(x_{I}',x_{J}')$ in  $V_{\mathcal{P}_C}$.
Since~$J=\emptyset$ implying that $ \mathcal{A}^{J}$ only contains an empty vector, the condition $x_J=x_J'$ in Definition~\ref{def-char-graph-k=1} is always satisfied.
With this, we  consider the edges induced by the conditions 1) and 2) in Definition~\ref{def-char-graph-k=1}, respectively.

\begin{itemize}

\item First,
by Example~1 in \cite{Guang_Improved_upper_bound},
 all the $(I,x_J)$-equivalence classes are
 \begin{equation}\label{ex-equiv-class}
\begin{aligned}
&\Cl_1= \big\{(0,0,0)\big\}, \quad \Cl_2=\big\{(0,0,1),(0,1,0),(1,0,0) \big\}, \\
&\Cl_3= \big\{ (0,1,1),(1,0,1),(1,1,0) \big\}, \quad \Cl_4=\big\{(1,1,1)\big\}.
\end{aligned}
\end{equation}
We remark that the vectors of  source messages with the same arithmetic sum are in the same equivalence class.
It follows from the condition~1) in Definition~\ref{def-char-graph-k=1} that the two vertices
  $(x_{I},x_{J})=x_{I}$ and $(x_{I}',x_{J}')=x_{I}'$
 are connected if $x_{I}$ and $x_{I}'$ are in  different  $(I,x_J)$-equivalence classes, or equivalently, $f(x_I)\neq f(x_I')$ (where  $I=S$).  All such
  edges in $E_{\mathcal{P}_C}$ (satisfying the condition~1) in Definition~\ref{def-char-graph-k=1})
   are depicted in Fig.\,\ref{fig:multi_char_G_XY} with solid lines.

\item Furthermore, we consider the edges in  $E_{\mathcal{P}_C}$ induced by the condition~2)  in Definition~\ref{def-char-graph-k=1}.
We write the two  vertices
  $x_{I}=\big(\{x_{I_1},x_{I_2}\},\,x_L\big)$ and $x_{I}'=\big(\{x_{I_1}',x_{I_2}'\},\,x_L'\big)$.
 If
  $x_{I}$ and~$x_{I}'$  satisfy the condition~2), then
   both of them must be $(I,x_J)$-equivalent and thus are in either $\Cl_2$ or $\Cl_3$, because  $\Cl_1$ and $\Cl_4$ are two singleton equivalence classes.

 \textbf{Case~1:} $x_{I}, x_{I}'\in \Cl_2$.

By the requirement of  $x_L=x_L'$ in the condition 2) and $L=\{\sigma_2\}$, it can only be  $x_2=x_2'=0$, and then both $x_I$ and $x_I'$ are $(0,0,1)$ and $(1,0,0)$. Without loss of generality, we let
 \begin{align*}
 x_{I}=(x_1=0,x_2=0,x_3=1)\quad\text{and}\quad
 x_{I}'=(x_1'=1,x_2'=0,x_3'=0).
 \end{align*}
By Example~2 in \cite{Guang_Improved_upper_bound}, all the $(I_1,x_L=0, {x}_J)$-equivalence classes are $\mathrm{cl}_{I_{1},1}=\{0\}$ and $\mathrm{cl}_{I_{1},2}=\{1\}$, which immediately implies that $x_{I_1}=x_1=0$ and  $x_{I_1}'=x_1'=1$ are not $(I_{1}, x_{L}=0, x_{J} )$-equivalent.
Then, the two  vertices  $x_{I}$ and $x_{I}'$ satisfy the condition~2) and thus
 are connected. See the dashed line between  $(0,0,1)$ and $(1,0,0)$ in Fig.\,\ref{fig:multi_char_G_XY}.

 \textbf{Case~2:} $x_{I},x_{I}'\in \Cl_3$.

Similar to Case~1, we have $x_L=x_L'=1$ (i.e., $x_2=x_2'=1$), and  both $x_I$ and $x_I'$ are $(0,1,1)$ and $(1,1,0)$, say
$
 x_{I}=(x_1=0,x_2=1,x_3=1)$ and $
 x_{I}'=(x_1'=1,x_2'=1,x_3'=0).
$
By Example~2 in \cite{Guang_Improved_upper_bound}, all the $(I_2,x_L=1, {x}_J)$-equivalence classes are $\mathrm{cl}_{I_{2},1}=\{0\}$ and $\mathrm{cl}_{I_{2},2}=\{1\}$, which immediately shows that $x_{I_2}=x_3=1$ and  $x_{I_2}'=x_3'=0$ are not $(I_{2}, x_{L}=1, x_{J} )$-equivalent. Thus, the two vertices $x_{I}$ and $x_{I}'$
 are connected. See the dashed line between  $(0,1,1)$ and $(1,1,0)$ in Fig.\,\ref{fig:multi_char_G_XY}.
\end{itemize}

We have thus  completed the specification of the characteristic graph $G_{\mathcal{P}_C}$.
\end{example}

%

With the characteristic graph as defined, we in the following present a general lower bound on the computing capacity, which can be applicable to arbitrary network topologies, arbitrary information sources and arbitrary target functions. The proof of the lower bound is deferred to Section~\ref{sec-pf-lower-bound1}.

\begin{theorem}\label{thm-low-bound}
Consider a  model of network function computation  $(\mathcal{N},X_{S},f)$.
Then,
\begin{align}\label{thm1-eq}
\mathcal{C}(\mathcal{N},X_S,f)\geq\max_{(C,\mathcal{P}_C)\in\Lambda(\mathcal{N})\times \textbf{\textup{P}}_C}
\frac{~H_{\omega}(G_{\mathcal{P}_C},X_{I_C\cup J_C})~}{|C|},
\end{align}
where $\textbf{\textup{P}}_C$ denotes the collection of all the strong partitions of a cut set $C\in\Lambda(\mathcal{N})$.
\end{theorem}

In the following example, we continue to consider the model $(\mathcal{N},X_S,f)$ discussed in Example~\ref{ex:G_X-G_XY_a_nontrival_examp} to specify the general  lower bound obtained in Theorem~\ref{thm-low-bound}.

\begin{example}\label{ex2}
Continuing from Example~\ref{ex:G_X-G_XY_a_nontrival_examp},
 we consider the probabilistic  graph $(G_{\mathcal{P}_C},X_{I\cup J})$, where we recall that
  $X_{I\cup J}=X_I=X_S=(X_1,X_2,X_3)$ with
  the probability distribution
  \begin{align*}
P_{X_S}(x_1,x_2,x_3)={\rm{Pr}}\big\{X_1=x_1,X_2=x_2,X_3=x_3\big\}=\frac{1}{8},
\qquad\forall~x_i\in\{0,1\}~\textup{for }i= 1,2,3.
\end{align*}
  We now calculate the clique entropy $H_{\omega}(G_{\mathcal{P}_C},X_{I\cup J})$, i.e., $H_{\omega}(G_{\mathcal{P}_C},X_S)$.
 We note that each $(I,x_J)$-equivalence class  $\Cl_i,i=1,2,3,4$  (cf.~\eqref{ex-equiv-class}) is an autonomous vertex subset in $G_{\mathcal{P}_C}$, and all of them constitute  a partition of $V_{\mathcal{P}_C}=\mathcal{A}^S$. Further, we see that all $\Cl_i,i=1,2,3,4$ are completely-connected in the characteristic graph $G_{\mathcal{P}_C}$. In the following, we write $G=(V,E)$ to replace $G_{\mP_C}=(V_{\mP_C},E_{\mP_C})$ for notational simplicity, and by
 Corollary~\ref{cor-clique-entropy-pro1}, we calculate that
\begin{align}
H_{\omega}(G,X_S)&=\sum_{i=1}^4 P_{X_S}(\Cl_i)\cdot\Big[ H_{\omega}\big(G|_{\Cl_i},X_S|_{\Cl_i}\big)-\log P_{X_S}(\Cl_i)\Big]\nonumber\\
&=\sum_{i=1}^4 P_{X_S}(\Cl_i)\cdot H_{\omega}\big(G|_{\Cl_i},X_S|_{\Cl_i}\big)-\sum_{i=1}^4 P_{X_S}(\Cl_i)\cdot\log P_{X_S}(\Cl_i)\nonumber\\
&=P_{X_S}(\Cl_2)\cdot H_{\omega}\big(G|_{\Cl_2},X_S|_{\Cl_2}\big)
+P_{X_S}(\Cl_3)\cdot H_{\omega}\big(G|_{\Cl_3},X_S|_{\Cl_3}\big)
+3\cdot\left(\!1-\frac{\log 3}{4}\!\right)\label{ex-illstr-3}\\
&=\frac{3}{8}\cdot H_{\omega}\big(G|_{\Cl_2},X_S|_{\Cl_2}\big)+
\frac{3}{8}\cdot H_{\omega}\big(G|_{\Cl_3},X_S|_{\Cl_3}\big)+3\cdot\left(1-\frac{\log 3}{4}\right),\label{ex-illstr-3.1}
\end{align}
where the equalities~\eqref{ex-illstr-3} and~\eqref{ex-illstr-3.1}  are explained as follows.   It follows from
 $|\Cl_1|=|\Cl_4|=1$ that $G|_{\Cl_1}$ and $G|_{\Cl_4}$ are two empty graphs. By Proposition~\ref{prop:clique-entro-property}, we immediately obtain that $$H_{\omega}\big(G|_{\Cl_1},X_S|_{\Cl_1}\big)
=H_{\omega}\big(G|_{\Cl_4},X_S|_{\Cl_4}\big)=0.$$
On the other hand, by a calculation of $P_{X_S}(\Cl_1)=P_{X_S}(\Cl_4)=1/8$ and $P_{X_S}(\Cl_2)=P_{X_S}(\Cl_3)=3/8$, we have
\begin{align*}
-\sum_{i=1}^4 P_{X_S}(\Cl_i)\cdot\log P_{X_S}(\Cl_i)=3\cdot\left(1-\frac{\log 3}{4}\right)
\end{align*}
and the equality~\eqref{ex-illstr-3.1}.

\begin{figure}
\centering
\begin{tikzpicture}
\tikzstyle{n0} = [circle,minimum size=2mm,text centered]
\tikzstyle{v} = [draw,circle,fill=gray!30,minimum size=6pt, inner sep=0pt]

\node (A) [n0] {};
\node (B)[n0,right of=A,xshift=2cm]{};
\node (C)[n0,below of=A,yshift=-1.2cm]{};
\node (D)[n0,right of=C,xshift=0.6cm]{};

\node(1) [v,above of=A,yshift=0.3cm]{};\node [above of=1,yshift=-0.6cm]{\small{$(1,0,0)$}};
\node(2) [v,above of=B,yshift=0.3cm]{};\node [above of=2,yshift=-0.6cm]{\small{$(0,0,1)$}};
    \node(3) [v,above of=D,yshift=0.3cm]{};\node [above of=3,yshift=-0.6cm]{\small{$(0,1,0)$}};
    \draw[thick] (1) -- (2); 
\end{tikzpicture}\vspace{-3em}
\caption{The  graph $G|_{\Cl_2}$.}\label{fig:multi_char_G__{Cl_2}}

\end{figure}
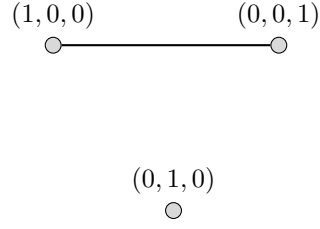

Furthermore, we are ready to calculate the clique entropies $H_{\omega}\big(G|_{\Cl_2},X_S|_{\Cl_2}\big) $ and $H_{\omega}\big(G|_{\Cl_3},X_S|_{\Cl_3}\big)$ in the following.
  We first consider the probabilistic  graph $(G|_{\Cl_2},X_S|_{\Cl_2})$, where the graph $G|_{\Cl_2}$ is the projection of $G$ onto $\Cl_2$ as depicted in Fig.\,\ref{fig:multi_char_G__{Cl_2}},
  and the probability distribution of $X_S|_{\Cl_2}$ is given by
  \begin{equation*}
P_{X_S|_{\Cl_2}}(x_1,x_2,x_3)=
\frac{P_{X_S}(x_1,x_2,x_3)}{P_{X_S}(\Cl_2)}=\frac{1}{3},\qquad\forall~(x_1,x_2,x_3)\in \Cl_2.
\end{equation*}
In the graph $G|_{\Cl_2}$, we partition the vertex set $\Cl_2$ into
\begin{equation*}
\Cl_{2,0}\triangleq\{(0,0,1),~(1,0,0)\}\quad\text{ and }\quad\Cl_{2,1}\triangleq\{(0,1,0)\}.
\end{equation*}
We can readily see that $\Cl_{2,0}$ and $\Cl_{2,1}$  are disjoint and autonomous in $G|_{\Cl_2}$.
We further see that both $\Cl_{2,0}$ and $\Cl_{2,1}$ are isolated in $G|_{\Cl_2}$. See  Fig.\,\ref{fig:multi_char_G__{Cl_2}}.
 With this, by Corollary~\ref{cor-clique-entropy-pro1}, we can calculate that
\begin{align}
&H_{\omega}\big(G|_{\Cl_2},X_S|_{\Cl_2}\big)\nonumber\\
&= P_{X_S|_{\Cl_2}}\big(\Cl_{2,0}\big)\cdot H_{\omega}\big(G|_{\Cl_2}|_{\Cl_{2,0}},X_S|_{\Cl_2}|_{\Cl_{2,0}}\big)
+P_{X_S|_{\Cl_2}}\big(\Cl_{2,1}\big)\cdot H_{\omega}\big(G|_{\Cl_2}|_{\Cl_{2,1}},X_S|_{\Cl_2}|_{\Cl_{2,1}}\big)
\nonumber\\
&=P_{X_S|_{\Cl_2}}\big(\Cl_{2,0}\big)\cdot H_{\omega}\big(G|_{\Cl_{2,0}},X_S|_{\Cl_{2,0}}\big)
+P_{X_S|_{\Cl_2}}\big(\Cl_{2,1}\big)\cdot H_{\omega}\big(G|_{\Cl_{2,1}},X_S|_{\Cl_{2,1}}\big)
\label{ex-illstr-4.1-00}\\
&=P_{X_S|_{\Cl_2}}\big(\Cl_{2,0}\big)\cdot H_{\omega}\big(G|_{\Cl_{2,0}},X_S|_{\Cl_{2,0}}\big)
\label{ex-illstr-4.1-0}\\
&=P_{X_S|_{\Cl_2}}\big(\Cl_{2,0}\big)\cdot H\big(X_S|_{\Cl_{2,0}}\big)=\frac{2}{3},
\label{ex-illstr-4.1}
\end{align}
where the equality~\eqref{ex-illstr-4.1-00} follows from the fact that
the projections of $G|_{\Cl_2}$  and  $X_S|_{\Cl_2}$ onto $\Cl_{2,0}$ (resp. $\Cl_{2,1}$) are the same as $G|_{\Cl_{2,0}}$ and  $X_S|_{\Cl_{2,0}}$
(resp. $G|_{\Cl_{2,1}}$ and  $X_S|_{\Cl_{2,1}}$);
the equality  \eqref{ex-illstr-4.1-0}  follows because $|\Cl_{2,1}|=1$ implies that $G|_{\Cl_{2,1}}$ is an empty graph and thus $H_{\omega}\big(G|_{\Cl_{2,1}},X_S|_{\Cl_{2,1}}\big)
=0$ by
 Proposition~\ref{prop:clique-entro-property};
and the  equalities in \eqref{ex-illstr-4.1} hold because
 $G|_{\Cl_{2,0}}$ is a complete graph and thus
 $$H_{\omega}\big(G|_{\Cl_{2,0}},X_S|_{\Cl_{2,0}}\big)=H(X_S|_{\Cl_{2,0}})=\frac{2}{3}$$
by Proposition~\ref{prop:clique-entro-property}.

 By the same argument, we  can also calculate
\begin{align}
H_{\omega}\big(G|_{\Cl_3},X_S|_{\Cl_3}\big)=\frac{2}{3}.
\label{ex-illstr-5.1}
\end{align}
 Combining  \eqref{ex-illstr-4.1} and \eqref{ex-illstr-5.1} with  \eqref{ex-illstr-3.1}, we obtain that
\begin{align*}
H_{\omega}(G,X_{I\cup J})=\frac{7}{2}-\frac{3}{4}\log 3.
\end{align*}

In fact, the pair $(C,\mathcal{P}_C)$ discussed here achieves
the maximum of the right hand side of  \eqref{thm1-eq}
over all pairs in $\Lambda(\mathcal{N})\times \textbf{\textup{P}}_C$, i.e.,
\begin{align*}
\max_{(C,\mathcal{P}_C)\in\Lambda(\mathcal{N})\times \textbf{\textup{P}}_C}
\frac{~H_{\omega}(G_{\mathcal{P}_C},X_{I_C\cup J_C})~}{|C|}=\frac{7}{4}-\frac{3}{8}\log 3.
\end{align*}
We have thus specified
the obtained lower bound for the model  $(\mathcal{N},X_S,f)$ as follows:
\begin{align}\label{ex:lower-bound-1}
\mathcal{C}(\mathcal{N},X_S,f)\geq\frac{7}{4}-\frac{3}{8}\log 3.
\end{align}
\end{example}

\subsection{Proof of Theorem~\ref{thm-low-bound}}\label{sec-pf-lower-bound1}

In this subsection, we  prove the general lower bound  in Theorem~\ref{thm-low-bound}. Before proving the theorem, we first  present three definitions including two equivalence
relations and the $k$-fold characteristic graph, which are the generalizations of Definitions~\ref{def:I_a_j_equiv'},~\ref{def:I_a_L_a_j_equiv'} and~\ref{def-char-graph-k=1}, respectively, from the case of $k=1$ to the general case of $k$.

 \begin{definition}[\!\!{\cite[Definition~1]{Guang_Improved_upper_bound}}]\label{def:k-shot-I_a_j_equiv'}
	Consider two disjoint sets $I,J\subseteq S$ and a fixed $\boldsymbol{a}_{J}\in \mathcal{A}^{k\times J}$. For any $\boldsymbol{b}_{ I}$ and $\boldsymbol{b}'_{ I}\in \mathcal{A}^{k\times I}$, we say $\boldsymbol{b}_{I}$ and $\boldsymbol{b}'_{I}$ are $(I,\boldsymbol{a}_{J})$-equivalent if
	\begin{equation}\nonumber
		f(\boldsymbol{b}_{I},\boldsymbol{a}_{J},\boldsymbol{d})=	f(\boldsymbol{b}'_{I},\boldsymbol{a}_{J},\boldsymbol{d}),\quad \forall~ \boldsymbol{d}\in\mathcal{A}^{k\times  S\setminus(I\cup J)}.
	\end{equation}
\end{definition}


 \begin{definition}[\!\!{\cite[Definition~4]{Guang_Improved_upper_bound}}]\label{def:k-shot-I_a_L_a_j_equiv'}
	Let $I$ and $J$ be two disjoint subsets of $S$. Let
	$I_{\ell},~\ell= 1, 2,\cdots, m,$ be~$m$ disjoint subsets of $I$ and let $L = I\setminus(\cup_{\ell=1}^{m}I_{\ell})$. For given $\boldsymbol{a}_{J}\in\mathcal{A}^{k\times J}$ and $\boldsymbol{a}_{L}\in\mathcal{A}^{k\times L}$, we say that
	$\boldsymbol{b}_{I_{\ell}}$	 and $\boldsymbol{b}_{I_{\ell}}'$	 in $\mathcal{A}^{k\times I_{\ell}}$	 are $(I_{\ell}, \boldsymbol{x}_{L}, \boldsymbol{x}_{J} )$-equivalent for $1\leq\ell \leq m$,
	if for each $\boldsymbol{c}_{I_{j}}\in\mathcal{A}^{k\times I_{j}} $ with $1\leq j\leq m$ and $j\neq \ell$,
	\begin{align*}
\big(\boldsymbol{b}_{I_{\ell}}\,,\boldsymbol{a}_{L},\,\{\boldsymbol{c}_{I_{j}},\,1\leq j\leq m,\, j\neq \ell\}\big)\in\mathcal{A}^{k\times I}\quad\text{and}\quad
	\big(\boldsymbol{b}_{I_{\ell}}',\,\boldsymbol{a}_{L},\,\{\boldsymbol{c}_{I_{j}},\,1\leq j\leq m,\, j\neq \ell\}\big)\in\mathcal{A}^{k\times I}\end{align*}
		are $(I, \boldsymbol{a}_{J})$-equivalent.
	\end{definition}

\begin{definition}\label{def-char-graph-k}
Let $C\in\Lambda(\mathcal{N})$ be a cut set and $\mathcal{P}_{C} = \{C_{1}, C_{2}, \cdots, C_{m}\}$ be a strong partition of $C$. Let $I=I_{C}$, $J=J_{C}$ and $I_{\ell}=I_{C_{\ell}}$ for $1\leq\ell \leq m$,  and accordingly $L=I\setminus(\cup_{\ell=1}^{m} I_{\ell})$.
Then, the $k$-fold  characteristic graph $G_{\mP_C,f}^{\,k}=(V_{\mP_C,f}^{\,k},E_{\mP_C,f}^{\,k})$ associated with the strong partition $\mathcal{P}_{C}$ and the target function~$f$ is defined to be an undirected graph, where the vertex set $V_{\mP_C,f}^{\,k}$
 consists of all the vectors of source messages in $\mathcal{A}^{k\times (I\cup J)}$, and
  two vertices
\begin{align*}
(\boldsymbol{x}_{I},\boldsymbol{x}_{J})=\big(\{\boldsymbol{x}_{I_\ell},\,1\leq \ell\leq m\},\,\boldsymbol{x}_L,\,\boldsymbol{x}_J\big)
\quad\text{and}\quad
(\boldsymbol{x}_{I}',\boldsymbol{x}_{J}')=\big(\{\boldsymbol{x}_{I_\ell}',\,1\leq \ell\leq m\},\,\boldsymbol{x}_L',\,\boldsymbol{x}_J'\big)
\end{align*}
 in $\mathcal{A}^{k\times (I\cup J)}$ are connected  if   $
    \boldsymbol{x}_J=\boldsymbol{x}_J'
    $ and  one of the following two conditions are satisfied:
\begin{enumerate}
   \item $\boldsymbol{x}_{I}$ and $\boldsymbol{x}_I'$ are not $(I,\boldsymbol{x}_{J})$-equivalent;

    \item    $\boldsymbol{x}_{I}$ and $\boldsymbol{x}_I'$ are  $(I,\boldsymbol{x}_{J})$-equivalent, but
      $\boldsymbol{x}_L=\boldsymbol{x}_L'$ and  there exists an index $\ell$ for $1\leq\ell\leq m$ such that $\boldsymbol{x}_{I_\ell}$ and $\boldsymbol{x}_{I_\ell}'$ are not $(I_{\ell}, \boldsymbol{x}_{L}, \boldsymbol{x}_{J} )$-equivalent.
      \end{enumerate}
 \end{definition}
Similarly, when $f$ is clear from the context,  we write $G_{\mP_C}^{\,k}=(V_{\mP_C}^{\,k},V_{\mP_C}^{\,k})$ to replace $G_{\mP_C,f}^{\,k}=(V_{\mP_C,f}^{\,k},E_{\mP_C,f}^{\,k})$ for notational simplicity. Note that  $(G_{\mP_C}^{\,k},\boldsymbol{X}_{I_C\cup J_C})$ is a probabilistic graph.
The following lemma gives a  relationship between the $k$-fold characteristic graph~$G_{\mP_C}^{\,k}$ and the $k$-fold AND product of the  characteristic graph $G_{\mP_C}$.
 \begin{lemma}\label{The Relation Between 123}
  Let $
G_{\mP_C}=(V_{\mP_C},E_{\mP_C})$ and $G_{\mP_C}^{\,k}=(V_{\mP_C}^{\,k},E_{\mP_C}^{\,k})$ be the characteristic graph and  $k$-fold characteristic graph associated with a strong partition $\mathcal{P}_{C}$ of a cut set $C$ and the target function~$f$, respectively, where $k$ is an arbitrary positive integer.
Let  $G_{\mP_C}^{\land k}=\big(V_{\mP_C}^{\land k},E_{\mP_C}^{\land k}\big)$  be the $k$-fold AND product  of~$G_{\mP_C}$.
Then,
\begin{align*}
 V_{\mP_C}^{\land k} = V^{\,k}_{\mP_C} \quad \text{and}\quad
E_{\mP_C}^{\land k} \subseteq E^{\,k}_{\mP_C}.
\end{align*}
The above relationship between  $G^{\,k}_{\mP_C}$ and $G_{\mP_C}^{\land k}$ is denoted by $
 G_{\mP_C}^{\land k} \subseteq G^{\,k}_{\mP_C}.$
\end{lemma}
\begin{IEEEproof}
 See Appendix~\ref{appendix:pf-The Relation Between 123}.
\end{IEEEproof}

In fact, Lemma~\ref{The Relation Between 123} can be enhanced further  when  the $k$-fold OR  product $G_{\mP_C}^{\vee k}=\big(V_{\mP_C}^{\vee k},E_{\mP_C}^{\vee k}\big)$ of the characteristic graph~$G_{\mP_C}$ is involved. We present this enhanced result in the following theorem. Nevertheless, we only need Lemma~\ref{The Relation Between 123} here to prove Theorem~\ref{thm-low-bound}.
\begin{theorem}\label{The-relationship-between=123-1}
Let $
G_{\mP_C}=(V_{\mP_C},E_{\mP_C})$ and $G_{\mP_C}^{\,k}=(V_{\mP_C}^{\,k},E_{\mP_C}^{\,k})$ be the characteristic graph and  $k$-fold characteristic graph associated with a strong partition $\mathcal{P}_{C}$ of a cut set $C$ and the target function~$f$, respectively, where $k$ is an arbitrary positive integer.
Let  $G_{\mP_C}^{\land k}=\big(V_{\mP_C}^{\land k},E_{\mP_C}^{\land k}\big)$ and $G_{\mP_C}^{\vee k}=\big(V_{\mP_C}^{\vee k},E_{\mP_C}^{\vee k}\big)$ be the $k$-fold AND product and $k$-fold OR product  of~$G_{\mP_C}$, respectively.
Then,$$
 G_{\mP_C}^{\land k} \subseteq G^{\,k}_{\mP_C}\subseteq G_{\mP_C}^{\vee k}.$$
More precisely,
$V_{\mP_C}^{\land k} = V^{\,k}_{\mP_C}
= V_{\mP_C}^{\vee k}$~~and~~$E_{\mP_C}^{\land k} \subseteq E^{\,k}_{\mP_C}
\subseteq  E_{\mP_C}^{\vee k}$.
\end{theorem}
For Theorem~\ref{The-relationship-between=123-1}, the proof of the part that $G^{\,k}_{\mP_C}\subseteq G_{\mP_C}^{\vee k}$ is similar to the proof of Lemma~\ref{The Relation Between 123} and so we omit it.

Now, we are ready to prove Theorem~\ref{thm-low-bound}. Let $\mbC$ be   an arbitrary admissible $k$-shot uniquely-decodable network  function-computing code     for  the model $(\mathcal{N},X_S,f)$, of which the global encoding functions are $g_e,\,e \in \mE $.
Consider a  cut set $C\in\Lambda(\mathcal{N})$, where we let $I=I_C$ and $J=J_C$. Since no directed path exist from any source node in $S\setminus(I\cup J)$ to any node in $\{\text{tail}(e):\,e\in C\}$, the vector of information sources  $\boldsymbol {X}_{S\setminus (I\cup J)}$ do not contribute to the values of $g_C(\boldsymbol {X}_S) = \big(g_e(\boldsymbol {X}_S):\, e\in C\big)$. Hence, we can write
$g_C(\boldsymbol {X}_I,\boldsymbol {X}_J,\boldsymbol {X}_{S\setminus (I\cup J)})$  as $g_C(\boldsymbol {X}_I,\boldsymbol {X}_J)$.
Next, we present the following lemma, which plays a crucial role to proving our general  lower  bound in Theorem~\ref{thm-low-bound}.
 \begin{lemma}\label{coloring=coding}
 Consider an admissible $k$-shot uniquely-decodable code  $\mbC$    for  the model $(\mathcal{N},X_S,f)$, of which the global encoding functions are $g_e,\,e \in \mE $. Let $G^{\,k}_{\mP_C}=(V^{\,k}_{\mP_C},E^{\,k}_{\mP_C})$ be the $k$-fold  characteristic graph associated with  a strong partition $\mathcal{P}_{C}$ of a cut set $C$ and the target function~$f$.
 Then, $g_{C}$ is  a coloring of $G^{\,k}_{\mP_C}$, namely that
\begin{align*}
g_{C}(\boldsymbol {x}_I,\boldsymbol {x}_J)\neq g_{C}(\boldsymbol {x}_I',\boldsymbol {x}_J')\quad \text{with}\quad I\triangleq I_{C}\,\, \text{and}\,\, J\triangleq J_{C}
\end{align*}
for any two connected vertices $(\boldsymbol {x}_I,\boldsymbol {x}_J)$ and $(\boldsymbol {x}_I',\boldsymbol {x}_J')$ in $V^{\,k}_{\mP_C}$, i.e., $$\big((\boldsymbol {x}_I,\boldsymbol {x}_J),\,(\boldsymbol {x}_I',\boldsymbol {x}_J')\big)\in E_{\mP_C}^{\,k}.$$
 \end{lemma}
  \begin{IEEEproof}
  We first consider a cut set $C\in\Lambda(\mathcal{N})$ and a strong partition $\mathcal{P}_{C} = \{C_{1}, C_{2}, \cdots, C_{m}\}$ of~$C$. For  notational simplicity, we let $I=I_{C}$, $J=J_{C}$ and $I_{\ell}=I_{C_{\ell}}$ for $1\leq\ell \leq m$,  and accordingly let $L=I\setminus(\cup_{\ell=1}^{m} I_{\ell})$. We further let $G^{\,k}_{\mP_C}=(V^{\,k}_{\mP_C},E^{\,k}_{\mP_C})$ be the $k$-fold  characteristic graph associated with  a strong partition $\mathcal{P}_{C}$ of a cut set $C$ and the target function~$f$.

Now, we let
 \begin{align*}
(\boldsymbol{x}_{I},\boldsymbol{x}_{J})=\big(\{\boldsymbol{x}_{I_\ell},\,1\leq \ell\leq m\},\,\boldsymbol{x}_L,\,\boldsymbol{x}_J\big)\quad\text{and}\quad
(\boldsymbol{x}_{I}',\boldsymbol{x}_{J}')=\big(\{\boldsymbol{x}_{I_\ell}',\,1\leq \ell\leq m\},\,\boldsymbol{x}_L',\,\boldsymbol{x}_J'\big)
\end{align*}
be two connected vertices in $V^{\,k}_{\mP_C}$ ($=\mathcal{A}^{k\times (I\cup J)}$). By the definition of $G^{\,k}_{\mP_C}$ (cf.~Definition~\ref{def-char-graph-k}),  it must be  $\boldsymbol {x}_J=\boldsymbol {x}_J'$. Next, we will prove  $g_{C}(\boldsymbol {x}_I,\boldsymbol {x}_J)\neq g_{C}(\boldsymbol {x}_I',\boldsymbol {x}_J)$ according to the two ways of edge-connection  corresponding to the conditions 1) and 2) in~Definition~\ref{def-char-graph-k}.

\textbf{Case 1:}  $\boldsymbol{x}_{I}$ and $\boldsymbol{x}_I'$ are not $(I,\boldsymbol{x}_{J})$-equivalent.

For this case,  by Definition~\ref{def:k-shot-I_a_j_equiv'}, there exists a vector of source messages  $\boldsymbol {d} \in \mathcal{A}^{ k\times S/(I\cup J)}$ such that
    \begin{equation}\label{pf-code-color-eq1}
    f(\boldsymbol {x}_I,\,\boldsymbol {x}_J,\,\boldsymbol {d}) \neq f(\boldsymbol {x}_I',\,\boldsymbol {x}_J,\,\boldsymbol {d}),
    \end{equation}
where we recall that  $\boldsymbol {x}_J=\boldsymbol {x}_J'$. We further consider the edge subset $D\triangleq\bigcup_{\sigma \in (S\setminus I)} \mathrm{Out} (\sigma)$.  Clearly, $K_D=S\setminus I$. Let $\widehat{C}\triangleq C \cup D$ and thus $\widehat{C}$ is a global cut set, i.e., $I_{\widehat{C}}=S$.
Since $g_{\mathrm{In}(\rho)}(\boldsymbol {X}_S)$  is a function of $g_{\widehat{C}}(\boldsymbol {X}_S)$ and the code $\mbC$ can compute $f(\boldsymbol {X}_S)$ with zero error, the inequality~\eqref{pf-code-color-eq1} implies that
    \begin{equation*}
    g_{\widehat{C}}(\boldsymbol {x}_I,\,\boldsymbol {x}_J,\,\boldsymbol {d})\neq g_{\widehat{C}}(\boldsymbol {x}_I',\,\boldsymbol {x}_J,\,\boldsymbol {d}). 
    \end{equation*}
 Together with $K_C=I\cup J$ and $K_D=S\setminus I$, we obtain that
  $$\big(g_{C}(\boldsymbol {x}_I,\,\boldsymbol {x}_J),g_{D}(\boldsymbol {x}_J,\,\boldsymbol {d})\big)=g_{\widehat{C}}(\boldsymbol {x}_I,\,\boldsymbol {x}_J,\,\boldsymbol {d})\neq g_{\widehat{C}}(\boldsymbol {x}_I',\,\boldsymbol {x}_J,\,\boldsymbol {d})=\big(g_{C}(\boldsymbol {x}_I',\,\boldsymbol {x}_J),g_{D}(\boldsymbol {x}_J,\,\boldsymbol {d})\big).$$
  This immediately implies that   \begin{align*}
g_{C}(\boldsymbol {x}_I,\boldsymbol {x}_J)\neq  g_{C}(\boldsymbol {x}_I',\boldsymbol {x}_J).
\end{align*}

\textbf{Case 2:} $\boldsymbol{x}_{I}$ and $\boldsymbol{x}_I'$ are  $(I,\boldsymbol{x}_{J})$-equivalent, but
      $\boldsymbol{x}_L=\boldsymbol{x}_L'$ and  there exists an index $\ell$ for $1\leq\ell\leq m$ such that $\boldsymbol{x}_{I_\ell}$ and $\boldsymbol{x}_{I_\ell}'$ are not $(I_{\ell}, \boldsymbol{x}_{L}, \boldsymbol{x}_{J} )$-equivalent.

      Without loss of generality, we assume  that  $\boldsymbol{x}_{I_{1}}$ and $\boldsymbol{x}_{I_{1}}'$ are not $(I_{{1}}, \boldsymbol{x}_{L}, \boldsymbol{x}_{J} )$-equivalent.
       By Definition~\ref{def:k-shot-I_a_L_a_j_equiv'},
       there exist  $\boldsymbol {c}_{I_j}\in \mathcal{A}^{k\times I_j}$, $2\leq j\leq m$ such that
       \begin{align*}
\big(\boldsymbol{x}_{I_{1}}\,,\boldsymbol{x}_{L},\,\{\boldsymbol{c}_{I_{j}},\,2\leq j\leq m\}\big)\quad\text{and}\quad
	\big(\boldsymbol{x}_{I_{1}}',\,\boldsymbol{x}_{L},\,\{\boldsymbol{c}_{I_{j}},\,2\leq j\leq m\}\big)\end{align*}
		in $\mathcal{A}^{k\times I}$ are not $(I, \boldsymbol{x}_{J})$-equivalent.
Furthermore, by Definition~\ref{def:k-shot-I_a_j_equiv'}, there exists   $\boldsymbol {d}' \in \mathcal{A}^{ k\times S/(I\cup J)}$, such that
\begin{align*}
&f\big(\boldsymbol {x}_{I_{1}},\,\boldsymbol {x}_L,\,\{\boldsymbol {c}_{I_j},\,2\leq j\leq m\},\,\boldsymbol {x}_J,\,\boldsymbol {d}'\big) \neq f\big(\boldsymbol {x}_{I_{1}}',\,\boldsymbol {x}_L,\,\{\boldsymbol {c}_{I_j},\,2\leq j\leq m\},\,\boldsymbol {x}_J,\,\boldsymbol {d}'\big).
\end{align*}
We still let $D\triangleq\bigcup_{\sigma \in (S\setminus I)} \mathrm{Out} (\sigma)$ and  $\widehat{C}\triangleq C \cup D$, and similarly, we have
    \begin{align*}
   & g_{\widehat{C}}\big(\boldsymbol {x}_{I_{1}},\,\boldsymbol {x}_L,\,\{\boldsymbol {c}_{I_j},\,2\leq j\leq m\},\,\boldsymbol {x}_J,\,\boldsymbol {d}'\big)\neq g_{\widehat{C}}\big(\boldsymbol {x}_{I_{1}}',\,\boldsymbol {x}_L,\,\{\boldsymbol {c}_{I_j},\,2\leq j\leq m\},\,\boldsymbol {x}_J,\,\boldsymbol {d}'\big),
    \end{align*}
or equivalently,
\begin{align*}
&\big(g_{C}(\boldsymbol {x}_{I_{1}},\,\boldsymbol {x}_L,\,\{\boldsymbol {c}_{I_j},\,2\leq j\leq m\},\,\boldsymbol {x}_J),g_D(\boldsymbol{x}_J,\boldsymbol{d}')\big)\\
&\neq\big(g_{C}(\boldsymbol {x}'_{I_{1}},\,\boldsymbol {x}_L,\,\{\boldsymbol {c}_{I_j},\,2\leq j\leq m\},\,\boldsymbol {x}_J),g_D(\boldsymbol{x}_J,\boldsymbol{d}')\big).
\end{align*}
This implies that
    \begin{align}
    g_{C}(\boldsymbol {x}_{I_{1}},\,\boldsymbol {x}_L,\,\{\boldsymbol {c}_{I_j},\,2\leq j\leq m\},\,\boldsymbol {x}_J)\neq g_{C}(\boldsymbol {x}'_{I_{1}},\,\boldsymbol {x}_L,\,\{\boldsymbol {c}_{I_j},\,2\leq j\leq m\},\,\boldsymbol {x}_J).\label{color-pf-eq4}
    \end{align}

    Recalling that $\mP_C=\{C_1,C_2,\cdots, C_m\}$ is a strong partition  of $C$, we obtain that
        \begin{align}\label{color-pf-eq6}
    &g_{C}\big(\boldsymbol {x}_{I_{1}},\,\boldsymbol {x}_L,\,\{\boldsymbol {c}_{I_j},\,2\leq j\leq m\},\,\boldsymbol {x}_J\big)=\big(g_{C_{1}}(\boldsymbol{x}_{I_{1}},\boldsymbol{x}_L,\boldsymbol{x}_J),
    g_{C_j}(\boldsymbol{c}_{I_j},\boldsymbol{x}_L,\boldsymbol{x}_J),2\leq j\leq m\big),
    \end{align}
    and
     \begin{align}\label{color-pf-eq6.1}
    &g_{C}\big(\boldsymbol {x}'_{I_{1}},\,\boldsymbol {x}_L,\,\{\boldsymbol {c}_{I_j},\,2\leq j\leq m\},\,\boldsymbol {x}_J\big)=\big(g_{C_{1}}(\boldsymbol{x}'_{I_{1}},\boldsymbol{x}_L,\boldsymbol{x}_J),
    g_{C_j}(\boldsymbol{c}_{I_j},\boldsymbol{x}_L,\boldsymbol{x}_J),2\leq j\leq m\big),
    \end{align}
    where  we note that $K_{C_j}\subseteq I_{C_j}\cup L\cup J$ for $1\leq j\leq m$.
    Combining \eqref{color-pf-eq4}, \eqref{color-pf-eq6} and \eqref{color-pf-eq6.1}, we have
\begin{align}
g_{C_{1}}(\boldsymbol{x}_{I_{1}},\boldsymbol{x}_L,\boldsymbol{x}_J)\neq g_{C_{1}}(\boldsymbol{x}_{I_{1}}',\boldsymbol{x}_L,\boldsymbol{x}_J)
.\label{color-pf-eq9}
\end{align}
Together with the equations
  \begin{align*}
    g_{C}(\boldsymbol {x}_{I},\boldsymbol {x}_{J})=g_{C}\big(\{\boldsymbol {x}_{I_j},\,1\leq j\leq m\},\,\boldsymbol {x}_L,\boldsymbol {x}_J\big)
    =\big(g_{C_j}(\boldsymbol{x}_{I_j},\boldsymbol{x}_L,\boldsymbol{x}_J),1\leq j\leq m\big)
    \end{align*}
    and
    \begin{align*}
     g_{C}(\boldsymbol {x}'_{I},\boldsymbol {x}_{J})=g_{C}\big(\{\boldsymbol {x}_{I_j}',\,1\leq j\leq m\},\,\boldsymbol {x}_L,\boldsymbol {x}_J\big)=\big(g_{C_j}(\boldsymbol{x}_{I_j}',\boldsymbol{x}_L,\boldsymbol{x}_J),1\leq j\leq m\big),
    \end{align*}
    the equation~\eqref{color-pf-eq9} immediately implies that $g_{C}(\boldsymbol {x}_I,\boldsymbol {x}_J)\neq g_{C}(\boldsymbol {x}_I',\boldsymbol {x}_J)$.

    Combining the above two cases, the lemma is proved.
\end{IEEEproof}
\bigskip

%
%

 We recall the $k$-shot  uniquely-decodable code~$\mbC$, of which the global encoding functions are $g_e,\,e \in \mE$. Let $C$ be an arbitrary cut set in $\Lambda(\mathcal{N})$  and  $\mathcal{P}_{C} = \{C_{1}, C_{2}, \cdots, C_{m}\}$ be an arbitrary  strong partition of~$C$.
 For the expected bit rate $R_{e}(\mbC)$ on each edge $e\in \mE$ (cf.~\eqref{rate-def-on-edge}), we have
\begin{equation*}
R_{e}(\mbC)=\frac{L_{e}(\mbC)}{k}
=\frac{E\big[\ell\big(g_{e}(\boldsymbol{X}_{S})\big)\big]}{k}
\geq \frac{H\big(g_{e}(\boldsymbol{X}_{S})\big)}{k},
\end{equation*}
where the inequality follows from the uniquely-decodable property of the code (cf. the definition in Section~\ref{Ud-code}).
For the cut set $C$, we further consider
\begin{align}
&\sum_{e\in C}R_{e}(\mbC)
\geq\frac{1}{k}\sum_{e\in C}H\big(g_{e}(\boldsymbol{X}_{S})\big)
\geq\frac{1}{k}H\big(g_{e}(\boldsymbol{X}_{S}),e\in C\big)\label{sum-rate-proof--1}\\
&=\frac{1}{k}H\big(g_{C}(\boldsymbol{X}_{S})\big)
=\frac{1}{k}H\big(g_{C}(\boldsymbol{X}_{I_C\cup J_C})\big),\label{sum-rate-proof-0}
\end{align}
where
 the second inequality in~\eqref{sum-rate-proof--1} follows from the independence bound of entropy and the second equality in~\eqref{sum-rate-proof-0} follows from the fact that  the vector of information sources $\boldsymbol {X}_{S\setminus (I_C\cup J_C)}$ do not contribute to the values of $g_C(\boldsymbol{X}_{S})$ (cf.  the paragraph immediately above Lemma~\ref{coloring=coding}).
It further follows from Lemma~\ref{coloring=coding} that  $g_{C}$ is a coloring of the $k$-fold characteristic graph $G_{\mP_C}^{\,k}$. Recalling that $(G_{\mP_C}^{\,k},\boldsymbol{X}_{I_C\cup J_C})$ is a probabilistic graph, we thus have $$H\big(g_{C}(\boldsymbol{X}_{I_C\cup J_C})\big)\geq H_{\mX}(G_{\mP_C}^{\,k},\boldsymbol{X}_{I_C\cup J_C}),$$
where $H_{\mX}(G_{\mP_C}^{\,k},\boldsymbol{X}_{I_C\cup J_C})$ is the
 chromatic entropy  of the probabilistic graph $(G_{\mP_C}^{\,k},\boldsymbol{X}_{I\cup J})$, i.e.,
  \begin{equation*}
H_{\chi}(G_{\mP_C}^{\,k}, \boldsymbol{X}_{I\cup J})=\min\big\{H\big(c(\boldsymbol{X}_{I\cup J})\big):\;\text{$c$ is a coloring of $G_{\mP_C}^{\,k}$}\big\}
\end{equation*}
(cf.~\eqref{def:chrom-entro} for the definition).
 Continuing from \eqref{sum-rate-proof-0},
  we obtain that
\begin{equation}\label{last-1}
\sum_{e\in C}R_{e}(\mbC)\geq \frac{1}{k}H\big(g_{C}(\boldsymbol{X}_{I_C\cup J_C})\big)\geq \frac{1}{k}H_{\mX}\big(G_{\mP_C}^{\,k},\boldsymbol{X}_{I_C\cup J_C}\big).
\end{equation}

Furthermore, we consider the $k$-fold AND product $G_{\mP_C}^{\land k}=(V_{\mP_C}^{\land k},E_{\mP_C}^{\land k})$ of $G_{\mP_C}$.
By Lemma~\ref{The Relation Between 123}, we obtain  that $
 V_{\mP_C}^{\land k} = V^{\,k}_{\mP_C}$ and $
E_{\mP_C}^{\land k} \subseteq E^{\,k}_{\mP_C}$. This implies that $(G_{\mP_C}^{\land k},\boldsymbol{X}_{I_C\cup J_C})$  is also a  probabilistic graph and a coloring of $G_{\mP_C}^{\,k}$ is also a coloring of $G_{\mP_C}^{\land k}$.
Thus we have
\begin{align*}
H_{\mX}\big(G_{\mP_C}^{\,k},\boldsymbol{X}_{I_C\cup J_C}\big) \geq H_{\mX}\big(G_{\mP_C}^{\land k},\boldsymbol{X}_{I_C\cup J_C}\big).
\end{align*}
By Lemmas~\ref{lemma:graph-chrom-entro-rel} and~\ref{lemma:AND-pro=sum-g-en}, we further have
\begin{align*}
H_{\mX}\big(G_{\mP_C}^{\land k},\boldsymbol{X}_{I_C\cup J_C}\big)
\geq H_{\omega}\big(G_{\mP_C}^{\land k},\boldsymbol{X}_{I_C\cup J_C}\big)=k H_{\omega}\big(G_{\mP_C},X_{I_C\cup J_C}\big).
\end{align*}
 We continue to consider~\eqref{last-1} and  obtain that
\begin{equation}\label{main-theorem-1}
\sum_{e\in C}R_{e}(\mbC) \geq H_{\omega}(G_{\mathcal{P}_C},X_{I_C\cup J_C}).
\end{equation}
For the admissible $k$-shot uniquely-decodable  code $\mbC$ thus considered,
we recall  the coding rate $R(\mbC)=\max_{e\in\mathcal{E}}R_e(\mbC)$ (cf.~\eqref{def-RC}).
Together with~\eqref{main-theorem-1}, we obtain that
\begin{align*}
|C|\cdot R(\mbC)\geq \sum_{e\in C}R_e(\mbC)\geq H_{\omega}(G_{\mathcal{P}_C},X_{I_C\cup J_C}),
\end{align*}
namely that
\begin{align}
R(\mbC)\geq \frac{~ H_{\omega}(G_{\mathcal{P}_C},X_{I_C\cup J_C})~}{|C|}.\label{pfthm1---eq2}
\end{align}
We note that the  inequality \eqref{pfthm1---eq2} is true for all pairs $(C,\mathcal{P}_C)\in\Lambda(\mathcal{N})\times \textbf{\textup{P}}_C$, and thus
\begin{align*}
R(\mbC)\geq \max_{(C,\mathcal{P}_C)\in\Lambda(\mathcal{N})\times \textbf{\textup{P}}_C}\frac{~ H_{\omega}(G_{\mathcal{P}_C},X_{I_C\cup J_C})~}{|C|}.
\end{align*}
Further, the above lower bound is valid for any positive integer $k$ and any admissible $k$-shot uniquely-decodable  code.
Hence, we have proved that
\begin{equation*}
\mathcal{C}(\mathcal{N},X_{S},f)\geq\max_{(C,\mathcal{P}_C)\in\Lambda(\mathcal{N})\times \textbf{\textup{P}}_C}\frac{~ H_{\omega}(G_{\mathcal{P}_C},X_{I_C\cup J_C})~}{|C|}.
\end{equation*}
The theorem is proved.

 \section{Improved  Lower Bound with Comparison on  Fixed-Length  Coding}\label{sec-improve-bound}

In this section, we prove an improved general lower bound on the computing capacity $\mathcal{C}(\mathcal{N},X_S,f)$ by introducing a new equivalence relation to refine the probability distribution of information sources. In addition, we  compare uniquely-decodable network function-computing codes and fixed-length network function-computing codes, which can be regarded as a special  type of uniquely-decodable network function-computing codes. This shows that the uniquely-decodable network function-computing coding  outperforms the fixed-length network function-computing coding in terms of the computing capacity.

\subsection{Improved  Lower Bound}\label{subsec-impr-bound}

We continue to consider the  model $(\mathcal{N},X_{S},f)$.
We first present an equivalence relation amongst   vectors of information sources
which plays an important role in  improving the lower bound on the computing capacity $\mathcal{C}(\mathcal{N},X_{S},f)$.

\begin{definition}\label{def:equivalent-pc}
Let $C\in\Lambda(\mathcal{N})$ be a cut set and $\mathcal{P}_{C} = \{C_{1}, C_{2}, \cdots, C_{m}\}$ be a strong partition of $C$. Let $I=I_{C}$, $J=J_{C}$ and $I_{\ell}=I_{C_{\ell}}$ for $1\leq\ell \leq m$,  and accordingly $L=I\setminus(\cup_{\ell=1}^{m} I_{\ell})$.
Consider two vectors of information sources $X_{I\cup J}=(X_i,\;\sigma_i\in I\cup J)$ and $\widehat{X}_{I\cup J}=(\widehat{X}_i,\;\sigma_i\in I\cup J)$ with the same support set  $\mathcal{A}^{I \cup J}$. We say that
  $X_{I\cup J}$ and $\widehat{X}_{I \cup J}$ are  $\mathcal{P}_{C}$-equivalent, written as
 $X_{I\cup J}\overset{\mathcal{P}_{C}}{\sim} \widehat{X}_{I \cup J}$,
 if
   for each $1\leq\ell\leq m$, $X_{I_{\ell}\cup L\cup J}=(X_i,\;\sigma_i\in I_{\ell}\cup L\cup J)$ and  $\widehat{X}_{I_{\ell}\cup L\cup J}=(\widehat{X}_i,\;\sigma_i\in I_{\ell}\cup L\cup J)$  have the identical probability distribution, i.e.,
\begin{align*}
  P_{X_{I_{\ell}\cup L\cup J}}(x_{I_\ell\cup L\cup J})= P_{\widehat{X}_{I_{\ell}\cup L\cup J}}(x_{I_\ell\cup L\cup J}),\quad\forall~x_{I_\ell\cup L\cup J}\in \mathcal{A}^{I_\ell\cup L\cup J}.
\end{align*}
\end{definition}
It is easy to see that the above $\mathcal{P}_{C}$-equivalence is  an equivalence relation.
We  remark that for each vector of information sources
 $\widehat{X}_{I\cup J}$ that is $\mathcal{P}_{C}$-equivalent to $X_{I\cup J}$, $(G_{\mathcal{P}_C},\widehat{X}_{I\cup J})$ is also a probabilistic graph.
 We
denote by $\textbf{\textup{H}}(G_{\mathcal{P}_C},X_{I\cup J})$
   the maximum value of the clique entropies $H_{\omega}\big(G_{\mathcal{P}_C},\widehat{X}_{I\cup J}\big)$
over all the vectors of information sources
$\widehat{X}_{I\cup J}$ that are $\mathcal{P}_{C}$-equivalent to $X_{I\cup J}$, i.e.,
\begin{align}\label{def:H_C(P_C)}
\textbf{\textup{H}}(G_{\mathcal{P}_C},X_{I\cup J})\triangleq\max_{\substack{\text{all}~\widehat{X}_{I \cup J}~\text{$\mathcal{P}_{C}$-equivalent to $X_{I\cup J}$}}}H_{\omega}(G_{\mathcal{P}_C},\widehat{X}_{I\cup J}).
\end{align}
With the definition of $\textbf{\textup{H}}(G_{\mathcal{P}_C},X_{I\cup J})$, we can improve the general lower bound on the computing capacity $\mathcal{C}(\mathcal{N},X_S,f)$ in Theorem~\ref{thm-low-bound} as stated in the following theorem.

\begin{theorem}\label{thm-impro-low-bound}
Consider the  model of network function computation  $(\mathcal{N},X_{S},f)$.
Then,
\begin{align}
\mathcal{C}(\mathcal{N},X_S,f)\geq\max_{(C,\mathcal{P}_C)\in\Lambda(\mathcal{N})\times \textbf{\textup{P}}_C}\frac{~\textbf{\textup{H}}(G_{\mathcal{P}_C},X_{I_C\cup J_C})~}{|C|}.\label{thm2-eq}
\end{align}
\end{theorem}

It follows from \eqref{def:H_C(P_C)} that
 \begin{equation*}
 \textbf{\textup{H}}(G_{\mathcal{P}_C},X_{I\cup J})\geq H_{\omega}(G_{\mathcal{P}_C},X_{I\cup J}),\quad\forall~(C,\mathcal{P}_C)\in \Lambda(\mathcal{N})\times \textbf{\textup{P}}_C.\label{remark2-eq}
  \end{equation*}
  Hence, the general lower bound in Theorem~\ref{thm-impro-low-bound} is an enhancement of the one  in Theorem~\ref{thm-low-bound}.

\begin{IEEEproof}[Proof of Theorem~\ref{thm-impro-low-bound}]
 Let $\mbC$ be   an arbitrary admissible $k$-shot uniquely-decodable network  function-computing code for  the model $(\mathcal{N},X_S,f)$, of which the global encoding functions are $g_e,\,e \in \mE $.
Consider a  cut set $C\in\Lambda(\mathcal{N})$ and a strong partition $\mathcal{P}_{C} = \{C_{1}, C_{2}, \cdots, C_{m}\}$ of $C$.
Throughout the proof, we still let $I=I_{C}$, $J=J_{C}$, $I_{\ell}=I_{C_{\ell}}$ for $1\leq\ell \leq m$,  and accordingly $L=I\setminus(\cup_{\ell=1}^{m} I_{\ell})$ for  notational simplicity.
 We note that  $K_{C_\ell}\subseteq I_{C_\ell}\cup L\cup J$ for each  $1 \leq \ell\leq m$, which implies that   no path exists from any source node in $S\setminus(I_{\ell}\cup L\cup J)$ to $\text{tail}(e)$ for each~$e\in C_\ell$. With this, for each $e\in C_\ell $ with $1\leq\ell \leq m$,  the expected bit rate $R_{e}(\mbC)$ (cf.~\eqref{rate-def-on-edge}) can be written as
\begin{align}\label{pf-impr-bound-eq2}
R_e(\mbC)=\frac{L_{e}(\mbC)}{k}=\frac{E\big[\ell\big(g_{e}(\boldsymbol{X}_S)\big)\big]}{k}
=\frac{E\big[\ell\big(g_e(\boldsymbol{X}_{I_{\ell}\cup L\cup J})\big)\big]}{k}.
\end{align}

Now, we consider a new model  $(\mathcal{N},\widehat{X}_{S},f)$ associated with the pair $(C,\mathcal{P}_C)$, where
  $\widehat{X}_S\triangleq (\widehat{X}_i,\;\sigma_i\in S)$ with $\widehat{X}_i$ being
 the information source  at the $i$th source node~$\sigma_i$  taking values in the finite set $\mathcal{A}$, and   the joint probability distribution $P_{\widehat{X}_S}$  satisfies
 \begin{equation*}
P_{\widehat{X}_S}(x_S)\triangleq{\rm{Pr}}\big\{\widehat{X}_S=x_S\big\}>0,\quad~\forall~ x_S\triangleq (x_1,x_2,\cdots,x_s)\in\mathcal{A}^s,
 \end{equation*}
  and  $\widehat{X}_{I \cup J}\overset{\mathcal{P}_{C}}{\sim} X_{I\cup J}$.
 We note that the admissibility of a uniquely-decodable  code for a network function computation model only depends  on the support set of the information sources and is independent of their particular probability distribution, while the computing capacity depends on the probability distribution of the information sources.
Hence, since  $X_{S}$ and $\widehat{X}_S$ have the same support set~$\mathcal{A}^S$,
 an  admissible uniquely-decodable  code  for  the model $(\mathcal{N},X_{S},f)$ is also admissible for the model $(\mathcal{N},\widehat{X}_{S},f)$, and vice versa.

We now continue to consider the code $\mbC$ and regard $\mbC$  as an admissible $k$-shot uniquely-decodable  code  for  the model $(\mathcal{N},\widehat{X}_{S},f)$. For each edge $e\in\mathcal{E}$, we use $\widehat{L}_{e}(\mbC)$ and $\widehat{R}_e(\mbC)$ to denote the expected number of bits transmitted on   $e$ and the expected bit rate on  $e$ by using the code $\mbC$ for model $(\mathcal{N},\widehat{X}_{S},f)$, respectively. Then, we have
 \begin{align*}
 \widehat{R}_e(\mbC)\triangleq \frac{\widehat{L}_{e}(\mbC)}{k}=\frac{E\big[\ell\big(g_{e}(\widehat{\boldsymbol{X}}_S)\big)\big]}{k},\quad\forall~
 e\in\mathcal{E}.
 \end{align*}
 In particular, for each $e\in C_\ell$ with $1\leq \ell \leq m$,
   similar to \eqref{pf-impr-bound-eq2}, we further have
\begin{align}\label{pf-impr-bound-eq3}
  \widehat{R}_e(\mbC)=\frac{E\big[\ell\big(g_e(\widehat{\boldsymbol{X}}_{I_{\ell}\cup L\cup J})\big)\big]}{k}.
\end{align}
By the same argument to prove \eqref{main-theorem-1},  we  obtain that
\begin{align}
\sum_{e\in C}\widehat{R}_e(\mbC)\geq H_{\omega}(G_{\mathcal{P}_C},\widehat{X}_{I\cup J}).\label{pf-impr-bound-eq4}
\end{align}

Furthermore, with the fact that $\widehat{X}_{I \cup J}\overset{\mathcal{P}_{C}}{\sim} X_{I\cup J}$,
$X_{I_{\ell}\cup L\cup J}$ and  $\widehat{X}_{I_{\ell}\cup L\cup J}$  have the same probability distribution for each $1\leq \ell \leq m$. This immediately implies that
\begin{equation}
E\Big[\ell\big(g_e(\boldsymbol{X}_{I_{\ell}\cup L\cup J})\big)\Big]=E\Big[\ell\big(g_e(\widehat{\boldsymbol{X}}_{I_{\ell}\cup L\cup J})\big)\Big],\quad\forall~e\in C_{\ell},~1\leq \ell\leq m.
\label{pf-impr-bound-eq4-11}
\end{equation}
Combining \eqref{pf-impr-bound-eq2}, \eqref{pf-impr-bound-eq3} and \eqref{pf-impr-bound-eq4-11}, we obtain that
\begin{equation*}
R_e(\mbC)=\widehat{R}_e(\mbC),\quad\forall~e\in C_{\ell},~1\leq \ell\leq m,
\end{equation*}
or equivalently,
\begin{equation}\nonumber
R_e(\mbC)=\widehat{R}_e(\mbC),\quad\forall~e\in C.
\end{equation}
Together with \eqref{pf-impr-bound-eq4}, we further obtain that
\begin{align}\label{pf-impr-bound-eq10}
\sum_{e\in C}R_e(\mbC)=\sum_{e\in C}\widehat{R}_e(\mbC)\geq H_{\omega}(G_{\mathcal{P}_C},\widehat{X}_{I\cup J}).
\end{align}
We note that the inequality \eqref{pf-impr-bound-eq10} holds for any vector of information sources~$\widehat{X}_{I\cup J}$ that is $\mathcal{P}_{C}$-equivalent to  $X_{I \cup J}$. Together with the definition of $\textbf{\textup{H}}(G_{\mathcal{P}_C},X_{I\cup J})$ (cf.~\eqref{def:H_C(P_C)}), we immediately have
\begin{align}\label{main-theorem-2}
\sum_{e\in C}R_e(\mbC)\geq\textbf{\textup{H}}(G_{\mathcal{P}_C},X_{I\cup J}).
\end{align}

For the admissible $k$-shot uniquely-decodable  code $\mbC$ considered here,
we recall that the coding rate $R(\mbC)=\max_{e\in\mathcal{E}}R_e(\mbC)$ (cf.~\eqref{def-RC}), and by~\eqref{main-theorem-2} we obtain that
\begin{align*}
|C|\cdot R(\mbC)\geq \sum_{e\in C}R_e(\mbC)\geq \textbf{\textup{H}}(G_{\mathcal{P}_C},X_{I\cup J}),
\end{align*}
namely that
\begin{align}
R(\mbC)\geq \frac{~ \textbf{\textup{H}}(G_{\mathcal{P}_C},X_{I\cup J})~}{|C|}.\label{pfthm1-eq2}
\end{align}
We note that the  inequality \eqref{pfthm1-eq2} is true for all pairs $(C,\mathcal{P}_C)\in\Lambda(\mathcal{N})\times \textbf{\textup{P}}_C$, and thus we have
\begin{align*}
R(\mbC)\geq \max_{(C,\mathcal{P}_C)\in\Lambda(\mathcal{N})\times \textbf{\textup{P}}_C}\frac{~ \textbf{\textup{H}}(G_{\mathcal{P}_C},X_{I\cup J})~}{|C|}.
\end{align*}
Furthermore, the above lower bound is valid for any positive integer $k$ and any admissible $k$-shot uniquely-decodable  code.
This thus implies that
\begin{equation*}
\mathcal{C}(\mathcal{N},X_{S},f)\geq\max_{(C,\mathcal{P}_C)\in\Lambda(\mathcal{N})\times \textbf{\textup{P}}_C}\frac{~ \textbf{\textup{H}}(G_{\mathcal{P}_C},X_{I\cup J})~}{|C|}.
\end{equation*}
The theorem is proved.
\end{IEEEproof}

In the following, we continue to use the specific model  discussed in Examples~\ref{ex:G_X-G_XY_a_nontrival_examp} and \ref{ex2} to  illustrate the improved  lower bound  in Theorem~\ref{thm-impro-low-bound} and show that this improved  bound
 is a strict enhancement of the lower bound  in
Theorem~\ref{thm-low-bound}.

\begin{example}\label{exam-3}
Recall the  specified model $(\mathcal{N},X_S,f)$ used in Example~\ref{ex:G_X-G_XY_a_nontrival_examp} and~\ref{ex2}, where we consider computing the binary arithmetic sum $f$ over the diamond network $\mathcal{N}$ as depicted in Fig.\,\ref{fig:N_1}. Here, $X_S=(X_1,X_2,X_3)$ and the information sources~$X_1, X_2, X_3$ are i.i.d. random variables according to the uniform distribution over the alphabet $\mathcal{A}=\{0,1\}$.
By recalling~\eqref{ex:lower-bound-1} in Example~\ref{ex2},
the lower bound in Theorem~\ref{thm-low-bound} for the model $(\mathcal{N},X_S,f)$  is
\begin{align}\label{ex-outbound-1}
\max_{(C,\mathcal{P}_C)\in\Lambda(\mathcal{N})\times \textbf{\textup{P}}_C}
\frac{~H_{\omega}(G_{\mathcal{P}_C},X_{I_C\cup J_C})~}{|C|}=\frac{~H_{\omega}(G_{\mathcal{P}_{C^*}},X_{I_{C^*}\cup J_{C^*}})~}{|C^*|}=\frac{7}{4}-\frac{3}{8}\log 3\approx 1.156,
\end{align}
where $C^*=\{e_5,e_6\}$ and $\mathcal{P}_{C^*}=\big\{C_1=\{e_5\},C_2=\{e_6\}\big\}$.
In Appendix~\ref{appendix:calculation-of-lower-bound}, we specify the improved lower bound  in Theorem~\ref{thm-impro-low-bound} to be
\begin{align}\label{ex-outbound-2}
\max_{(C,\mathcal{P}_C)\in\Lambda(\mathcal{N})\times \textbf{\textup{P}}_C}\frac{~\textbf{\textup{H}}(G_{\mathcal{P}_C},X_{I_C\cup J_C})~}{|C|}=\frac{~\textbf{\textup{H}}(G_{\mathcal{P}_{C^*}},X_{I_{C^*}\cup J_{C^*}})~}{|C^*|}= \frac{1}{2}\log5 \approx 1.161,
\end{align}
which makes approximately $4\%$ improvement ($(1.161-1.156)/1.156\approx4\%$) for this quite simple model.
This also shows that the improved lower bound obtained is a  strict enhancement of the one  in
Theorem~\ref{thm-low-bound} for the model  $(\mathcal{N},X_S,f)$.
\end{example}

\subsection{Comparison on  Fixed-Length Network Function-Computing Coding}

The fixed-length network function-computing coding for the   model $(\mathcal{N},X_S,f)$ has been investigated persistently in the literature~\cite{Huang_Comment_cut_set_bound,
Appuswamy13,Ramamoorthy-Langberg-JSAC13-sum-networks,
Appuswamy14,Tripathy-Ramamoorthy-IT18-sum-networks,Appuswamy11,
Guang_Improved_upper_bound,Kowshik12,Li_Xu_vector_linear_diamond,
Yao-Jafar-3user,Guang_Zhang_Arithmetic_sum_Sel_Areas}, 
Therein, the fundamental computing capacity has been investigated, which is defined as the maximum average number of times that the target function can
 be computed with zero error for transmitting at most one bit on every edge in the network, and denoted by $\adjustedbar{\mathcal{C}}{}(\mathcal{N},X_S,f)$.
Because fixed-length codes are considered here, with the assumption  that the support set of $X_S$ is the set $\mathcal{A}^S$, the computing capacity $\adjustedbar{\mathcal{C}}{}(\mathcal{N},X_S,f)$ is independent of the particular  probability distribution of $X_S$.\footnote{This is the reason why  the  computing capacity  is written as $\mathcal{C}(\mathcal{N},f)$ in the literature~\cite{Huang_Comment_cut_set_bound,Appuswamy11,
Guang_Improved_upper_bound,Li_Xu_vector_linear_diamond,Guang_Zhang_Arithmetic_sum_Sel_Areas},
but  in the current paper,  we still write the computing capacity as $\adjustedbar{\mathcal{C}}{}(\mathcal{N},X_S,f)$ for notational consistency.} This computing capacity $\adjustedbar{\mathcal{C}}{}(\mathcal{N},X_S,f)$ is considered from the viewpoint of the network throughput, while if it is considered from the same viewpoint of data compression as the setup in the current paper, it is equivalent to consider $1/\,\adjustedbar{\mathcal{C}}{}(\mathcal{N},X_S,f)$  which we denote by $\adjustedbar{\mathcal{C}}{}^{\bot}(\mathcal{N},X_S,f)$.
We further note that
an admissible
 fixed-length network function-computing code can be also regarded as  an admissible
 uniquely-decodable network function-computing code for the same model. This immediately shows that
 \begin{align}\label{cap-rel-fix-var-cod}
 \mathcal{C}(\mathcal{N},X_S,f)\leq \adjustedbar{\mathcal{C}}{}^{\perp }(\mathcal{N},X_S,f).
 \end{align}
 Intuitively, $ \adjustedbar{\mathcal{C}}{}^{\perp }(\mathcal{N},X_S,f)$  measures
 the efficiency of computing the target function over the network in the \emph{worst case}, while
 $\mathcal{C}(\mathcal{N},X_S,f)$  measures
 the efficiency of computing the target function over the network in the \emph{average case}.
 Several general upper bounds on the computing capacity $\adjustedbar{\mathcal{C}}{}(\mathcal{N},X_S,f)$ have been obtained  in  \cite{Appuswamy11,Huang_Comment_cut_set_bound,Guang_Improved_upper_bound}, where ``general'' means that the upper bounds are applicable to arbitrary network topologies and arbitrary target functions. The best known upper bound is the one proved by Guang~\emph{et al.}~\cite{Guang_Improved_upper_bound} in using the approach of the cut-set strong partition. With this, we immediately obtain the best known lower bound on $\adjustedbar{\mathcal{C}}{}^{\perp }(\mathcal{N},X_S,f)$ as stated in the  theorem below. Following the theorem, we will prove  that our obtained lower bounds  in Theorems~\ref{thm-low-bound} and \ref{thm-impro-low-bound} on $\mathcal{C}(\mathcal{N},X_S,f)$ are  not larger than the best known lower bound on $\adjustedbar{\mathcal{C}}{}^{\perp }(\mathcal{N},X_S,f)$, which is consistent with the relationship between the two computing capacities as presented in \eqref{cap-rel-fix-var-cod}.

 \begin{theorem}
  \label{thm:upper_bound}
  Consider the  model of network function computation  $(\mathcal{N},X_{S},f)$.
Then,
\begin{align*}
\adjustedbar{\mathcal{C}}{}^{\perp }(\mathcal{N},X_S,f)\geq \max_{C\in\Lambda(\mN)}\dfrac{\log n_{C,f}}{|C|}=\max_{(C,\mathcal{P}_C)\in\Lambda(\mathcal{N})\times \textbf{\textup{P}}_C}\dfrac{\log n_{C}(\mathcal{P}_C)}{|C|},
\end{align*}
where $n_{C,f}$ and $n_{C}(\mathcal{P}_C)$ will become clear later (also cf.~\cite{Guang_Improved_upper_bound}).
\end{theorem}

To clarify the  notation $n_{C,f}$ and $n_{C}(\mathcal{P}_C)$, we consider a cut set $C\in\Lambda(\mathcal{N})$ and a strong partition $\mathcal{P}_{C} = \{C_{1}, C_{2}, \cdots, C_{m}\}$ of $C$, and let $I=I_{C}$, $J=J_{C}$ and $I_{\ell}=I_{C_{\ell}}$ for $1\leq \ell\leq m$, and accordingly $L=I\setminus(\cup_{\ell=1}^{m} I_{\ell})$.
For each $(I_{\ell}, a_{L}, a_{J} )$-equivalence class $\mathrm{cl}_{I_{\ell}}$, $1\leq \ell\leq m$, we define the set
\begin{equation}
\langle\mathrm{cl}_{I_{1}},\mathrm{cl}_{I_{2}},\cdots,\mathrm{cl}_{I_{m}},a_L\rangle
\triangleq\Big\{(x_{I_1},x_{I_2},\cdots,x_{I_m},a_L):~x_{I_\ell}
\in\mathrm{cl}_{I_\ell}~\text{for}~1\leq \ell\leq m \Big\}.\label{def-set-cl-a_L}
\end{equation}
For each $(I,{a}_J)$-equivalence class $\Cl[{a}_J]$, we define
\begin{align}
 N\big({a}_{L}, \Cl[{a}_J]\big)& \triangleq \#\Big\{ \big( \cl_{I_1}, \cl_{I_2}, \cdots, \cl_{I_m} \big):\
       \cl_{I_\ell} \text{ is an $(I_\ell, {a}_{L}, {a}_J)$-equivalence class, }
       1\leq \ell\leq m ,\nonumber\\
&\qquad \qquad \qquad \qquad \qquad \quad \quad ~\textrm{and}~\big\langle \cl_{I_1}, \cl_{I_2}, \cdots, \cl_{I_m}, {a}_{L} \big\rangle \subseteq \Cl[{a}_J]\Big\},\label{no_finer_eq_cl}
\end{align}
where ``$\#\{\cdot\}$'' stands for the size of the set.
We further let
\begin{align}\label{equ:N_Cl}
N\big(\Cl[{a}_J]\big) \triangleq \max_{a_{L}\in \mA^{L}} N\big(a_{L}, \Cl[{a}_J]\big),
\end{align}
and consider the summation of $N\big(\Cl[{a}_J]\big)$ over all the $(I,{a}_J)$-equivalence classes
\begin{align*}
\sum_{\text{all }\Cl[{a}_J]} N\big(\Cl[{a}_J]\big).
\end{align*}
Now, we let
\begin{align}\label{n_C_Parti_1st}
n_C(\mP_C) \triangleq \max_{{a}_J\in \mA^{J}} \sum_{\text{all }\Cl[{a}_J]} N\big(\Cl[{a}_J]\big),
\end{align}
and
\begin{align}\label{n_C_f_1st}
n_{C,f}= \max_{\mathcal{P}_C\in \textbf{\textup{P}}_C}  n_C(\mP_C).
\end{align}

To show that the relationship between the lower bounds, we re-understand $n_C(\mP_C)$ in a graph-theoretic way, which is in fact the clique number $\omega(G_{\mathcal{P}_C})$ of the characteristic graph $G_{\mathcal{P}_C}$, i.e., the number of vertices in a maximum clique in $G_{\mathcal{P}_C}$.

\begin{theorem}\label{ncpc=w(Gpc)}
Let $C\in\Lambda(\mathcal{N})$ be a cut set and $\mathcal{P}_{C}$ be a strong partition of $C$.
 Further, let $
G_{\mP_C}$ be the characteristic graph associated with the strong partition $\mathcal{P}_{C}$ and the target function~$f$.
  Then, $$n_C(\mP_C)= \omega(G_{\mathcal{P}_C}).$$
\end{theorem}
\begin{IEEEproof}
See Appendix~\ref{pf-thm-ncpc=w(Gpc)}.
\end{IEEEproof}

With \eqref{n_C_f_1st} and Theorem~\ref{ncpc=w(Gpc)}, we rewrite
\begin{equation*}
n_{C,f}=\max_{\mathcal{P}_C\in \textbf{\textup{P}}_C}  \omega(G_{\mathcal{P}_C}).
\end{equation*}
Recalling  the lower bounds presented in Theorems~\ref{thm-low-bound}, \ref{thm-impro-low-bound} and~\ref{thm:upper_bound}, we have
\begin{align}
\frac{~H_{\omega}(G_{\mathcal{P}_C},X_{I_C\cup J_C})~}{|C|}\leq \frac{~\textbf{\textup{H}}(G_{\mathcal{P}_C},X_{I_C\cup J_C})~}{|C|}\leq
\frac{\log \omega(G_{\mathcal{P}_C})}{|C|},\label{remark3-eq1}
\end{align}
where the two inequalities in~\eqref{remark3-eq1} follow from the definition of $\textbf{\textup{H}}(G_{\mathcal{P}_C},X_{I_C\cup J_C})$ and Lemma~\ref{lemma-cli-upper-bound}.
 We formally  state this relationship amongst the lower bounds in the following corollary.
 \begin{cor}
  Consider the  model of network function computation  $(\mathcal{N},X_{S},f)$.
Then,
\begin{align*}
\max_{(C,\mathcal{P}_C)\in\Lambda(\mathcal{N})\times \textbf{\textup{P}}_C}\frac{~H_{\omega}(G_{\mathcal{P}_C},X_{I_C\cup J_C})~}{|C|}&\leq \max_{(C,\mathcal{P}_C)\in\Lambda(\mathcal{N})\times \textbf{\textup{P}}_C} \frac{~\textbf{\textup{H}}(G_{\mathcal{P}_C},X_{I_C\cup J_C})~}{|C|}\leq\max_{C\in\Lambda(\mathcal{N})}
\frac{\log  n_{C,f} }{|C|}.
\end{align*}
 \end{cor}

Next, we continue  with the specific model used in Examples~\ref{ex:G_X-G_XY_a_nontrival_examp}, \ref{ex2} and~\ref{exam-3} to  illustrate  the uniquely-decodable network function-computing coding scheme indeed outperforming
the fixed-length network function-computing coding scheme.

\begin{example}
Recall the  specified model $(\mathcal{N},X_S,f)$ used in Examples~\ref{ex:G_X-G_XY_a_nontrival_examp}, \ref{ex2} and \ref{exam-3}, where we consider computing the binary arithmetic sum $f$ over the diamond network $\mathcal{N}$ as depicted in Fig.\,\ref{fig:N_1}. Here, $X_S=(X_1,X_2,X_3)$ and the information sources~$X_1, X_2, X_3$ are i.i.d. random variables according to the uniform distribution over the alphabet $\mathcal{A}=\{0,1\}$. Following from \cite{Appuswamy11} and \cite{Guang_Improved_upper_bound}, we can obtain  that the minimum coding rate  achieved by admissible fixed-length network function-computing codes for $(\mathcal{N},X_S,f)$ is $(1+\log 3)/2 $, or equivalently,
\begin{align*}
\adjustedbar{\mathcal{C}}{}^{\perp }(\mathcal{N},X_S,f)=\frac{1+\log 3}{2}\approx  1.293.
\end{align*}
In the following, we will show that admissible uniquely-decodable network function-computing codes can achieve coding rates  strictly smaller than $1.293$.

We first consider an optimal fixed-length coding scheme for $(\mathcal{N},X_S,f)$, which asymptotically achieves the computing capacity $\adjustedbar{\mathcal{C}}{}^{\perp }(\mathcal{N},X_S,f)$. Let $k$ be an even number and consider a $(k,n)$ fixed-length network function-computing code for $(\mathcal{N},X_S,f)$, where the positive integer $n$ is the fixed length of the code, i.e., the maximum number of bits transmitted on all  edges; and all the global encoding functions~$\bar{g}_e,e\in\mathcal{E}$ are given as follows:
\begin{align*}
\bar{g}_{e_1}(\boldsymbol{x}_S)&=\bar{g}_{e_1}(\boldsymbol{x}_1)=\boldsymbol{x}_1;\qquad \bar{g}_{e_2}(\boldsymbol{x}_S)=\bar{g}_{e_2}(\boldsymbol{x}_2)=(x_{2,1}, x_{2,2}, \cdots,
x_{2,k/2})^{\top};\\
\bar{g}_{e_3}(\boldsymbol{x}_S)&=\bar{g}_{e_3}(\boldsymbol{x}_2)=(x_{2,k/2+1}, x_{2,k/2+2}, \cdots ,
x_{2,k})^{\top};\qquad
\bar{g}_{e_4}(\boldsymbol{x}_S)=\bar{g}_{e_4}(\boldsymbol{x}_3)=\boldsymbol{x}_3;\\
\bar{g}_{e_5}(\boldsymbol{x}_S)&=\bar{g}_{e_5}(\boldsymbol{x}_1,\boldsymbol{x}_2)=\begin{bmatrix}
 x_{1,i}+x_{2,i}: & 1\leq i\leq k/2 \\
 x_{1,i}: & k/2< i\leq k\\
 \end{bmatrix};\\
\bar{g}_{e_6}(\boldsymbol{x}_S)&=\bar{g}_{e_6}(\boldsymbol{x}_2,\boldsymbol{x}_3)=\begin{bmatrix}
 x_{3,i}: & 1\leq i\leq k/2 \\
 x_{2,i}+ x_{3,i}: & k/2< i\leq k\\
 \end{bmatrix};
\end{align*}
where $\boldsymbol{x}_S=(\boldsymbol{x}_1,\boldsymbol{x}_2,\boldsymbol{x}_3)$ with $\boldsymbol{x}_i=(x_{i,1}, x_{i,2}, \cdots ,
x_{i,k})^{\top}\in\mathcal{A}^k$ being the source message generated by the source node $\sigma_i$ for $i=1,2,3$.
Clearly, the coding scheme works, or the code is admissible, for any pair $(k,n)$ such that
$
2^n\geq 3^{k/2}\cdot 2^{k/2}.
$
This  implies that the maximum bit rate over all edges satisfies
\begin{align*}
\frac{n}{k}\geq \frac{1+\log 3}{2}.
\end{align*}

Based on the above  fixed-length coding scheme, we can construct an  admissible
 $k$-shot uniquely-decodable    code for the model  $(\mathcal{N},X_S,f)$ by encoding the random variables $\bar{g}_e(\boldsymbol{X}_S)$ transmitted on all edges $e\in\mathcal{E}$ (i.e., $
 \bar{g}_{e_1}(\boldsymbol{X}_1) ,$ $\bar{g}_{e_2}(\boldsymbol{X}_2) ,$ $\bar{g}_{e_3}(\boldsymbol{X}_2) ,$ $ \bar{g}_{e_4}(\boldsymbol{X}_3), $ $\bar{g}_{e_5}(\boldsymbol{X}_1,\boldsymbol{X}_2)$ and $ \bar{g}_{e_6}(\boldsymbol{X}_2,\boldsymbol{X}_3)
$)
 into uniquely-decodable codes, denoted by $\big\{g_{e_i}(\boldsymbol{x}_S):\forall~\boldsymbol{x}_S\in\mathcal{A}^S\big\}$ for $1\leq i\leq 6$, such that
 \begin{align*}
 E\Big[\ell\big(g_{e_i}(\boldsymbol{X}_S)\big)\Big]\leq H\big(g_{e_i}(\boldsymbol{X}_S)\big)+1,\quad\forall~1\leq i\leq 6.
 \end{align*}
The above inequalities can be achieved by using, for example, Huffman codes or Shannon codes. Further, we denote by $\mbC$ the constructed uniquely-decodable code and thus for each $e\in\mathcal{E}$,
 \begin{align*}
R_{e}(\mbC)=\frac{L_{e}(\mbC)}{k}=\frac{E\big[\ell\big(g_{e}(\boldsymbol{X}_S)\big)\big]}{k}\leq
\frac{H\big(g_{e}(\boldsymbol{X}_S)\big)+1}{k}.
\end{align*}
After simple  calculations of $R_e(\mbC)$ for all $e\in\mathcal{E}$, we have
\begin{align*}
R_{e_1}(\mbC)=R_{e_4}(\mbC)\leq 1+\frac{1}{k},\quad
R_{e_2}(\mbC)=R_{e_3}(\mbC)\leq \frac{1}{2}+\frac{1}{k},\quad\text{and}\quad
R_{e_5}(\mbC)=R_{e_6}(\mbC)\leq \frac{5}{4}+\frac{1}{k}.
\end{align*}
Thus, the coding rate $R(\mbC)$   satisfies
\begin{equation}\label{eq:constr-code-com-rate-lower-bound}
\mathcal{C}(\mathcal{N},X_S,f)\leq R(\mbC)=\max_{e\in\mathcal{E}}\;R_e(\mbC)\leq \frac{5}{4}+\frac{1}{k}.
\end{equation}
We note that the above inequality \eqref{eq:constr-code-com-rate-lower-bound} is valid for any (even) positive integer $k$, and hence
\begin{equation*}
\mathcal{C}(\mathcal{N},X_S,f)\leq\lim_{k\to\infty}  \Big(\frac{5}{4}+\frac{1}{k}\Big)= 1.25<1.293\approx\adjustedbar{\mathcal{C}}{}^{\perp }(\mathcal{N},X_S,f).
\end{equation*}
This shows that in network function computation,  the uniquely-decodable network function-computing coding outperforms
the fixed-length network function-computing coding in terms of the computing capacity.
\end{example}

An interesting open question here is whether  better uniquely-decodable codes (of smaller coding rates) can be constructed rather than the construction  by transforming optimal fixed-length codes to uniquely-decodable codes as used in the above example.

\section{Discussion}\label{sec-discussion}

In this paper, we considered uniquely-decodable  coding for  zero-error network function computation and investigated  the computing capacity. We first proved some new results on clique entropy. With them, we proved a  lower bound on the computing capacity associated with clique entropies of the induced characteristic graphs, where the obtained lower bound is applicable to  arbitrary network topologies, arbitrary  information sources, and arbitrary target functions.
By refining the probability distribution of information sources, we further proved an improved  lower bound, which strictly enhances the  lower bound we have obtained. Subsequently, we  compared uniquely-decodable network function-computing coding and fixed-length network function-computing coding, and showed that the former indeed outperforms the latter in terms of the computing capacity.

In order to specify the lower bounds obtained in Theorems~\ref{thm-low-bound} and  \ref{thm-impro-low-bound}, we need to calculate  the clique entropy associated with the induced characteristic graph. However, the calculation of  the clique entropy for a general graph is highly challenging.
However,  all the  graphs used in the  paper  have particular structural properties that the vertex set  can be partitioned into either isolated or completely-connected blocks. Motivated by the observation, we can develop an approach to  hierarchically  decompose the  characteristic graph so that we can efficiently calculate  the clique entropy. In fact, this hierarchically  decomposing approach is similar to  the argument used to prove Theorem~\ref{ncpc=w(Gpc)}.

In zero-error network function computation, the characterization problem of coding rate region   was also considered for fixed-length codes, uniquely-decodable codes, and variable-length codes in the literature, e.g., \cite{Kowshik12,Ramamoorthy-vari-length-NFC16,Ramamoorthy-vari-length-NFC18,Ramamoorthy-vari-length-arxiv}, where the \emph{coding rate region}   is defined as the set  of all achievable rate tuples of the edges in the network. Clearly, a fixed-length code is a uniquely-decodable code, and a uniquely-decodable code is a variable-length code. However,  a uniquely-decodable code is allowed to use multiple times to guarantee that the target function is always computed with zero error, while the zero-error computation of the target function  may be not  guaranteed when using a variable-length code  multiple times.
 In fact,  for the same type of codes, the characterization problem of  the coding rate region   can be regarded as the dual problem of the computing capacity characterization  as studied in the current  paper.

Tripathy and  Ramamoorthy~\cite{Ramamoorthy-vari-length-NFC16,Ramamoorthy-vari-length-NFC18} considered the characterization problem of the coding rate region of the variable-length codes for computing an arbitrary target function with   i.i.d. information sources over a so-called diamond network, which is a degenerate one  of the network as depicted in Fig.\,\ref{fig:N_1}. By applying the approach of the cut-set strong partition developed by Guang~\emph{et al.}~\cite{Guang_Improved_upper_bound}, a  step-by-step procedure was proposed to compute
an outer bound on the coding rate region for variable-length codes in \cite{Ramamoorthy-vari-length-NFC18}.\footnote{We refer the reader to the full version \cite{Ramamoorthy-vari-length-arxiv} for more details.} This implicit outer bound on the coding rate region does not enable the determination  of whether a given rate tuple satisfies the outer bound or not.
Subsequently, by using the proposed procedure, the outer bound was illustrated for a classical model of computing the arithmetic sum of three i.i.d.  binary information sources  over the diamond network. The  diamond network has a very simple but typical ``non-tree'' structure, and  many well-known multi-terminal information-theoretic problems were considered over the diamond network (e.g.,~\cite{Dia-netw-1,Dia-netw-2,Dia-netw-4}).
In fact, by applying  the clique entropy approach developed in the current  paper, we can obtain an explicit outer bound in closed form on  the coding rate region  for uniquely-decodable codes (which also applies to the case of  variable-length codes). Further, this explicit outer bound is ``general'', which is  applicable to arbitrary network topologies, arbitrary information sources, and arbitrary target functions.
In particular,  by specifying our outer bound  for the mentioned model in~\cite{Ramamoorthy-vari-length-NFC16,Ramamoorthy-vari-length-NFC18} of computing the arithmetic sum of three i.i.d.  binary information sources  over the diamond network,  our outer bound strictly improves the one obtained  in~\cite{Ramamoorthy-vari-length-NFC18}, even though we here focus on uniquely-decodable codes.

 Kowshik and Kumar~\cite{Kowshik12}  obtained an outer bound on the coding rate region for uniquely-decodable codes,
which in fact is only valid under certain constraints on either the network
topology or the target function.
Nevertheless, we can modify their outer bound to a general one  applicable to  arbitrary network topologies, arbitrary  information sources and arbitrary target functions. Even though, our outer bound as mentioned above considerably improves the modified outer bound.
All the details of the discussions in the above three paragraphs on the characterization of the coding rate region will be found in a forthcoming paper.

\numberwithin{theorem}{section}
\appendices

\section{Proof of ~Proposition~\ref{prop:clique-entro-property}}\label{appendix-pf-prop:clique-entro-property}
We first consider $G=(V,E)$ to be an empty graph. Evidently, $E=\emptyset$.
Let $W$ be an arbitrary random variable such that
$Z\in W\in\Omega(G),$
where we note that $\Omega(G)=\big\{\{z\}:\,z\in V\big\}$ because $G$ is an empty set.
Together with $Z\in W\in\Omega(G),$ namely that
$$P_{ZW}(z,w)=0,\quad~\forall~(z,w)\in V\times\Omega(G)\;\text{with}\; z\notin w,$$
we have obtained that for each $z\in V$,
\begin{align*}
P_{ZW}(z,w)=
\begin{cases}
1 & \text{if $w=\{z\}$, }  \\
0 & \text{otherwise.}
\end{cases}
\end{align*}
This implies that $H(Z|W)=0$.
Further, by the definition of $H_{\omega}(G,Z)$ (cf.~\eqref{def:clique-entro}), $H(Z|W)=0$ is true for all random variables $W$ with $Z\in W\in\Omega(G)$. We have thus  proved that  $$H_{\omega}(G,Z)=\max_{Z\in W\in\Omega(G)} H(Z|W)=0.$$

Next, we consider $G=(V,E)$ to be a complete graph. Clearly, we have
$$\Omega(G)=2^V\setminus\{\emptyset\}\triangleq \big\{\text{all nonempty subsets of } V \big\},$$
and in particular, $V\in \Omega(G)$.
We take the random variable $W^*$ such that $${\rm{Pr}}(W^*=V)=1\quad~\text{ and}\quad~{\rm{Pr}}(W^*=w)=0,~\forall~w\in\Omega(G)\setminus V.$$
 We can readily verify that
$$P_{ZW}(z,w)=0,\quad~\forall~(z,w)\in V\times\Omega(G)\;\text{with}\; z\notin w, $$
implying that $W^*$ satisfies the condition $Z\in W^*\in\Omega(G)$.
By the definition of $H_{\omega}(G,Z)$, we have
$$H_{\omega}(G,Z)=\max_{Z\in W\in\Omega(G)} H(Z|W)\geq H(Z|W^*)=H(Z),$$
where the last equality holds because $W^*$ is a constant.
Together with the fact that $H_{\omega}(G,Z)\leq H(Z)$ by definition, we have thus proved that $H_{\omega}(G,Z)= H(Z)$.

\section{Proof of~Lemma~\ref{lemma-cli-upper-bound}}\label{pf-lemma-cli-upper-bound}
Consider the clique entropy $$H_{\omega}(G,Z)=\max_{Z\in W\in\Omega(G)} H(Z|W).$$ We let $W$ be a random variable that satisfies
the condition $Z\in W\in\Omega(G)$ and consider
\begin{align}
H(Z|W)&=\sum_{w\in \Omega(G)}\;\sum_{z\in V}P_{ZW}(z,w)\cdot\log \frac{1}{P_{Z|W}(z|w)}\nonumber\\
&=\sum_{w\in \Omega(G)}\;\sum_{z\in w}P_{ZW}(z,w)\cdot\log \frac{1}{P_{Z|W}(z|w)}\label{pf-lemma-cli-upper-bound-0}\\
&=\sum_{w\in \Omega(G)}\;\sum_{z\in w}P_W(w)\cdot P_{Z|W}(z|w)\cdot\log \frac{1}{P_{Z|W}(z|w)}\nonumber\\
&=\sum_{w\in \Omega(G)}P_W(w)\cdot\sum_{z\in w}P_{Z|W}(z|w)\cdot\log \frac{1}{P_{Z|W}(z|w)}\nonumber\\
&\leq \sum_{w\in \Omega(G)}P_W(w)\cdot \log|w|\label{pf-lemma-cli-upper-bound-1}
\\
&\leq \sum_{w\in \Omega(G)}P_W(w)\cdot \log \omega(G)= \log \omega(G),\label{pf-lemma-cli-upper-bound-3}
\end{align}where the equality \eqref{pf-lemma-cli-upper-bound-0} follows from
 the fact that
\begin{equation*}
P_{ZW}(z,w)=0,\quad~\forall~(z,w)\in V\times\Omega(G)\;\text{with}\; z\notin w
\end{equation*}
by the condition $Z\in W\in\Omega(G)$ (cf.~\eqref{cond-X-in-W-Omega(G)});
the inequality~\eqref{pf-lemma-cli-upper-bound-1} follows from the cardinality bound, i.e.,
 $$\sum_{z\in w}P_{Z|W}(z|w)\cdot\log \frac{1}{P_{Z|W}(z|w)}\leq \log|w|;$$
and the inequality in~\eqref{pf-lemma-cli-upper-bound-3} follows from the definition of the clique number
$\omega(G)$. With this, we note that $H(Z|W)\leq \log \omega(G)$ is true for all random variables~$W$
  satisfying the condition $Z\in W\in\Omega(G)$.
The lemma is thus proved.

\section{Proof of~Lemma~\ref{lemma:clique_entropy_grouping_property}}
\label{appendix:pf-clique_entropx_grouping_property}

Consider the  probabilistic graph $(G,Z)$, and use  $G^{\textup{c}}$ to  denote the complement graph of $G$. We can readily see that  $(G^{\textup{c}},Z)$ is also a probabilistic graph, and a vertex subset $U\subseteq V$  is autonomous in $G$ if and only if
 $U$  is autonomous in $G^{\textup{c}}$. Let  $U\subseteq V$ be such an autonomous vertex subset in~$G$ and $u$ be a new  vertex not in $V$.
 We can verify that
  \begin{align}
  \big(G|_{ U \rightarrow u}\big)^{\textup{c}}=G^{\textup{c}}|_{ U \rightarrow u}\quad\text{and}\quad
  (G|_{ U })^{\textup{c}}=G^{\textup{c}}|_{ U }.\label{eq:clique_entropx_grouping_propertx_1-1}
  \end{align}
Together with the substitution lemma (cf.~Lemma~\ref{lemma:graph_entropy_grouping_property}), we have
\begin{align}
&H_{\kappa}(G^{\textup{c}},Z)=H_{\kappa}\big(G^{\textup{c}}|_{ U \rightarrow u},Z|_{ U \rightarrow u}\big)+P_{Z}( U )\cdot H_{\kappa}\big(G^{\textup{c}}|_{ U },Z|_{ U }\big)\nonumber\\
&=H_{\kappa}\big((G|_{ U \rightarrow u})^{\textup{c}},Z|_{ U \rightarrow u}\big)+P_{Z}( U )\cdot H_{\kappa}\big((G|_{ U})^{\textup{c}},Z|_{ U}\big)\label{eq:clique_entropx_grouping_propertx_2-1}\\
&=H(Z|_{ U \rightarrow u})-
H_{\omega}\big(G|_{ U \rightarrow u},Z|_{ U \rightarrow u}\big)+P_{Z}( U )\cdot \big(H(Z|_{ U})-
H_{\omega}(G|_{ U},Z|_{ U})\big)\label{eq:clique_entropx_grouping_propertx_2-2}\\
&=H(Z|_{ U \rightarrow u})+P_{Z}( U )\cdot H(Z|_{ U})-H_{\omega}\big(G|_{ U \rightarrow u},Z|_{ U \rightarrow u}\big)-P_{Z}( U )\cdot H_{\omega}
(G|_{ U },Z|_{ U }),\label{eq:clique_entropx_grouping_propertx_2-3}
\end{align}
where the equality~\eqref{eq:clique_entropx_grouping_propertx_2-1} follows from \eqref{eq:clique_entropx_grouping_propertx_1-1} and the equality~\eqref{eq:clique_entropx_grouping_propertx_2-2} follows from Lemma~\ref{lemma:graph_clique_entropy_relation}. Further, we claim that
\begin{align}
H(Z|_{ U \rightarrow u})+P_{Z}( U )\cdot H(Z|_{ U})=H(Z)\label{eq:clique_entropx_grouping_propertx_3}
\end{align}
due to the grouping property of the notion of entropy. To be specific, we have
 \begin{align*}
 H(Z)&=\sum_{z\in V}-P_Z(z)\cdot\log P_Z(z)=\sum_{z\in V\setminus U}-P_Z(z)\cdot\log P_Z(z)+\sum_{z\in U}-P_Z(z)\cdot\log P_Z(z)\\
 &=\Bigg[\sum_{z\in V\setminus U}-P_Z(z)\cdot\log P_Z(z)-P_{Z}( U )\cdot \log P_{Z}( U )\Bigg]+P_{Z}( U )\cdot\Bigg[\,\sum_{z\in U}-\frac{P_Z(z)}{P_{Z}( U )}\cdot \log \frac{P_Z(z)}{P_{Z}( U )}\Bigg]\\
 &=H\big(Z|_{ U \rightarrow u}\big)+P_{Z}( U )\cdot H\big(Z|_{ U }\big),
 \end{align*}
where the last equality follows by recalling the probability distributions of the random variables $Z|_{ U \rightarrow u}$ and $Z|_{ U }$ in \eqref{def:random-var-X_U-u} and \eqref{def:random-var-X_U}, respectively.
Combining \eqref{eq:clique_entropx_grouping_propertx_3} with \eqref{eq:clique_entropx_grouping_propertx_2-3}, we obtain that
\begin{align}
H_{\kappa}(G^{\textup{c}},Z)
&=H(Z)-H_{\omega}\big(G|_{ U \rightarrow u},Z|_{ U \rightarrow u}\big)-P_{Z}( U )\cdot H_{\omega}
(G|_{ U },Z|_{ U }).\label{eq:clique_entropx_grouping_propertx_3.2}
\end{align}

It further follows from  Lemma \ref{lemma:graph_clique_entropy_relation} that
\begin{align}
H_{\kappa}(G^{\textup{c}},Z)=H(Z)-H_{\omega}(G,Z).
\label{eq:clique_entropx_grouping_propertx_2.1}
\end{align}
Combining \eqref{eq:clique_entropx_grouping_propertx_3.2} and \eqref{eq:clique_entropx_grouping_propertx_2.1}, we thus obtain that
\begin{align*}
H(Z)-H_{\omega}(G,Z)=H(Z)-H_{\omega}\big(G|_{ U \rightarrow u},Z|_{ U \rightarrow u}\big)-P_{Z}( U )\cdot H_{\omega}
(G|_{ U },Z|_{ U }),
\end{align*}
namely that
\begin{align*}
H_{\omega}(G,Z)=H_{\omega}\big(G|_{ U \rightarrow u},Z|_{ U \rightarrow u}\big)+P_{Z}( U )\cdot H_{\omega}\big(G|_{ U },Z|_{ U }\big).
\end{align*}
We have thus  proved the lemma.

\section{Proof of~Lemma~\ref{lemma:AND-pro=sum-g-en}}
\label{appendix:pf-AND-pro=sum-g-en}
Consider the $k$ probabilistic graphs $(G_1,Z_1), \,(G_2,Z_2), \,\cdots,\, (G_k,Z_k)$.
We note that the random variable $\boldsymbol{Z}$ (defined by $(Z_{1},Z_{2},\cdots,Z_{k})^{\top}$)
takes all the vertices of the AND product $\boldsymbol{G}_\land\triangleq{\land}^{k}_{i=1}G_{i}$ (resp.~the OR product $\boldsymbol{G}_\vee\triangleq{\vee}^{k}_{i=1}G_{i}$) as values. This immediately implies that $(\boldsymbol{G}_\land,\boldsymbol{Z})$
(resp.~$(\boldsymbol{G}_\vee,\boldsymbol{Z})$) is also a probabilistic graph.
Together with the fact that $(G^{\textup{c}},Z)$ is a probabilistic graph if $(G,Z)$ is a probabilistic graph,
we obtain that $({\vee}^{k}_{i=1}G_{i}^{\textup{c}},\boldsymbol{Z})$
(resp.~$({\land}^{k}_{i=1}G_{i}^{\textup{c}},\boldsymbol{Z})$) is also a probabilistic graph.

By Lemma~\ref{lemma:graph_clique_entropy_relation}, we have
\begin{align}
H_{\omega}(\boldsymbol{G}_{\land},\boldsymbol{Z})
&=H(\boldsymbol{Z})-H_{\kappa}(\boldsymbol{G}_{\land}^{\textup{c}},\boldsymbol{Z})
=H(\boldsymbol{Z})-
H_{\kappa}({\vee}^{k}_{i=1}G_{i}^{\textup{c}},\boldsymbol{Z}),
\label{use-OR-pro-grp-entr=sum-g-en}
\end{align}
where $\boldsymbol{G}_{\land}^{\textup{c}}$ stands for the complement graph of $\boldsymbol{G}_{\land}$, i.e.,
$\boldsymbol{G}_{\land}^{\textup{c}}=\big({\land}^{k}_{i=1}G_{i}\big)^{\textup{c}}$,
 and the last equality in \eqref{use-OR-pro-grp-entr=sum-g-en} follows from the fact that $\big({\land}^{k}_{i=1}G_{i}\big)^{\textup{c}}={\vee}^{k}_{i=1}G_{i}^{\textup{c}}$ by the definitions of the AND product and OR product. By Lemma~\ref{lemma:OR-pro-grp-entr=sum-g-en}, since $Z_{i},1\leq i\leq k$ are mutually independent, we further have
\begin{align*}
H_{\kappa}\big({\vee}^{k}_{i=1}G_{i}^{\textup{c}},\boldsymbol{Z}\big)
=\sum_{i=1}^k H_{\kappa}(G_{i}^{\textup{c}},Z_i).
\end{align*}
With this, we continue \eqref{use-OR-pro-grp-entr=sum-g-en} and have
\begin{align}
H_{\omega}(\boldsymbol{G}_{\land},\boldsymbol{Z})&=H(\boldsymbol{Z})-\sum_{i=1}^k H_{\kappa}(G_{i}^{\textup{c}},Z_i)\nonumber
\\
&=H(\boldsymbol{Z})-\sum_{i=1}^k\Big(H(Z_i)- H_{\omega}(G_{i},Z_i)\Big)\label{eq:Rhs-eq1}\\
&=H(\boldsymbol{Z})-\sum_{i=1}^k H(Z_i)+ \sum_{i=1}^k H_{\omega}(G_{i},Z_i)\nonumber\\
&=\sum_{i=1}^k H_{\omega}(G_{i},Z_i),
\label{eq:Rhs-eq2}
\end{align}
where the equality  \eqref{eq:Rhs-eq1}  follows from Lemma~\ref{lemma:graph_clique_entropy_relation}, and the equality  \eqref{eq:Rhs-eq2}  follows from the condition that~$Z_{i},1\leq i\leq k$ are mutually independent. The lemma is proved.

\section{Proof of~Lemma~\ref{The Relation Between 123}}\label{appendix:pf-The Relation Between 123}
Let $C$ be a cut set  in $\Lambda(\mN)$ and  $\mP_C=\{C_1,C_2,\cdots, C_m\}$ be a strong partition of $C$. We let $I=I_C$, $J=J_C$, $I_{\ell}=I_{C_{\ell}}$ for $1\leq\ell\leq m,$  and  accordingly, $L=I\setminus (\cup^m_{\ell=1}I_{\ell})$ for notational simplicity.
Clearly, we have
 \begin{equation*}
V_{\mP_C}^{\land k}=V^{\,k}_{\mP_C} =\mathcal{A}^{k \times (I\cup J)}.
 \end{equation*}
  Hence, it remains to prove $
E_{\mP_C}^{\land k} \subseteq E^{\,k}_{\mP_C}.
$

Consider two arbitrary vertices  that are connected in $G_{\mP_C}^{\land k}$:
$$
(\boldsymbol{x}_{I},\boldsymbol{x}_{J})=\big(\{\boldsymbol{x}_{I_\ell},\,1\leq \ell\leq m\},\,\boldsymbol{x}_L,\,\boldsymbol{x}_J\big)\in \mathcal{A}^{k \times (I\cup J)}$$ and $$
(\boldsymbol{x}_{I}',\boldsymbol{x}_{J}')=\big(\{\boldsymbol{x}_{I_\ell}',\,1\leq \ell\leq m\},\,\boldsymbol{x}_L',\,\boldsymbol{x}_J'\big)\in \mathcal{A}^{k \times (I\cup J)}.
$$
We write $\boldsymbol{x}_{I}=(\boldsymbol{x}_{I,1},\boldsymbol{x}_{I,2},
\cdots,\boldsymbol{x}_{I,k})^\top$, and
 $\boldsymbol{x}_{I}'=(\boldsymbol{x}_{I,1}',\boldsymbol{x}_{I,2}',
 \cdots,\boldsymbol{x}_{I,k}')^\top$, where $\boldsymbol{x}_{I,p},\boldsymbol{x}'_{I,p}\in \mathcal{A}^{I}$ for $1\leq p\leq k$ are the rows of $\boldsymbol{x}_{I}$ and $\boldsymbol{x}_{I}'$, respectively. Similarly, we also write $\boldsymbol{x}_{J}=(\boldsymbol{x}_{J,1},\boldsymbol{x}_{J,2},\cdots,\boldsymbol{x}_{J,k})^\top$
and $\boldsymbol{x}_{J}'=(\boldsymbol{x}_{J,1}',\boldsymbol{x}_{J,2}',\cdots,\boldsymbol{x}_{J,k}')^\top$  with  $\boldsymbol{x}_{J,p},\boldsymbol{x}'_{J,p}\in \mathcal{A}^{J}$ for $1\leq p \leq k$.

 Further, for each  $1\leq p \leq k$, we write
 \begin{align*}
(\boldsymbol{x}_{I,p},\boldsymbol{x}_{J,p})=\big(\{\boldsymbol{x}_{I_\ell,p},\,1\leq \ell\leq m\},\,\boldsymbol{x}_{L,p},\,\boldsymbol{x}_{J,p}\big)\end{align*}
and
\begin{align*}
(\boldsymbol{x}_{I,p}',\boldsymbol{x}_{J,p}')=\big(\{\boldsymbol{x}_{I_\ell,p}',\,1\leq \ell\leq m\},\,\boldsymbol{x}_{L,p}',\,\boldsymbol{x}_{J,p}'\big).
\end{align*}
By the definition of the $k$-fold AND product $G_{\mP_C}^{\land k}$ of $G_{\mP_C}$, $(\boldsymbol{x}_{I},\boldsymbol{x}_{J})$ and $
(\boldsymbol{x}_{I}',\boldsymbol{x}_{J}')$ are connected in $G_{\mP_C}^{\land k}$
if and only
if $(\boldsymbol {x}_{I,p},\boldsymbol {x}_{J,p})\in \mathcal{A}^{I\cup J}$ and $(\boldsymbol {x}_{I,p}',\boldsymbol {x}_{J,p}')\in \mathcal{A}^{I\cup J}$, regarded as two vertices in  $G_{\mP_C}$, are connected in $G_{\mP_C}$ for all $1\leq p\leq k$ with  $(\boldsymbol {x}_{I,p},\boldsymbol {x}_{J,p})\neq (\boldsymbol {x}_{I,p}',\boldsymbol {x}_{J,p}')$.
 By the definition of the characteristic graph $G_{\mP_C}$, such two connected vertices $(\boldsymbol {x}_{I,p},\boldsymbol {x}_{J,p})$ and $(\boldsymbol {x}_{I,p}',\boldsymbol {x}_{J,p}')$ satisfy
$    \boldsymbol{x}_{J,p}=\boldsymbol{x}_{J,p}'
    $     and  one of the following two conditions:
\begin{enumerate}
   \item $\boldsymbol{x}_{I,p}$ and $\boldsymbol{x}_{I,p}'$ are not $(I,\boldsymbol{x}_{J,p})$-equivalent;
   \item $\boldsymbol{x}_{I,p}$ and $\boldsymbol{x}_{I,p}'$ are  $(I,\boldsymbol{x}_{J,p})$-equivalent, but
      $\boldsymbol{x}_{L,p}=\boldsymbol{x}_{L,p}'$ and  there exists an index $\ell$ for $1\leq\ell\leq m$ such that $\boldsymbol{x}_{I_\ell,p}$ and $\boldsymbol{x}_{I_\ell,p}'$ are not $(I_{\ell}, \boldsymbol{x}_{L,p}, \boldsymbol{x}_{J,p} )$-equivalent.
      \end{enumerate}

First, it follows from $\boldsymbol {x}_{J,p}=\boldsymbol {x}_{J,p}'$ for all $1\leq p\leq k$ that $\boldsymbol {x}_{J}=\boldsymbol {x}_{J}'.$ Since all different $(\boldsymbol {x}_{I,p},\boldsymbol {x}_{J,p})$ and $(\boldsymbol {x}_{I,p}',\boldsymbol {x}_{J,p}')$ for $1\leq p\leq k$ are connected in $G_{\mP_C}$,  we first consider the following case  according to the above condition 1).
 For the two vertices $(\boldsymbol {x}_{I},\boldsymbol {x}_{J})$ and $(\boldsymbol {x}_{I}',\boldsymbol {x}_{J}')$ (where $\boldsymbol {x}_{J}=\boldsymbol {x}_{J}'$), there exists  $p$ with $1\leq p\leq k$
   such that $\boldsymbol {x}_{I,p}$ and $\boldsymbol {x}_{I,p}'$ satisfy the condition 1).
Without loss of generality, we assume  that  $\boldsymbol{x}_{I,1}$ and $\boldsymbol {x}_{I,1}'$ are not $(I,\boldsymbol {x}_{J,1})$-equivalent.
  By  Definition~\ref{def:I_a_j_equiv'}, there exists  $d \in \mathcal{A}^{S/(I\cup J)}$ such that
    \begin{align}\label{graph-case-2-def-ineq_1}
    f(\boldsymbol {x}_{I,1},\boldsymbol {x}_{J,1},d) \neq f(\boldsymbol {x}_{I,1}',\boldsymbol {x}_{J,1},d).
    \end{align}
    We  take   $\boldsymbol{d}=(\boldsymbol{d}_1,\boldsymbol{d}_2,
    \cdots,\boldsymbol{d}_k)^\top\in \mathcal{A}^{k\times S/(I\cup J)}$ such that $\boldsymbol{d}_1=d$ and all other $\boldsymbol{d}_p\in \mathcal{A}^{S/(I\cup J)}$ for $2\leq p \leq k$ are arbitrary. Then,  the  equation~\eqref{graph-case-2-def-ineq_1} implies that 
  \begin{align*}
    f(\boldsymbol {x}_I,\boldsymbol {x}_J,\boldsymbol{d})=\big(f(\boldsymbol {x}_{I,p},\boldsymbol {x}_{J,p},\boldsymbol{d}_p), 1\leq p \leq k\big)^\top
     \neq \big(f(\boldsymbol {x}'_{I,p},\boldsymbol {x}_{J,p},\boldsymbol{d}_p), 1\leq p \leq k\big)^\top=f(\boldsymbol {x}_I',\boldsymbol {x}_J,\boldsymbol{d}),
    \end{align*}
    namely that, $\boldsymbol{x}_{I}$ and $\boldsymbol{x}'_{I}$ are not $(I,\boldsymbol{x}_{J})$-equivalent by Definition~\ref{def:k-shot-I_a_j_equiv'}. Together with Definition~\ref{def-char-graph-k}, we have proved that the two vertices  $(\boldsymbol {x}_I,\boldsymbol {x}_J)$ and $(\boldsymbol {x}_I',\boldsymbol {x}_J)$ are also connected in $G^{\,k}_{\mP_C}$.

Next, we consider the other case according to the condition 2).
     Against  the above case, for
   the two vertices $(\boldsymbol {x}_{I},\boldsymbol {x}_{J})$ and $(\boldsymbol {x}_{I}',\boldsymbol {x}_{J}')$ (where $\boldsymbol {x}_{J}=\boldsymbol {x}_{J}'$), we assume that $\boldsymbol {x}_{I,p}$ and $\boldsymbol {x}_{I,p}'$ are
   $(I,\boldsymbol{x}_{J,p})$-equivalent for all $1\leq p\leq k$.
    By the definition of $k$-fold AND product of a graph and the condition 2) mentioned above, for each $1\leq p\leq k$, either  $(\boldsymbol {x}_{I,p},\boldsymbol {x}_{J,p})=(\boldsymbol {x}_{I,p}',\boldsymbol {x}_{J,p}')$ or
      $\boldsymbol{x}_{L,p}=\boldsymbol{x}_{L,p}'$ and  there exists    $1\leq\ell\leq m$ such that $\boldsymbol{x}_{I_\ell,p}$ and $\boldsymbol{x}_{I_\ell,p}'$ are not $(I_{\ell}, \boldsymbol{x}_{L,p}, \boldsymbol{x}_{J,p} )$-equivalent.
       With this, we easily see that
        \begin{align*}
        \boldsymbol{x}_{L}=
        (\boldsymbol{x}_{L,1},\boldsymbol{x}_{L,2},\cdots,\boldsymbol{x}_{L,k})^\top
        =(\boldsymbol{x}_{L,1}',\boldsymbol{x}_{L,2}',\cdots,\boldsymbol{x}_{L,k}')^\top
        =\boldsymbol{x}_{L}'.
        \end{align*}
        Further, we assume without of generality that
  $\boldsymbol {x}_{I_1,1}$ and $\boldsymbol {x}_{I_1,1}'$ (i.e., $\ell=1$ and $p=1$) are not $(I_{1}, \boldsymbol {x}_{L,1}, \boldsymbol {x}_{J,1} )$-equivalent.
   By Definition~\ref{def:I_a_L_a_j_equiv'},
   there exist  $c_{I_j}\in \mathcal{A}^{I_j}$ for $2\leq j\leq m$ such that
   	\begin{align*}
   \big(\boldsymbol{x}_{I_{1},1}\,,\boldsymbol{x}_{L,1},\,\{c_{I_{j}},\,2\leq j\leq m\}\big)\in\mathcal{A}^{I}\quad\text{and}\quad
	\big(\boldsymbol{x}_{I_{1},1}'\,,\boldsymbol{x}_{L,1},\,\{c_{I_{j}},\,2\leq j\leq m\}\big)\in\mathcal{A}^{I}\end{align*}
		are not $(I, \boldsymbol{x}_{J,1})$-equivalent. By  Definition~\ref{def:I_a_j_equiv'}, there further exists  $d \in \mathcal{A}^{S/(I\cup J)}$ such that
\begin{align}\label{graph-case-2-def-ineq}
&f\big(\boldsymbol {x}_{I_1,1},\,\boldsymbol {x}_{L,1},\,\{c_{I_j},\,2\leq j\leq m\},\,\boldsymbol{x}_{J,1},\,d\big) \neq f\big(\boldsymbol {x}_{I_1,1}',\,\boldsymbol {x}_{L,1},\,\{c_{I_j},\,2\leq j\leq m\},\,\boldsymbol {x}_{J,1},\,d\big).
\end{align}
We  take   $\boldsymbol{d}=(\boldsymbol{d}_1,\boldsymbol{d}_2,
    \cdots,\boldsymbol{d}_k)^\top\in \mathcal{A}^{k\times S/(I\cup J)}$ such that $\boldsymbol{d}_1=d$ and all other $\boldsymbol{d}_p\in \mathcal{A}^{S/(I\cup J)}$, $2\leq p \leq k$ are arbitrary. Similarly, for each $2\leq j \leq m$, we take $\boldsymbol{c}_{I_j}=(\boldsymbol{c}_{I_j,1},\boldsymbol{c}_{I_j,2},
    \cdots,\boldsymbol{c}_{I_j,k})^\top\in \mathcal{A}^{k\times I_j}$ such that $\boldsymbol{c}_{I_j,1}=c_{I_j}$ and all other  $\boldsymbol{c}_{I_j,p}\in \mathcal{A}^{I_j}$ for $2\leq p \leq k$ are arbitrary.  Then,  the  equation~\eqref{graph-case-2-def-ineq} implies that
\begin{align*}
&f(\boldsymbol {x}_{I_1},\,\boldsymbol {x}_L,\,\{\boldsymbol{c}_{I_j},\,2\leq j\leq m\},\,\boldsymbol{x}_J,\,\boldsymbol{d})\\
&=\big(f(\boldsymbol {x}_{I_1,p},\,\boldsymbol {x}_{L,p},\,\{\boldsymbol{c}_{I_j,p},\,2\leq j\leq m\},\,\boldsymbol{x}_{J,p},\,\boldsymbol{d}_p),1\leq p\leq k\big)^\top\\
&\neq\big(f(\boldsymbol {x}'_{I_1,p},\,\boldsymbol {x}_{L,p},\,\{\boldsymbol{c}_{I_j,p},\,2\leq j\leq m\},\,\boldsymbol{x}_{J,p},\,\boldsymbol{d}_p),1\leq p\leq k\big)^\top\\
&= f\big(\boldsymbol {x}_{I_1}',\,\boldsymbol {x}_L,\,\{\boldsymbol{c}_{I_j},\,2\leq j\leq m\},\,\boldsymbol {x}_J,\,\boldsymbol{d}\big),
\end{align*}
 which shows that $\boldsymbol{x}_{I_1}$ and $\boldsymbol{x}_{I_1}'$ are not $(I_{1}, \boldsymbol{x}_{L}, \boldsymbol{x}_{J} )$-equivalent.
Together with  $\boldsymbol {x}_L=\boldsymbol {x}_L'$ and $\boldsymbol {x}_J=\boldsymbol {x}_J'$, we obtain that  $(\boldsymbol {x}_I,\boldsymbol {x}_J)$ and $(\boldsymbol {x}_I',\boldsymbol {x}_J')$ are connected in the $k$-fold characteristic graph $G^{\,k}_{\mP_C}$ by Definition~\ref{def-char-graph-k}.

Based on all the above discussions, we have shown that all connected two vertices
$(\boldsymbol {x}_I,\boldsymbol {x}_J)$ and $(\boldsymbol {x}_I',\boldsymbol {x}_J')$ in $G_{\mP_C}^{\land k}$ are also connected in the $k$-fold characteristic graph $G^{\,k}_{\mP_C}$, i.e.,
$E_{\mP_C}^{\land k} \subseteq E^{\,k}_{\mP_C}$.
The lemma is proved.

 \section{Specifying the Improved Lower Bound in Example~\ref{exam-3}}\label{appendix:calculation-of-lower-bound}

We now   specify the improved lower bound  in Theorem~\ref{thm-impro-low-bound} for the specific model $(\mathcal{N},X_S,f)$  in Example~\ref{exam-3}. For notational  consistency with Example~\ref{ex2}, we write the  $C=\{e_5,e_6\}$ and the strong partition $\mathcal{P}_C=\big\{C_1=\{e_5\},C_2=\{e_6\}\big\}$ to replace $C^*$ and $\mathcal{P}_{C^*}$ in \eqref{ex-outbound-1} and \eqref{ex-outbound-2}, respectively.
We still let $I\triangleq I_{C}=S=\{\sigma_1,\sigma_2,\sigma_3\}$, $J\triangleq J_{C}=\emptyset$, $I_{1}\triangleq I_{C_1}=\{\sigma_1\}$, $I_{2}\triangleq I_{C_2}=\{\sigma_3\}$,  and accordingly let $L\triangleq I\setminus(I_{1}\cup I_2)=\{\sigma_2\}$.

In the following,
we calculate the value of  $\textbf{\textup{H}}(G_{\mathcal{P}_C},X_{I\cup J})$ with the characteristic graph $G_{\mathcal{P}_C}$ as depicted  in Fig.\,\ref{fig:multi_char_G_XY} and $X_{I\cup J}=X_S=(X_1,X_2,X_3)$.
We consider
another vector of information sources $\widehat{X}_{I\cup J}=\widehat{X}_S=(\widehat{X}_1,\widehat{X}_2,\widehat{X}_3)$ with the support set  $\mathcal{A}^{I\cup J}=\mathcal{A}^{S}=\{0,1\}^3$  and $\widehat{X}_{I \cup J}\overset{\mathcal{P}_{C}}{\sim} X_{I\cup J}$, where the probability distribution is given by
  \begin{align} P_{\widehat{X}_S}(x_1,x_2,x_3)={\rm{Pr}}\big\{\widehat{X}_1=x_1,\widehat{X}_2=x_2,\widehat{X}_3=x_3\big\}
  \triangleq p_{x_1x_2x_3},
\quad\forall~x_i\in\{0,1\}~\textup{for }i= 1,2,3.\label{pro-eq1}
\end{align}
First, we  claim that $\widehat{X}_{I \cup J}\overset{\mathcal{P}_{C}}{\sim} X_{I\cup J}$ if and only if the following three conditions are satisfied:
\begin{subnumcases} {}
 p_{x_1x_2x_3}>0,  & $\forall~x_i\in\{0,1\}$ for $i= 1,2,3$,\label{ex3-p_{ijk}-conditions-1}\\
p_{x_1x_20}+p_{x_1x_21}=1/4, & $\forall~x_i\in\{0,1\}$ for $i= 1,2$,\label{ex3-p_{ijk}-conditions-2}\\
p_{0x_2x_3}+p_{1x_2x_3}=1/4,&$\forall~x_i\in\{0,1\}$ for $i= 2,3$,
\label{ex3-p_{ijk}-conditions-3}
\end{subnumcases}
To see this, we first assume that $\widehat{X}_{I \cup J}\overset{\mathcal{P}_{C}}{\sim} X_{I\cup J}$.

  \begin{itemize}
    \item Since the  support set of $\widehat{X}_{I \cup J}$ ($=\widehat{X}_{S}$) is $\mathcal{A}^{S}$, by \eqref{pro-eq1} we immediately obtain that
$p_{x_1x_2x_3}>0 $ for all $x_i\in\{0,1\}$, $i= 1,2,3.$ Then the condition~\eqref{ex3-p_{ijk}-conditions-1} is satisfied.

    \item  Recall that $J=\emptyset$, $I_{1}=\{\sigma_1\}$, $I_{2}=\{\sigma_3\}$,  and $L=\{\sigma_2\}$.
        By the definition of $\mathcal{P}_{C}$-equivalence (cf.~Definition~\ref{def:equivalent-pc}),
         $\widehat{X}_{I_{1}\cup L\cup J}=(\widehat{X}_1,\widehat{X}_2)$ and $X_{I_{1}\cup L\cup J}=(X_1,X_2)$ are identically distributed, and  also $\widehat{X}_{I_{2}\cup L\cup J}=(\widehat{X}_2,\widehat{X}_3)$ and $X_{I_{2}\cup L\cup J}=(X_2,X_3)$ are identically distributed.
   This implies that for any $x_1,x_2\in\{0,1\}$,
    \begin{align}
&p_{x_1x_20}+p_{x_1x_21}={\rm{Pr}}\big\{\widehat{X}_1=x_1,\widehat{X}_2=x_2,\widehat{X}_3=0
\big\}+{\rm{Pr}}\big\{\widehat{X}_1=x_1,\widehat{X}_2=x_2,\widehat{X}_3=1
\big\}\nonumber\\&= {\rm{Pr}}\big\{\widehat{X}_1=x_1,\widehat{X}_2=x_2\big\}=  {\rm{Pr}}\big\{X_1=x_1,X_2=x_2\big\}=\frac{1}{4},\label{cal-ex-eq20}
    \end{align}
    where the last  equality in \eqref{cal-ex-eq20} follows from the fact that
 $X_1,X_2,X_3$ are  i.i.d. random variables according to the uniform distribution over the alphabet $\mathcal{A}=\{0,1\}$.
Similarly, for any $x_2,x_3\in\{0,1\}$,
 \begin{align*}
 & p_{0x_2x_3}+p_{1x_2x_3} ={\rm{Pr}}\big\{\widehat{X}_1=0,\widehat{X}_2=x_2,\widehat{X}_3=x_3
\big\}+{\rm{Pr}}\big\{\widehat{X}_1=1,\widehat{X}_2=x_2,\widehat{X}_3=x_3
\big\}\\&= {\rm{Pr}}\big\{\widehat{X}_2=x_2,\widehat{X}_3=x_3\big\}=  {\rm{Pr}}\big\{X_2=x_2,X_3=x_3\big\}=\frac{1}{4}.
    \end{align*}
    Thus, the conditions \eqref{ex3-p_{ijk}-conditions-2} and \eqref{ex3-p_{ijk}-conditions-3} have been proved to be satisfied.
  \end{itemize}

  On the other hand, we assume that the three conditions \eqref{ex3-p_{ijk}-conditions-1}, \eqref{ex3-p_{ijk}-conditions-2} and \eqref{ex3-p_{ijk}-conditions-3} are satisfied.
  The condition \eqref{ex3-p_{ijk}-conditions-1} implies that $\widehat{X}_{I \cup J}$ has the same  support set $\mathcal{A}^{I\cup J}$ as $X_{I \cup J}$. The conditions \eqref{ex3-p_{ijk}-conditions-2} and \eqref{ex3-p_{ijk}-conditions-3} implies that
  \begin{align*}
 P_{\widehat{X}_{I_{1}\cup L\cup J}}(x_{I_1\cup L\cup J})
&=P_{\widehat{X}_1\widehat{X}_2}(x_1,x_2)=p_{x_1x_2 0}+p_{x_1x_2 1}=\frac{1}{4}
\\&=P_{X_1X_2}(x_1,x_2)=P_{X_{I_{1}\cup L\cup J}}(x_{I_1\cup L\cup J}),\quad\forall~
x_{I_1\cup L\cup J}=(x_1,x_2)\in\mathcal{A}^2,
\end{align*}
and similarly,
  \begin{align*}
 P_{\widehat{X}_{I_{2}\cup L\cup J}}(x_{I_2\cup L\cup J})
&=P_{\widehat{X}_2\widehat{X}_3}(x_2,x_3)=p_{0x_2x_3}+p_{1x_2x_3}=\frac{1}{4}
\\&=P_{X_2X_3}(x_2,x_3)=P_{X_{I_{2}\cup L\cup J}}(x_{I_2\cup L\cup J}),\quad\forall~
x_{I_2\cup L\cup J}=(x_2,x_3)\in\mathcal{A}^2.
\end{align*}
This thus  implies that $\widehat{X}_{I \cup J}\overset{\mathcal{P}_{C}}{\sim} X_{I\cup J}$.

Next, we calculate the clique entropy $H_{\omega}(G_{\mP_C},\widehat{X}_{I\cup J})$, i.e., $H_{\omega}(G_{\mP_C},\widehat{X}_S)$. Before the calculation,
   we write $G=(V,E)$ to replace $G_{\mP_C}=(V_{\mP_C},E_{\mP_C})$ for notational simplicity.
   We use the same  approach used to calculate $H_{\omega}(G,X_S)$ in Example~\ref{ex2} to calculate $H_{\omega}(G,\widehat{X}_S)$.
    We recall that all the four $(I,x_J)$-equivalence classes  $\Cl_i,i=1,2,3,4$  (cf.~\eqref{ex-equiv-class}) are  autonomous vertex subsets in $G$ and  constitute  a partition of $V_{\mathcal{P}_C}$ ($=\mathcal{A}^S$). Further, each $\Cl_i$ is completely-connected in $G$.
By
 Corollary~\ref{cor-clique-entropy-pro1}, we  calculate that
  \begin{align}
 H_{\omega}(G,\widehat{X}_S)&=\sum_{i=1}^4 P_{\widehat{X}_S}(\Cl_i)\cdot\Big[ H_{\omega}\big(G|_{\Cl_i},\widehat{X}_S|_{\Cl_i}\big)-\log P_{\widehat{X}_S}(\Cl_i)\Big]\nonumber\\
&=\sum_{i=1}^4 P_{\widehat{X}_S}(\Cl_i)\cdot H_{\omega}\big(G|_{\Cl_i},\widehat{X}_S|_{\Cl_i}\big)-\sum_{i=1}^4 P_{\widehat{X}_S}(\Cl_i)\cdot\log P_{\widehat{X}_S}(\Cl_i)\nonumber\\
&=P_{X_S}(\Cl_2)\cdot H_{\omega}\big(G|_{\Cl_2},\widehat{X}_S|_{\Cl_2}\big)+P_{X_S}(\Cl_3)\cdot H_{\omega}\big(G|_{\Cl_3},\widehat{X}_S|_{\Cl_3}\big)\nonumber
\\
&~~~~-\sum_{i=1}^4 P_{\widehat{X}_S}(\Cl_i)\cdot\log P_{\widehat{X}_S}(\Cl_i),\label{ex3-eq1-11}
 \end{align}
 where, similarly, the equality~\eqref{ex3-eq1-11} follows from $H_{\omega}\big(G|_{\Cl_1},\widehat{X}_S|_{\Cl_1}\big)
=H_{\omega}\big(G|_{\Cl_4},\widehat{X}_S|_{\Cl_4}\big)=0$ because $|\Cl_1|=|\Cl_4|=1$ and so $G|_{\Cl_1}$ and $G|_{\Cl_4}$ are two empty graphs (cf.~Proposition~\ref{prop:clique-entro-property}). We further calculate that
\begin{align*}
P_{X_S}(\Cl_2)=p_{001}+p_{010}+p_{100},\quad
P_{X_S}(\Cl_3)=p_{011}+p_{101}+p_{110},
\end{align*}
and
 \begin{align*}
 \sum_{i=1}^4 P_{\widehat{X}_S}(\Cl_i)\cdot\log P_{\widehat{X}_S}(\Cl_i)= &\; p_{000}\cdot\log p_{000}+(p_{001}+p_{010}+p_{100})\cdot\log(p_{001}+p_{010}+p_{100})\nonumber\\
&\; +(p_{011}+p_{101}+p_{110})\cdot\log(p_{011}+p_{101}+p_{110})+p_{111}\cdot\log p_{111}.
 \end{align*}
Then,
 \begin{align}
 H_{\omega}(G,\widehat{X}_S)
=&\;(p_{001}+p_{010}+p_{100})\cdot H_{\omega}\big(G|_{\Cl_2},\widehat{X}_S|_{\Cl_2}\big)+
(p_{011}+p_{101}+p_{110})\cdot H_{\omega}\big(G|_{\Cl_3},\widehat{X}_S|_{\Cl_3}\big)\nonumber\\
&\;-p_{000}\cdot\log p_{000}-(p_{001}+p_{010}+p_{100})\cdot\log(p_{001}+p_{010}+p_{100})\nonumber\\
&\;-
(p_{011}+p_{101}+p_{110})\cdot\log(p_{011}+p_{101}+p_{110})-p_{111}\cdot\log p_{111},\label{ex3-eq1-0011}
 \end{align}
where the equality \eqref{ex3-eq1-0011} follows from the same way to prove \eqref{ex-illstr-3} and~\eqref{ex-illstr-3.1}.

Next, we calculate the two clique entropies $H_{\omega}\big(G|_{\Cl_2},\widehat{X}_S|_{\Cl_2}\big) $ and $H_{\omega}\big(G|_{\Cl_3},\widehat{X}_S|_{\Cl_3}\big)$.
  We first consider the probabilistic  graph $(G|_{\Cl_2},\widehat{X}_S|_{\Cl_2})$, where the graph $G|_{\Cl_2}$ is the projection of $G$ onto~$\Cl_2$ as depicted in Fig.\,\ref{fig:multi_char_G__{Cl_2}},
  and the probability distribution of $\widehat{X}_S|_{\Cl_2}$ is given by
  \begin{align*}
P_{\widehat{X}_S|_{\Cl_2}}(x_1,x_2,x_3)=
\frac{P_{\widehat{X}_S}(x_1,x_2,x_3)}{P_{\widehat{X}_S}(\Cl_2)}&=
\frac{p_{x_1x_2x_3}}{p_{001}+p_{010}+p_{100}},\\
&\quad\forall~(x_1,x_2,x_3)\in \Cl_2=\big\{(0,0,1),(0,1,0),(1,0,0) \big\}.
\end{align*}
We recall in Example~\ref{ex2} that
in the graph $G|_{\Cl_2}$, we partition the vertex set $\Cl_2$ into
\begin{equation*}
\Cl_{2,0}\triangleq\{(0,0,1),~(1,0,0)\}\quad\text{ and }\quad\Cl_{2,1}\triangleq\{(0,1,0)\},
\end{equation*}
where $\Cl_{2,0}$ and $\Cl_{2,1}$  are disjoint,  autonomous, and isolated in $G|_{\Cl_2}$ (See  Fig.\,\ref{fig:multi_char_G__{Cl_2}}).
It thus follows from Corollary~\ref{cor-clique-entropy-pro1} that
\begin{align}
&H_{\omega}\big(G|_{\Cl_2},\widehat{X}_S|_{\Cl_2}\big)\nonumber\\
&= P_{\widehat{X}_S|_{\Cl_2}}\big(\Cl_{2,0}\big)\cdot H_{\omega}\big(G|_{\Cl_2}|_{\Cl_{2,0}},\widehat{X}_S|_{\Cl_2}|_{\Cl_{2,0}}\big)
+P_{\widehat{X}_S|_{\Cl_2}}\big(\Cl_{2,1}\big)\cdot H_{\omega}\big(G|_{\Cl_2}|_{\Cl_{2,1}},\widehat{X}_S|_{\Cl_2}|_{\Cl_{2,1}}\big)
\nonumber\\
&=P_{\widehat{X}_S|_{\Cl_2}}\big(\Cl_{2,0}\big)\cdot H_{\omega}\big(G|_{\Cl_{2,0}},\widehat{X}_S|_{\Cl_{2,0}}\big)
+P_{\widehat{X}_S|_{\Cl_2}}\big(\Cl_{2,1}\big)\cdot H_{\omega}\big(G|_{\Cl_{2,1}},\widehat{X}_S|_{\Cl_{2,1}}\big)
\label{ex3-illstr-4.1-00}\\
&=P_{\widehat{X}_S|_{\Cl_2}}\big(\Cl_{2,0}\big)\cdot H_{\omega}\big(G|_{\Cl_{2,0}},\widehat{X}_S|_{\Cl_{2,0}}\big)
\label{ex3-illstr-4.1-0}\\
&=P_{\widehat{X}_S|_{\Cl_2}}\big(\Cl_{2,0}\big)\cdot H\big(\widehat{X}_S|_{\Cl_{2,0}}\big)\label{ex3-illstr-4.1}\\
&=\frac{p_{001}+p_{100}}{p_{001}+p_{010}+p_{100}}\cdot
\bigg(\frac{-p_{001}}{p_{001}+p_{100}}\cdot\log \frac{p_{001}}{p_{001}+p_{100}}+\frac{-p_{100}}{p_{001}+p_{100}}\cdot\log \frac{p_{100}}{p_{001}+p_{100}}\bigg)\nonumber\\
&=\frac{-p_{001}}{p_{001}+p_{010}+p_{100}}\cdot\log \frac{p_{001}}{p_{001}+p_{100}}+\frac{-p_{100}}{p_{001}+p_{010}+p_{100}}\cdot\log \frac{p_{100}}{p_{001}+p_{100}},\label{ex3-illstr-4.2}
\end{align}
where the equalities \eqref{ex3-illstr-4.1-00}, \eqref{ex3-illstr-4.1-0} and \eqref{ex3-illstr-4.1}  follow from the same way to prove \eqref{ex-illstr-4.1-00}, \eqref{ex-illstr-4.1-0} and \eqref{ex-illstr-4.1} in Example~\ref{ex2}.
By the same argument, we   also calculate
\begin{align}
&H_{\omega}\big(G|_{\Cl_3},\widehat{X}_S|_{\Cl_3}\big)\nonumber\\&=\frac{-p_{011}}{p_{011}+p_{101}+p_{110}}\cdot\log \frac{p_{011}}{p_{011}+p_{110}}+\frac{-p_{110}}{p_{011}+p_{101}+p_{110}}\cdot\log \frac{p_{110}}{p_{011}+p_{110}}.
\label{ex3-illstr-5.1}
\end{align}
Combining  \eqref{ex3-illstr-4.2} and \eqref{ex3-illstr-5.1} with  \eqref{ex3-eq1-11}, we obtain that
 \begin{align*}
 H_{\omega}(G,\widehat{X}_S)=&-p_{001}\cdot\log \frac{p_{001}}{p_{001}+p_{100}}-p_{100}\cdot\log \frac{p_{100}}{p_{001}+p_{100}}\nonumber\\
&-p_{011}\cdot\log \frac{p_{011}}{p_{011}+p_{110}}-p_{110}\cdot\log \frac{p_{110}}{p_{011}+p_{110}}\nonumber\\
&-p_{000}\cdot\log p_{000}-(p_{001}+p_{010}+p_{100})\cdot\log(p_{001}+p_{010}+p_{100})\nonumber\\
&-
(p_{011}+p_{101}+p_{110})\cdot\log(p_{011}+p_{101}+p_{110})-p_{111}\cdot\log p_{111}.
 \end{align*}

Now, we can consider
\begin{align}
\textbf{\textup{H}}(G,X_{I\cup J})&=\max_{\substack{\text{all}~\widehat{X}_{I \cup J}~\text{$\mathcal{P}_{C}$-equivalent to $X_{I\cup J}$}}}H_{\omega}\big(G_{\mathcal{P}_C},\widehat{X}_{I\cup J}\big)\nonumber\\&=\max_{\substack{\text{all}~p_{x_1x_2x_3},\;x_1,x_2,x_3\in\{0,1\}\\[1mm]\text{that satisfy the conditions \eqref{ex3-p_{ijk}-conditions-1}-\eqref{ex3-p_{ijk}-conditions-3}}}}
\;H_{\omega}(G,\widehat{X}_{S})\label{ex3-illstr-11-2}\\
&=\log5,\label{ex3-illstr-11-3}
\end{align}
where the  equality  \eqref{ex3-illstr-11-2} follows from
 the claim immediately below \eqref{pro-eq1}, and with the aid of a computer, the maximum value is $\log5$ and can be achieved by taking
\begin{align*}
p_{000}=p_{010}=p_{101}=p_{111}=0.1\quad\text{and}\quad p_{001}=p_{011}=p_{100}=p_{110}=0.15.
\end{align*}

In fact, the pair $(C,\mathcal{P}_C)$ discussed here achieves
the maximum of the right hand side of  \eqref{thm2-eq}
over all pairs in $\Lambda(\mathcal{N})\times \textbf{\textup{P}}_C$, i.e.,
\begin{align*}
\max_{(C,\mathcal{P}_C)\in\Lambda(\mathcal{N})\times \textbf{\textup{P}}_C}
\frac{~\textbf{\textup{H}}(G_{\mathcal{P}_C},X_{I_C\cup J_C})~}{|C|}=\frac{1}{2}\log5.
\end{align*}
We have already specified
the improved lower bound for the model  $(\mathcal{N},X_S,f)$ as follows:
\begin{align*}
\mathcal{C}(\mathcal{N},X_S,f)\geq\frac{1}{2}\log5.
\end{align*}

\section{Proof of Theorem~\ref{ncpc=w(Gpc)}}\label{pf-thm-ncpc=w(Gpc)}
Consider a cut set $C\in\Lambda(\mathcal{N})$ and a strong partition $\mathcal{P}_{C} = \{C_{1}, C_{2}, \cdots, C_{m}\}$ of $C$.
For notational simplicity, we similarly
let $I=I_{C}$, $J=J_{C}$ and $I_{\ell}=I_{C_{\ell}}$ for $1\leq\ell \leq m$, and accordingly let $L=I\setminus(\cup_{\ell=1}^{m} I_{\ell})$.
We further write $G=(V,E)$ to replace the characteristic graph $G_{\mathcal{P}_C}=(V_{\mathcal{P}_C},E_{\mathcal{P}_C})$, and use $\mathcal{M}\,a_J$ to stand for the set $\big\{(x_I,a_J):\,\text{all}~x_I\in \mathcal{M}\big\}$ for a subset $\mathcal{M}\subseteq\mathcal{A}^I$ and  a vector of source messages $a_J \in \mathcal{A}^J$.

First, we note  that in the characteristic graph $G$, $V=\mathcal{A}^{I\cup J}$ with $I\cap J=\emptyset$. Then, $\mathcal{A}^I a_J$ for all $a_{J}\in\mathcal{A}^J$  constitute  a partition of the vertex set $V=\mathcal{A}^{I\cup J}$, i.e.,
 \begin{align*}
 \bigcup_{a_J\in\mathcal{A}^J} \mathcal{A}^I a_J=\mathcal{A}^{I\cup J}.
 \end{align*}
 By the definition of the characteristic graph $G$ (cf.~Definition~\ref{def-char-graph-k=1}), it is not difficult to see that
 for each $a_{J}\in\mathcal{A}^J$, the vertex subset $\mathcal{A}^I  a_J$
is autonomous and isolated in $G$.
This thus implies that after operations of  replacing each vertex subset $\mathcal{A}^Ia_J$  by a new-added vertex  $u_{a_J}$, the resulting graph
$$G|_{\mathop{\uplus}\limits_{a_J\in \mathcal{A}^J} \mathcal{A}^I a_J\to u_{a_J}}$$
is an empty graph. Hence, we obtain that
\begin{align}\label{clique-entropy-comp-1}
\omega(G)=&\max_{a_J\in\mathcal{A}^J}\omega\big(G|_{\mathcal{A}^I a_J}\big),
\end{align}
where we recall that $G|_{\mathcal{A}^I a_J}$ is the projection of $G$ onto $\mathcal{A}^I a_J$.

Next, we consider the graph $G|_{\mathcal{A}^I a_J}$ for a vector of source messages  $a_J\in\mathcal{A}^J$.
Since all
the $(I,{a}_{J})$-equivalence classes constitute  a partition of $\mathcal{A}^{I}$, $\mathrm{Cl}\, a_J$ for all
the $(I,{a}_{J})$-equivalence classes $\mathrm{Cl}$
constitute  a partition of $\mathcal{A}^I a_J$. We further see that  each vertex subset
$\mathrm{Cl}\,a_J$ is autonomous and completely-connected in $G|_{\mathcal{A}^I  a_J}$  because two vertices $(x_I,a_J)\in\mathrm{Cl}_1  a_J$ and $(x_I',a_J)\in\mathrm{Cl}_2  a_J$ for  two distinct $(I,{a}_{J})$-equivalence classes  $\mathrm{Cl}_1$ and $\mathrm{Cl}_2$  satisfy  the condition~1) in Definition~\ref{def-char-graph-k=1}.
 This thus implies that
$$G|_{\mathcal{A}^I a_J}|_{\mathop{\uplus}\limits_{\mathrm{Cl}\in \textup{\textbf{P}}_{I}(a_J)}\mathrm{Cl}\,a_J\to u_{\mathrm{Cl}}}$$
is a complete graph, where $\textup{\textbf{P}}_{I}(a_J)$ is the collection of all the $(I,{a}_{J})$-equivalence classes, and $u_{\mathrm{Cl}}$ is the new-added vertex associated with the $(I,{a}_{J})$-equivalence class $\mathrm{Cl}$.
  Then, we have
\begin{align}
&\omega(G|_{\mathcal{A}^I a_J})=\sum_{\mathrm{Cl}\in\textup{\textbf{P}}_{I}(a_J)}
\omega\big(G|_{\mathcal{A}^I  a_J}|_{\mathrm{Cl}\,a_J}\big)=\sum_{\mathrm{Cl}\in\textup{\textbf{P}}_{I}(a_J)}
\omega\big(G|_{\mathrm{Cl}\, a_J}\big),
\label{clique-entropy-comp-2}
\end{align}
where  the last equality  follows from the fact that
the projection of $G|_{\mathcal{A}^I a_J}$  onto $\mathrm{Cl}\,a_J$ is the same as the projection of $G$ onto   $\mathrm{Cl}\,a_J$, i.e., $G|_{\mathcal{A}^I  a_J}|_{\mathrm{Cl}\,a_J}=G|_{\mathrm{Cl}\, a_J}$. Combining \eqref{clique-entropy-comp-1} and \eqref{clique-entropy-comp-2}, we immediately obtain that
\begin{align}
\omega(G)=\max_{a_J\in\mathcal{A}^J}\sum_{\mathrm{Cl}\in\textup{\textbf{P}}_{I}(a_J)}
\omega\big(G|_{\mathrm{Cl}\, a_J}\big).\label{clique-entropy-comp-2-1}
\end{align}

 We further consider
 the graph $G|_{\mathrm{Cl}\,a_J}$ for an $(I,{a}_{J})$-equivalence class $\mathrm{Cl}$.
We first partition the vertex subset $\mathrm{Cl}\, a_J$ into $\mathrm{Cl}_{a_L} a_J$ for all $a_L\in\mathcal{A}^L$, where
\begin{align*}
\mathrm{Cl}_{a_L}\triangleq \Big\{(x_{I_1},\,x_{I_2},\,\cdots,\,x_{I_m},\,x_L)\in \mathrm{Cl}:~x_L=a_L\Big\}.
\end{align*}
Clearly, $\mathrm{Cl}_{a_L}\subseteq \mathrm{Cl}.$
It follows from Definition~\ref{def-char-graph-k=1}, in particular, the condition~2) that     the vertex subset $\mathrm{Cl}_{a_L} a_J$ for each $a_L\in\mathcal{A}^L$
is autonomous and isolated in $G|_{\mathrm{Cl}\, a_J}$.
 This implies that
 $$G|_{\mathrm{Cl}_{a_J}}|_{\mathop{\uplus}\limits_{a_L\in\mathcal{A}^L}
 \mathrm{Cl}_{a_L} a_J\to u_{a_L}}$$
 is an empty graph, where $u_{a_L}$ is the new added vertex associated with  $\mathrm{Cl}_{a_L} a_J$. With this, we obtain that
\begin{align}
\omega\big(G|_{\mathrm{Cl}\, a_J}\big)=
\max_{a_L\in\mathcal{A}^L}\omega\big(G|_{\mathrm{Cl}\, a_J}|_{\mathrm{Cl}_{a_L} a_J}\big)=
\max_{a_L\in\mathcal{A}^L}
\omega\big(G|_{\mathrm{Cl}_{a_L} a_J}\big).\label{clique-entropy-comp-3}
\end{align}
Combining \eqref{clique-entropy-comp-2-1} and \eqref{clique-entropy-comp-3}, we obtain that
\begin{align}
\omega(G)=\max_{a_J\in\mathcal{A}^J}\sum_{\mathrm{Cl}\in\textup{\textbf{P}}_{I}(a_J)}
\max_{a_L\in\mathcal{A}^L}
\omega\big(G|_{\mathrm{Cl}_{a_L} a_J}\big).\label{clique-entropy-comp-3-1}
\end{align}

In the following, we focus on the graph $G|_{\mathrm{Cl}_{a_L} a_J}$.
Before discussing further, we need two lemmas below, where the proof of the latter is deferred to the end of the appendix.
 \begin{lemma}[\!\!{\cite[Lemma~1]{Guang_Improved_upper_bound}}]\label{lemma:equiv-relat}
 For any set of $(I_{\ell}, a_L, a_J)$-equivalence classes $\mathrm{cl}_{I_{\ell}}$, $\ell = 1, 2, \cdots, m$, all source messages $(x_{I_1}, x_{I_2}, \cdots, x_{I_m}, a_L)$ in $\langle \mathrm{cl}_{I_1}, \mathrm{cl}_{I_2}, \cdots, \mathrm{cl}_{I_m}, a_L \rangle$ are $(I, a_J)$-equivalent.\footnote{Here, we recall \eqref{def-set-cl-a_L} that
 $
\langle\mathrm{cl}_{I_{1}},\mathrm{cl}_{I_{2}},\cdots,\mathrm{cl}_{I_{m}},a_L\rangle
\triangleq\big\{(x_{I_1},x_{I_2},\cdots,x_{I_m},a_L):~x_{I_\ell}
\in\mathrm{cl}_{I_\ell} \text{ for }1\leq \ell\leq m \big\}.$
 } In other words, there exists an $(I, a_J)$-equivalence class $\mathrm{Cl}$ such that
\begin{align*}
\langle \mathrm{cl}_{I_1}, \mathrm{cl}_{I_2}, \cdots, \mathrm{cl}_{I_m}, a_L \rangle \subseteq \mathrm{Cl}.
\end{align*}
\end{lemma}
\begin{lemma}\label{lemma4-cliq-entro-comp}
Consider an  $(I,{a}_{J})$-equivalence class $\mathrm{Cl}$ and a vector of source messages
$a_{L}\in\mathcal{A}^L$.
Then, all the sets $\langle\mathrm{cl}_{I_{1}},\mathrm{cl}_{I_{2}},\cdots,
\mathrm{cl}_{I_{m}},a_L\rangle$ with $\cl_{I_\ell}$ being an $(I_\ell, {a}_{L}, {a}_J)$-equivalence class for $1\leq \ell\leq m$  that satisfy
$\big\langle \cl_{I_1}, \cl_{I_2}, \cdots, \cl_{I_m}, {a}_{L} \big\rangle \subseteq \Cl$ constitute  a partition of $\mathrm{Cl}_{a_L}$,  denoted by $ \textup{\textbf{P}} $, i.e.,
\begin{align}
  \textup{\textbf{P}}\triangleq &\Big\{ \langle\mathrm{cl}_{I_{1}},\mathrm{cl}_{I_{2}},\cdots,\mathrm{cl}_{I_{m}},a_L\rangle :\
       \cl_{I_\ell} \text{ is an $(I_\ell, {a}_{L}, {a}_J)$-equivalence class, }
       1\leq \ell\leq m,\nonumber\\
&~~\qquad \qquad \qquad \qquad \qquad \quad \textrm{and}~\big\langle \cl_{I_1}, \cl_{I_2}, \cdots, \cl_{I_m}, {a}_{L} \big\rangle \subseteq \Cl \Big\}.\label{def:P(Cl_(x_L))}
\end{align}
\end{lemma}

With Lemma~\ref{lemma4-cliq-entro-comp}, we partition  the vertex set $\mathrm{Cl}_{a_L} a_J$ of $G|_{\mathrm{Cl}_{a_L} a_J}$ into $\boldsymbol{\mathrm{cl}}\, a_J$ for all blocks  $\boldsymbol{\mathrm{cl}}\triangleq\langle\mathrm{cl}_{I_{1}},\mathrm{cl}_{I_{2}},\cdots,\mathrm{cl}_{I_{m}},a_L\rangle$
of the partition $\textup{\textbf{P}}$ of $\mathrm{Cl}_{a_L}$. For a block $\boldsymbol{\mathrm{cl}}$, we similarly see that  the vertex subset
$\boldsymbol{\mathrm{cl}}\, a_J$ of the graph $G|_{\mathrm{Cl}_{a_L} a_J}$ is autonomous and completely-connected.
 This thus implies that
$$G|_{\mathrm{Cl}_{a_L} a_J}|_{\mathop{\uplus}\limits_{\boldsymbol{\mathrm{cl}}\in \textup{\textbf{P}}}\boldsymbol{\mathrm{cl}}\, a_J\to u_{\boldsymbol{\mathrm{cl}}}}$$
is a complete graph, where  $u_{\boldsymbol{\mathrm{cl}}}$ is the new-added vertex associated with the $(I_\ell, {a}_{L}, {a}_J)$-equivalence class~$\boldsymbol{\mathrm{cl}}$.
With this, we obtain that
\begin{align}
\omega\big(G|_{\mathrm{Cl}_{a_L} a_J}\big)=
\sum_{
\text{$\boldsymbol{\mathrm{cl}}\in\textup{\textbf{P}}$}}
\omega\big(G|_{\mathrm{Cl}_{a_L}  a_J}|_{\boldsymbol{\mathrm{cl}}\, a_J}\big)
&=\sum_{
\text{$\boldsymbol{\mathrm{cl}}\in\textup{\textbf{P}}$}}
\omega\big(G|_{\boldsymbol{\mathrm{cl}} \,a_J}\big),
\label{clique-entropy-comp-4}
\end{align}
where the last equality  follows from the fact that
the projection of $G|_{\mathrm{Cl}_{a_L}  a_J}$  onto $\boldsymbol{\mathrm{cl}}\,a_J$ is the same as the projection of $G$ onto   $\boldsymbol{\mathrm{cl}}\,a_J$, i.e., $G|_{\mathrm{Cl}_{a_L}  a_J}|_{\boldsymbol{\mathrm{cl}} \,a_J}=G|_{\boldsymbol{\mathrm{cl}} \,a_J}$.
Furthermore, we can readily see that $G|_{\boldsymbol{\mathrm{cl}}\,a_J}$ is an empty graph
for each $\boldsymbol{\mathrm{cl}}\in\textup{\textbf{P}}$ because any two vertices in $\boldsymbol{\mathrm{cl}}\,a_J$ do not satisfy the conditions of Definition~\ref{def-char-graph-k=1}. This immediately implies that
$
 \omega\big(G|_{\boldsymbol{\mathrm{cl}}\,a_J}\big)=1
$ for each block
$\boldsymbol{\mathrm{cl}}\in\textup{\textbf{P}}$.
 Together with \eqref{clique-entropy-comp-4}, we  obtain that
 \begin{align}
&\omega\big(G|_{\mathrm{Cl}_{a_L}  a_J}\big)=\sum_{
\text{$\boldsymbol{\mathrm{cl}}\in\textup{\textbf{P}}$}}
1=\big|\textup{\textbf{P}}\big|= N\big({a}_{L}, \Cl[{a}_J]\big),
\label{clique-entropy-comp-4.1}
\end{align}
where the last equality   follows because \eqref{no_finer_eq_cl} is the number of the set in \eqref{def:P(Cl_(x_L))}.

We  continue  considering the equality \eqref{clique-entropy-comp-3-1} that
\begin{align}
\omega(G)&=\max_{a_J\in\mathcal{A}^J}\sum_{\mathrm{Cl}\in\textup{\textbf{P}}_{I}(a_J)}
\max_{a_L\in\mathcal{A}^L}
\omega\big(G|_{\mathrm{Cl}_{a_L} a_J}\big)\nonumber\\
&=\max_{a_J\in\mathcal{A}^J}\sum_{\mathrm{Cl}\in\textup{\textbf{P}}_{I}(a_J)}
\max_{a_L\in\mathcal{A}^L}
N\big({a}_{L}, \Cl[{a}_J]\big)\label{clique-entropy-comp-5.1}\\
&=\max_{a_J\in\mathcal{A}^J}\sum_{\mathrm{Cl}\in\textup{\textbf{P}}_{I}(a_J)}
N\big(\Cl[{a}_J]\big)\label{clique-entropy-comp-5.2}\\
&=n_C(\mP_C),\label{clique-entropy-comp-5.3}
\end{align}
where the equalities \eqref{clique-entropy-comp-5.1}, \eqref{clique-entropy-comp-5.2} and \eqref{clique-entropy-comp-5.3} follows from \eqref{clique-entropy-comp-4.1}, \eqref{equ:N_Cl}  and
 \eqref{n_C_Parti_1st}, respectively.


To complete the proof of Theorem~\ref{ncpc=w(Gpc)}, it remains to prove Lemma~\ref{lemma4-cliq-entro-comp}.
\begin{IEEEproof}[Proof of Lemma~\ref{lemma4-cliq-entro-comp}]
Consider  the  $(I,{a}_{J})$-equivalence class $\mathrm{Cl}$ and the vector of source messages
$a_{L}\in\mathcal{A}^L$. First,  we can readily see that all the sets in $\textup{\textbf{P}}$ (cf.~\eqref{def:P(Cl_(x_L))}) are disjoint and
$ \bigcup_{\boldsymbol{\mathrm{cl}}\in\textup{\textbf{P}}}\boldsymbol{\mathrm{cl}}\subseteq \mathrm{Cl}_{a_L}.
$
To complete the proof, it suffices to prove that
$ \bigcup_{\boldsymbol{\mathrm{cl}}\in\textup{\textbf{P}}}\boldsymbol{\mathrm{cl}}\supseteq \mathrm{Cl}_{a_L}.$
We consider an arbitrary vector of source messages $x_I=(x_{I_1},\,x_{I_2},\,\cdots,\,x_{I_m},\,a_L)$ in $\mathrm{Cl}_{a_L}$. Clearly, $x_I\in \mathrm{Cl}$. Next, we will prove $x_I\in \boldsymbol{\mathrm{cl}}$ for some $\boldsymbol{\mathrm{cl}}\in \textup{\textbf{P}}$.
First, we recall that for each $1\leq \ell\leq m$, $\mathcal{A}^{I_\ell}$ can be partitioned into $(I_{\ell}, a_{L}, a_{J} )$-equivalence classes. Hence, $x_{I_\ell}$ is in some $(I_{\ell}, a_{L}, a_{J} )$-equivalence class, say~$\mathrm{cl}_{I_{\ell}}$, and then
\begin{align}
x_I=(x_{I_1},\,x_{I_2},\,\cdots,\,x_{I_m},\,a_L)\in  \boldsymbol{\mathrm{cl}}\triangleq\langle\mathrm{cl}_{I_{1}},\mathrm{cl}_{I_{2}},\cdots,\mathrm{cl}_{I_{m}},a_L\rangle.
\label{pflemma4-cliq-entro-comp-eq2}
 \end{align}
 By Lemma~\ref{lemma:equiv-relat},
there always exists an $(I, x_J)$-equivalence class $\mathrm{Cl}'$ such that
\begin{align}
\boldsymbol{\mathrm{cl}}=\langle \mathrm{cl}_{I_1}, \mathrm{cl}_{I_2}, \cdots, \mathrm{cl}_{I_m}, a_L \rangle \subseteq \mathrm{Cl}',\label{pflemma4-cliq-entro-comp-eq3}
\end{align}
and further $\boldsymbol{\mathrm{cl}}\subseteq \mathrm{Cl}'_{a_L}$.
Combining \eqref{pflemma4-cliq-entro-comp-eq2} and \eqref{pflemma4-cliq-entro-comp-eq3}, we  have
$
x_I\in \mathrm{Cl}'_{a_L}\subseteq\mathrm{Cl}'.
$ Together with $x_I\in\mathrm{Cl}_{a_L}\subseteq \mathrm{Cl}$, we immediately obtain that $\mathrm{Cl}=\mathrm{Cl}'$ and also $\mathrm{Cl}_{a_L}'=\mathrm{Cl}_{a_L}$. This thus implies that $\boldsymbol{\mathrm{cl}}\subseteq \mathrm{Cl}_{a_L}$, i.e.,~$\boldsymbol{\mathrm{cl}}\in \textup{\textbf{P}}$, and hence $
 x_I\in\bigcup_{\boldsymbol{\mathrm{cl}}\in\textup{\textbf{P}}}
 \boldsymbol{\mathrm{cl}}.
$
As such, we have proved
 $\mathrm{Cl}_{a_L}\subseteq\bigcup_{\boldsymbol{\mathrm{cl}}\in\textup{\textbf{P}}}\boldsymbol{\mathrm{cl}}$ and the lemma.
\end{IEEEproof}


\begin{thebibliography}{99}

\bibitem{Appuswamy11}
R.~Appuswamy, M.~Franceschetti, N.~Karamchandani, and K.~Zeger, ``Network coding for computing: cut-set bounds,''
 \textit{IEEE Trans. Inf. Theory}, vol.~57, no.~2, pp.~1015--1030, Feb.~2011.




\bibitem{Kowshik12}
H.~Kowshik and P.~Kumar, ``Optimal function computation in directed and undirected graphs,'' \textit{IEEE Trans. Inf. Theory}, vol.~58, no.~6, pp. 3407--3418, June 2012.

\bibitem{Appuswamy13}
R.~Appuswamy, M.~Franceschetti, N.~Karamchandani, and K.~Zeger, ``Linear codes, target function classes, and network computing capacity,''
\textit{IEEE Trans. Inf. Theory}, vol.~59, no.~9, pp.~5741--5753, Sept.~2013.


\bibitem{Ramamoorthy-Langberg-JSAC13-sum-networks}
A.~Ramamoorthy and M.~Langberg, ``Communicating the sum of sources over a network,''
\textit{IEEE J. Sel. Areas Commun.}, vol.~31, no.~4, pp.~655--665, April 2013.

\bibitem{Appuswamy14}
R.~Appuswamy and M.~Franceschetti, ``Computing linear functions by linear coding over networks,'' \textit{IEEE Trans. Inf. Theory}, vol.~60, no.~1, pp. 422--431, Jan. 2014.



\bibitem{Huang_Comment_cut_set_bound}
 C. Huang, Z. Tan, S. Yang, and X. Guang, ``Comments on cut-set bounds on network function computation,''
 \textit{IEEE Trans. Inf. Theory}, vol.~64, no.~9, pp.~6454--6459, Sept.~2018.






\bibitem{Tripathy-Ramamoorthy-IT18-sum-networks}
A.~Tripathy and A.~Ramamoorthy, ``Sum-networks from incidence structures: construction and capacity analysis,''
\textit{IEEE Trans. Inf. Theory}, vol.~64, no.~5, pp.~3461--3480, May~2018.

\bibitem{Guang_Improved_upper_bound}
X. Guang, R. W. Yeung, S. Yang and C. Li, ``Improved upper bound on the network function computing capacity,''
\textit{IEEE Trans. Inf. Theory}, vol.~65, no.~6, pp.~3790--3811, June~2019.




\bibitem{Li_Xu_vector_linear_diamond}
D. Li and Y. Xu, ``Computing vector-linear functions on diamond network,''
\textit{IEEE Commun. Lett.}, vol.~26, no.~7, pp.~1519--1523, July~2022.


\bibitem{Yao-Jafar-3user}
Y. Yao and S. A. Jafar, ``The capacity of 3 user linear computation broadcast,'' \textit{IEEE Trans. Inf. Theory}, vol.~70, no.~6, pp.~4414--4438, June~2024.

\bibitem{Guang_Zhang_Arithmetic_sum_Sel_Areas}
R. Zhang, X. Guang, S. Yang, X. Niu, and B. Bai, ``Computation of binary arithmetic sum over an asymmetric diamond network,''
\textit{IEEE J. Sel. Areas Inf. Theory}, vol.~5, pp.~585--596, 2024.



\bibitem{Shannon48}
C.~E.~Shannon, ``A mathematical theory of communication,''
\textit{Bell Syst. Tech. Journal}, 27: 379-423, 623--656, 1948.




\bibitem{Alon-Orlitskx_source_cod_graph_entropies}
N. Alon and A. Orlitsky, ``Source coding and graphs entropies,''
    \textit{IEEE Trans. Inf. Theory}, vol.~42, pp.~1329--1339, Sept.~1996.




\bibitem{Koulgi_zero-error-cod_cor_inf_sour}
P. Koulgi, E. Tuncel, S. Regunathan, and K. Rose, ``On zero-error coding of correlated sources,''
    \textit{IEEE Trans. Inf. Theory}, vol.~49, pp.~2856--2873, Nove.~2003.


\bibitem{Witsenhausen-IT-76}
H.~Witsenhausen, ``The zero-error side information problem and chromatic numbers,'' \textit{IEEE Trans. Inf. Theory}, vol.~22, pp.~592--593, Sept. 1976.

\bibitem{Orlitsky-Roche_general_side_inf_model_rat_reg}
A. Orlitsky and J. R. Roche, ``Coding for computing,''
     \textit{IEEE Trans. Inf. Theory}, vol.~47, no.~3, pp.~903--917, Mar.~2001.

\bibitem{Malak-fun-compre-side-infor}
D. Malak, ``Fractional graph coloring for functional compression with side information,''
 in \emph{Proc. IEEE Inf. Theory Workshop (ITW)}, Mumbai, India, 2022, pp.~750--755.



\bibitem{Slepian-Wolf-IT73}
D.~Slepian and J.~K.~Wolf, ``Noiseless coding of correlated information sources,'' \textit{IEEE Trans. Inf. Theory}, vol.~19, no.~4, pp.~471--480, July~1973.


\bibitem{Korner-Marton-IT73}
J.~K\"{o}rner and K.~Marton, ``How to encode the modulo-two sum of binary
sources (Corresp.)'', \textit{IEEE Trans. Inf. Theory}, vol.~25, no.~2, pp.~219--221,
Mar.~1979.


\bibitem{Doshi_fun_comp_graph_color_sch}
V. Doshi, D. Shah, M. M{\'e}dard, and M. Effros, ``Functional compression through graph coloring,''
    \textit{IEEE Trans. Inf. Theory}, vol.~56, no.~8, pp.~3901--3917, Aug.~2010.


\bibitem{Feizi-Medard}
S. Feizi and M. M{\'e}dard, ``On network functional compression,''
\textit{IEEE Trans. Inf. Theory}, vol.~60, no.~9, pp.~5387--5401, Sept.~2014.

\bibitem{Guang_Zero-Fun-Compre} X. Guang and  R. Zhang, ``Zero-error distributed compression of binary
arithmetic sum,'' \textit{IEEE Trans. Inf. Theory},
vol.~70, no.~5, pp. 3100--3117,
May 2024.

\bibitem{Korner_graph-entropx_def} J. K{\"o}rner, ``Coding of an information source having ambiguous alphabet and the entropy of graphs,'' in \textit{Proc. 6th Prague Conf on Information Theory}, 1973, pp. 411--425.

\bibitem{Korner-group-property-of_graph-entropy} J. K{\"o}rner, G. Simonyi, and Z. Tuza, ``Perfect couples of graphs,'' \textit{Combinatorica}, vol. 12, no. 2, pp. 179--192, 1992.


\bibitem{Ramamoorthy-vari-length-NFC16}
A.~Tripathy and A.~Ramamoorthy, ``On computation rates for arithmetic
sum,'' in \textit{Proc. IEEE Int. Symp. Inf. Theory}, Barcelona, Spain, 2016, pp.~2354--2358.


\bibitem{Ramamoorthy-vari-length-NFC18}
A. Tripathy and A. Ramamoorthy, ``Zero-error function computation on a directed acyclic network,'' in \emph{Proc. IEEE Inf. Theory Workshop (ITW)}, Guangzhou, China, 2018, pp. 1--5.

\bibitem{Ramamoorthy-vari-length-arxiv}
A. Tripathy and A. Ramamoorthy, ``Zero-error function computation on a directed acyclic network,'' \url{https://arxiv.org/pdf/1805.03730.pdf}


\bibitem{Dia-netw-1}
A. H. Kaspi and T. Berger, ``Rate-distortion for correlated sources with partially separated encoders,'' \textit{IEEE Trans. Inf. Theory}, vol.~28, no.~6, pp. 828--840, Nov. 1982.

\bibitem{Dia-netw-2}
B.~Schein, ``Distributed coordination in network information theory,'' Ph.D. dissertation, Massachusetts Inst. Technol., Cambridge, 2001.



\bibitem{Dia-netw-4}
W. Kang and S. Ulukus, ``Capacity of a class of diamond channels,'' \textit{IEEE Trans. Inf. Theory}, vol.~57, no.~8, pp.~4955--4960, Aug.~2011.








\end{thebibliography}
\end{document}